\documentclass[runningheads]{llncs}
\usepackage{graphicx}
\usepackage{hyperref}
\usepackage{caption}
\usepackage{microtype}
\usepackage{booktabs}
\usepackage{longtable} 
\usepackage{multirow} 
\usepackage{marvosym}
\usepackage{apacite}
\usepackage[section]{placeins}
\usepackage{amssymb}
\usepackage{pifont}
\newcommand{\cmark}{\ding{51}}

\usepackage{pdflscape}

\begin{document}
\title{Trust, distrust, and appropriate reliance in (X)AI: a survey of empirical evaluation of user trust
\thanks{We gratefully acknowledge funding from the Deutsche Forschungsgemeinschaft (DFG, German Research Foundation) for grant TRR 318/1 2021 – 438445824. \\ Parts of this article have been accepted for publication in the Springer CCIS series of the 1st xAI world conference (Lisbon, Portugal; 2023). A preprint version prior to the peer-review is available at https://arxiv.org/abs/2307.13601.}} 
%
\titlerunning{Trust, distrust, and appropriate reliance in (X)AI}
%
\author{Roel Visser\inst{1}\orcidID{0009-0006-3067-5545}$^{\dag}$ \and 
        Tobias M. Peters\inst{2}\orcidID{0009-0008-5193-6243}$^{\dag}$ \and
        Ingrid Scharlau\inst{2}\orcidID{0000-0003-2364-9489} \and
        Barbara Hammer\inst{1}\orcidID{0000-0002-0935-5591}}
\authorrunning{Visser et al.}
\institute{CITEC - Cognitive Interaction Technology\\Bielefeld University, 33619 Bielefeld, Germany\\
\email{\{rvisser, bhammer\}@techfak.uni-bielefeld.de}\\
\and 
Psychology\\Paderborn University, 33098 Paderborn, Germany\\
\email{\{tobias.peters, ingrid.scharlau\}@uni-paderborn.de}}
\maketitle              
\def\thefootnote{\dag}\footnotetext{These authors contributed equally to this work and share first authorship.}
\addtocounter{footnote}{1}
\def\thefootnote{\arabic{footnote}}
\begin{abstract}
A current concern in the field of Artificial Intelligence (AI) is to ensure the trustworthiness of AI systems. The development of explainability methods is one prominent way to address this, which has often resulted in the assumption that the use of explainability will lead to an increase in the trust of users and wider society. However, the dynamics between explainability and trust are not well established and empirical investigations of their relation remain mixed or inconclusive. 

In this paper we provide a detailed description of the concepts of user trust and distrust in AI and their relation to appropriate reliance. For that we draw from the fields of machine learning, human-computer interaction, and the social sciences.
Furthermore, we have created a survey of existing empirical studies that investigate the effects of AI systems and XAI methods on user (dis)trust.
With clarifying the concepts and summarizing the empirical investigations, we aim to provide researchers, who examine user trust in AI, with an improved starting point for developing user studies to measure and evaluate the user's attitude towards and reliance on AI systems. 

\end{abstract}
\newpage
\section{Introduction}

Intelligent systems and decision making supported by Artificial Intelligence (AI) are becoming ever more present and relevant within our everyday lives. Especially their use in high-stakes applications like medical diagnosis, credit scoring, and parole and bail decisions has led to concerns about the AI models \cite{rudin2019stop}. This includes concerns about the AI's transparency, interpretability, accountability, and fairness \cite{Guidotti2019, Arrieta2020, Mohseni2021}. 

These concerns, which are further enforced by the EU's General Data Protection Regulation (GDPR, Art. 15) increased the research interest in making AI systems more trustworthy and reliable. A number of different guidelines have been set out to ensure the trustworthiness of AI (for an overview see \citeA{Thiebes.2021}), which should help increase users’ or stakeholders’ trust.
One prominent way to approach this is for modern (blackbox) AI methods to be able to explain their outputs \cite{Arrieta2020}. This led to a surge in the development of explainable AI (XAI) over the last years for a host of different applications, domains, and data types \cite{Guidotti2019, Arrieta2020, Samek2021}. As a result, explainability is often considered as a means to increase user trust \cite{Kastner.2021}. Problematically, the dynamics between explainability and trust are far from being clarified, with both terms lacking precise definitions \cite{ferrario2022} and results from empirical investigations of their relation remaining mixed and inconclusive \cite{Kastner.2021}.

In the following, we approach these problems by, first, summarizing the insights from psychological trust research that are already being employed in some of the work related to automation, AI, and human-computer interaction and clarifying the involved terminology, which is partially based on the work covered in \citeA{peters2023importance}.
From these insights we conclude that both trust and distrust may be of similar relevance for the interaction between explainability and appropriate reliance.

Secondly, we give on overview of recent work that studies the question of trust in the AI and XAI context broadly, as well as works that perform empirical studies on the evaluation on the effects that (X)AI methods have on user trust. We provide an extensive summary of the typical application domains, (X)AI methods, and outcome measurements to evaluate the targeted AI objectives.
With these contributions we want to support future research on user trust and distrust in the (X)AI context by identifying important considerations for studying trust in AI, giving recommendations for setting up empirical studies, and identifying gaps in the current research.

\section{Trust in AI}

Trust in AI can be defined as an attitude that a stakeholder has towards a system \cite{Kastner.2021}, while trustworthiness is a property of the system that justifies to trust the system \cite{Toreini2020}. 
When looking at literature related to (X)AI, it is \textbf{important to make this distinction between trust and trustworthiness}.
In some cases trust and trustworthiness are not clearly differentiated, or rather used interchangeably.
For example, \citeA{BarredoArrieta.2020} describe trustworthiness as the confidence that a model will act as intended when facing a given problem, which is a fitting description of trust.
The differentiation is critical, because apart from the system’s trustworthiness there are other factors that can influence trust \cite{Toreini2020}.

According to research on trust between humans (so-called trustor and trustee), trustworthiness is characterized by the trustee's ability, i.e. competence or expertise in the relevant context, the trustee's benevolence towards the trustor, and the trustee's integrity towards principles that the trustor finds acceptable \cite{Mayer.1995}. 
A high level of these three factors of trustworthiness does not necessarily lead to trust, and trust can also go along with little trustworthiness \cite{Mayer.1995}.

For the AI context a meta-analysis \cite{Kaplan2023} identified the expertise and personality traits of a trustor interacting with AI as significant predictors for trust in AI.
Other relevant influences on trust, e.g., cultural differences, the type of technology, or perceived risk were identified by previous research and are discussed in Section \ref{sec:trustFactors}.
Overall, trustworthiness influences trust but does not fully determine it. 
Even the most trustworthy model will not be trusted in every case by every person. Vice versa, people may -- and often do -- trust an untrustworthy model \cite{Chen2023}.

In complex interactions in which multiple outcomes with varying truth are possible trust is essential for humans \cite{luhmann2009a}.
By trusting a person engages in the interaction as if only certain interpretations are possible (e.g., taking things at face value) and thus rendering the interaction less complex \cite{luhmann2009a}. 
In analogy, trust is also important in human-AI interactions, because of the involved risk caused by the complexity and non-determinism of AI \cite{Glikson.2020}. Similarly, \citeA{hoff2015trust} argue that trust is not only important to interpersonal relations but can also be defining for the way people interact with technology. In other words, you can trust an AI to be correct in its recommendation or prediction, even though the AI could err and you may not be able to comprehend or retrace the way the AI came to its outputs. 

The Integrative Model of Organizational Trust by \citeA{Mayer.1995} is a prominent basis for trust in AI and automation \cite{Stanton.2021}. \citeA{Mayer.1995} define trust as ``[...] the willingness of a party to be vulnerable to the actions of another party based on the expectation that the other will perform a particular action important to the trustor, irrespective of the ability to monitor or control that other party" \cite[p. 712]{Mayer.1995}. Based on this definition they differentiate between factors that contribute to trust, trust itself, the role of risk, and the outcomes of trust. 
The authors highlight that for trust to be of concern, the situation of an interaction must involve some form of risk and vulnerability for the person who trusts.
If there is no vulnerability involved, cooperation can occur without trust, and if there is no risk present, it is situation of confidence, not of trust \cite{Mayer.1995}. 

Drawing from \citeA{Mayer.1995}, definitions of trust in automation also consider the necessity of uncertainty (i.e., risk) \cite{hoff2015trust, lee2004trust} and vulnerability \cite{lee2004trust, Kohn.2021}. Trust in automated systems ``plays a leading role in determining the willingness of humans to rely on automated systems in situations characterized by uncertainty" \cite[p. 407]{hoff2015trust}, and is defined as ``[...] the attitude that an agent will help achieve an individual’s goals in a situation characterized by uncertainty and vulnerability" \cite[p. 54]{lee2004trust}. 
Trust in a system and trust defined by \citeA{Mayer.1995} share that they influence the willingness to rely and the situational requirements of risk and vulnerability for them to be of importance. \citeA{Muir1996} observed that people used automated systems they trust but not those they do not trust. \citeA{Lee1994} report that operators did not use automation systems if their trust in them was less than their own self-confidence.

Trust is dynamic and can increase or decrease during interactions \cite{guo2021modeling, Glikson.2020, hoffman2009dynamics}. Furthermore, trust is not only influenced by the interactions and their trajectories but also by other factors outside the interaction \cite{hoff2015trust}. \citeA{hoff2015trust} identified dispositional, situational and initially learned factors of trust that lie outside the interaction.
Such a distinction is especially relevant for the XAI context, as an explanation can be viewed as working on a micro-level (during an interaction) and including a macro-level (context of the interaction) \cite{Rohlfing.2021}.

To conclude, different perspectives on trust should be distinguished. One could be interested in trust in a certain decision, a certain type of interaction, a certain type of (X)AI, or in trust in AI in general. Thus, different time points to assess indicators of trust are conceivable. This entails repeated measurements of trust during multiple interactions and measurements before and after the interactions, while it can also be sensible to assess the attitude towards AI in general. These distinction are visible in the different types of assessment that are employed for measuring trust, which will be discussed in Section \ref{sec:trustMeasurement}.

\section{Appropriate reliance \& appropriate trust}
\textbf{Reliance in the AI context} can be understood as a human decision or action that takes into consideration the decision or recommendation of an AI.
Trust is an attitude that benefits the decision to rely, as it has a critical role for human reliance on automation \cite{hoff2015trust}. 
Ideally, one would only rely on an (automated) system if it is correct. Two types of problems can occur here, namely overtrust and disuse. Disuse describes the situation in which it would be correct to rely but one does not, and overtrust describes the situation in which it would be wrong to rely but one still does. Preventing both disuse and overtrust would ensure appropriate reliance.

The problems of disuse and overtrust are well discussed in the field of trust in automation. In XAI literature they are sometimes explicitly employed as well \cite{Jacovi.2021, Mohseni2021}. More often they are only implicitly targeted when aiming for and discussing appropriate trust (see Section \ref{sec:desideratumApproprTrust}), which is why in the following we first draw the connection to previous, more well-founded work on trust in automation.

In the field of trust in automation the prevention of disuse and overtrust has been targeted by ensuring appropriate trust or calibrated trust.
\citeA{McBride2010} as well as \citeA{McGuirl2006} define appropriate trust as the alignment between perceived and actual performance of an automated system. This relates to a user's ability to recognize when the system is correct or incorrect and adjust their reliance on it accordingly.
\citeA{Han2020} describe calibrated trust as the alignment between actual trustworthiness and user trust.
Within their model for trust calibration in human-robot teams, \citeA{Visser.2020} define calibrated trust as given when a team member's perception of trustworthiness of another team member matches the actual trustworthiness of that team member. 
If this is not given, either `undertrust', which leads to disuse, or `overtrust' can occur \cite{Visser.2020, Parasuraman1997}.

Trust calibration in \citeauthor{Visser.2020}'s sense aims to assure a healthy level of trust and to avoid unhealthy trust relationships.
Thereto, they establish a process of trust calibration which accompanies collaboration by establishing and continuously re-calibrating trust between the team members. To prevent people from overtrusting, so-called, trust dampening methods are to be applied.
According to the authors these methods are especially worthwhile in interactions with machines and robots, as humans have a tendency to expect too much from automation \cite{Visser.2020}. The authors recommend to present human with exemplary failures, performance history, likelihood alarms, or provide information about the system's limitations.
Moreover, they make the connection to the expanding field of XAI arguing that explanation activities can help with calibrating trust. 
This idea is present in much work on XAI, but with an emphasis on preventing disuse and a neglect of the mitigation of overtrust.

\subsection{Explainability and trust}

A multitude of XAI studies implicitly or explicitly assume explainability to facilitate, or increase, trust \cite{Kastner.2021, ferrario2022}.
In their summary of current XAI studies concerned with user trust, \citeA{Kastner.2021} call this the \textbf{explainability-trust hypothesis}.
By connecting explainability with the facilitation of trust authors focus on one of two utilities of explanations, namely the explanation's utility to indicate correct predictions.
Yet, explanations can also indicate false predictions and, thus, sometimes another utility of explanations is also identified, e.g., not trusting predictions \cite{Ribeiro2016}, critical reflection \cite{Ehsan2020}, or enabling distrust \cite{Jacovi.2021}.
This second utility of explanations is entertained by \citeA{Kastner.2021} when discussing potential reasons for the mixed results of empirical investigations of the explainability-trust hypothesis. They state that explanations could actually reveal problems of the system that may have otherwise gone unnoticed and could lead a user not to trust the AI. 

The utility of an explanation to reveal problems of an AI is also targeted in the paper on the LIME algorithm, one of the first popular XAI methods which has been used for deep models \cite{Ribeiro2016}.
According to \citeA{Ribeiro2016} explanations are not only helpful for deciding when one should trust a prediction, but also beneficial in identifying when not to trust one. Thereby, they differentiate between the explanation's utility for trusting and not trusting, demonstrating the latter in an example where an explanation reveals a wrong causal relation underlying an automated decision. Yet, when generally discussing the benefit of explanations, \citeA{Ribeiro2016} argue that ``[...] explaining predictions is an important aspect in getting humans to trust and use machine learning effectively, if the explanations are faithful and intelligible" (pp. 1135-1136). 
The aim of getting humans to trust sets the focus on the utility of explanations to identify correct predictions, while the utility of explanation to identify wrong predictions falls short.

\subsection{Desideratum of appropriate trust}
\label{sec:desideratumApproprTrust}

Even though the two utilities of explanations are identified in the literature, when speaking in broad terms explainability is connected to a facilitation of trust, and thus, the mitigation of disuse. More careful formulations can also be observed, which target appropriate trust instead \cite{Kastner.2021}. The added appropriateness acknowledges that it would not be correct to trust in every case but it lacks clarity on what this entails. Implied by the appropriateness of trust is that neither blind trust leading to overtrust, nor blind distrust leading to disuse is wanted. 

\begin{figure}[h!]
    \centering
    \includegraphics[width=\textwidth]{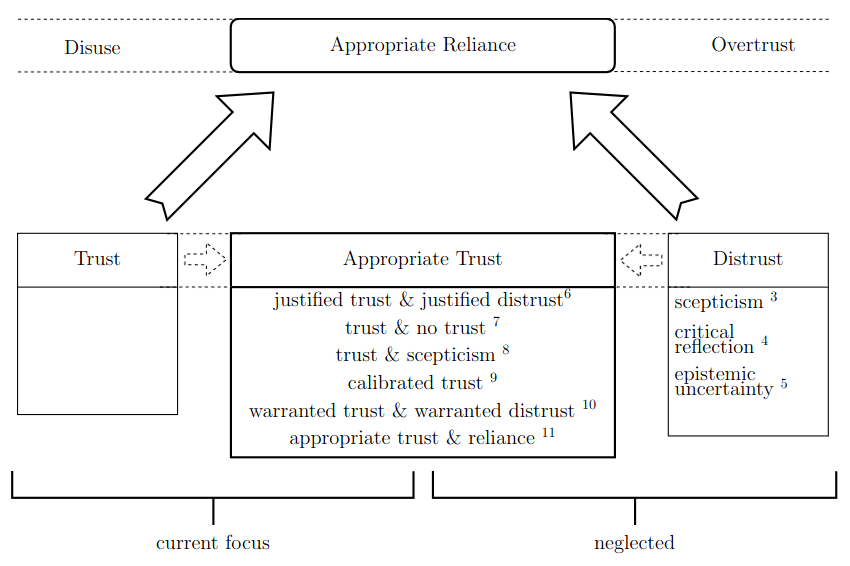}
    \caption{Desideratum of appropriate trust in AI and the relation of trust and distrust to appropriate reliance with the goal of preventing both disuse and overtrust.}
    \label{fig:approprTrust-Reliance}
\end{figure}

In other words, current improvements in automated systems, like XAI methods, are regarded as beneficial for \textbf{appropriate reliance} by preventing disuse and overtrust. Ideally, appropriate reliance should be achieved by fostering appropriate trust.  
Several formulations of this underlying notion of appropriate trust can be observed across the literature, which often entail trust and terms that can be summarized under distrust (see Fig. \ref{fig:approprTrust-Reliance}).

\def\thefootnote{3-12}\footnotetext{\citeA{Hoffman.2018}, \citeA{Ribeiro2016}, \citeA{Bansal2021}, \citeA{Visser.2020}, \citeA{Jacovi.2021}, \citeA{Gunning.2019}, \citeA{Gaube.2021}, \citeA{Ehsan2020}, \citeA{Jiang2022}}\def\thefootnote{\arabic{footnote}}

So, beneath the desideratum of increased appropriate trust lies the desideratum of increased appropriate reliance.
Yet, how can such an appropriate trust, its influencing factors, and the relation to appropriate reliance be conceptualized?
To rely appropriately one would consider correct decisions or recommendations of an AI and would disregard false ones. 
Trust does not lead to this because trust is not only influenced by the correctness, i.e. the performance of an AI. 
According to \citeA{hoff2015trust}, the performance of an automated system is similar to the trustee's ability in interpersonal trust, and the process and purpose of an automated system are analogous to benevolence and integrity. On top of that and as mentioned before, trust is not fully determined by trustworthiness.

\citeauthor{Mayer.1995}'s influential work on trust demonstrates the difference between trust and trustworthiness, but for the mitigation of overtrust their model does not provide a basis to proceed. 
Fostering trust, i.e., increasing the willingness to rely, mitigates the problem of disuse. However, for mitigating overtrust, not an absence of the willingness to rely, but the ability to identify reasons not to rely is needed. Trust in \citeauthor{Mayer.1995}'s model does not entail this, as they define trust “[…] irrespective of the ability to monitor or control that other party” (p. 712).
While it may be true that a lower willingness to rely, a lower trust, would decrease the likelihood of overtrust, it would also lead to less reliance overall. To this point, to mitigate overtrust reliance should be prevented if, and only if, it would be wrong to rely. 

More recent trust research highlights a distinction that might be of interest here. Several researchers provided evidence that trust and distrust are two related, yet separate dimensions \cite{Lewicki.1998, Vaske.2016, HarrisonMcKnight.2001, Benamati.2006}. A recent preprint also argues in favor of considering trust and distrust in the XAI context \cite{scharowski2023distrust}.
We assume that this separation of trust and distrust might help solving the conceptual issue and propose that \textbf{we need an understanding of both trust and distrust}. Psychological insights on the potential benefit of considering distrust will be detailed in the next section.

\section{Distrust as a separate dimension}
\label{sec:distrustSeparate}

While distrust is often regarded as the opposite of trust, the concept of a one-dimensional view of trust and distrust is being questioned and not widely accepted \cite{Schweer.2009, Lewicki.1998, Guo.2017, Schoorman.2007}.  
In the two-dimensional approach, by definition, low trust is not the same as high distrust, and low distrust is not equal to high trust \cite{Lewicki.1998}. This allows the coexistence of trust and distrust. 
Among others, trust is characterized by hope, faith, or assurance, and distrust by scepticism, fear, suspicion or vigilance \cite{Benamati.2006, Cho.2006, Lewicki.1998}. 

\citeA{Lewicki.1998} exemplify the separation of trust and distrust by contrasting low trust with high distrust. The authors regard expectations of beneficial actions being absent or present as antecedent to trust, and expectations of harmful actions being absent or present as antecedent to distrust. 
If the former is absent, low trust is expressed by a lack of hope, faith, and confidence. If the later is present, high distrust is expressed by vigilance, scepticism, and wariness. The combination of high trust and high distrust is described by the authors as a relationship in which opportunities are pursued while risks and vulnerabilities are monitored. 

When reviewing research that draws from two-dimensional approaches, concepts and terms like ``critical trust", ``trust but verify", and ``healthy distrust" are used \cite{Poortinga.2003, Lewicki.1998, Vaske.2016}.
These align well with the problem of mitigating overtrust. Yet, little consideration of the two-dimensional view on trust and distrust can be found when trust is considered in the technology context.

One, at least partial, reason for this is found in the field of organisational psychology. Even in this field, in which the conceptual critique towards the one-dimensional approach is the most visible, applied work still relies mostly on the model by \citeauthor{Mayer.1995} \cite{Vaske.2016}.
Discussing the trajectory of the conceptual debate on trust and distrust, \citeA{Vaske.2016} describes that most of the early work on trust are one-dimensional approaches. From the mid 80s onward these approaches were considered too simplistic.

Yet, efforts to resolve this debate and empirically test it remain scarce \cite{Rusk.2018}.
Instead of providing empirical evidence, work on the two-dimensional approach mostly reproduces common-sense assumptions \cite{Vaske.2016}. Moreover, only the concept trust is well researched and has a good theoretical background, while distrust remains in the state of conceptual debate and is receives little research attention \cite{Vaske.2016}.

The field of trust in XAI inherited this focus on trust and neglect of distrust, because it took prominent work on trust in automation \cite{lee2004trust, hoff2015trust} as a starting point \cite{Thiebes.2021}, which drew from \citeA{Mayer.1995}'s model.
Distrust is often connotated negatively \cite{Lewicki.1998, Vaske.2016} and sometimes explicitly considered something to be avoided \cite{Frison.2019, Seckler.2015}, or at least implied to be avoided. 
Yet considering the imperfection of contemporary ML models, distrust towards erroneous predictions and towards explanations that indicate them is not to be avoided, but fostered.  
Otherwise, a neglect of distrust remains, which is serious because it renders potential positive consequences of distrust invisible.

In a study by \citeA{McKnight.2004} the disposition to distrust predicted high-risk perceptions better than the disposition to trust did. For their study context of online expert advice sites they suggest that future research should study dispositional trust and also dispositional distrust.
By identifying positive consequences of distrust, psychological studies also point to the benefit of considering distrust.
Distrust or suspicion led, for example, to an increase of creativity \cite{Mayer.2011} or a reduction of the correspondence bias \cite{Fein.1996}. 
Moreover, a series of studies by \citeA{Posten.2021} showed an increase of memory performance. 
Finally, \citeA{Vaske.2016} identified a potential of distrust to improve critical reflection and innovation in the context of working in an organisational setting. 

Looking at potential underlying mechanisms of distrust, \citeauthor{Mayo.2015}'s \cite{Mayo.2015} review introduces a so-called distrust mindset as an explanation for the positive effects of distrust. The distrust mindset leads to an activation of incongruent and alternative associations, which aligns well with the increase of creativity, reflection, and innovation.
According to \citeA{Posten.2021}, trust triggers a perception focus on similarities that makes it harder to remember single entities. Distrust shifts the perception focus towards differences and, therefore, increases memory performance. 
Interestingly, in one of their studies \citeA{Posten.2021} observe a higher acceptance of misinformation in a trust condition, underlining the potential problem of the current trust focus in the (X)AI context and the danger of overtrust. 

A conceptual example of how trust and distrust can be targeted is provided by \citeA{Hoffman.2018} in their work on measuring trust in XAI.
They advocate that people experience a mixture of justified and unjustified trust, as well as justified and unjustified mistrust. Ideally, the user would develop the ability to trust the machine in certain task, goals, and problems, and also to appropriately distrust the machine in other tasks, goals, and problems.
This requires them to be able to decide when to trust or to correctly distrust, when scepticism is warranted.

In sum, although often connotated negatively, distrust also has positive consequences and merits separate from those of trust. Some work that considered both trust and distrust can be observed across different sub-fields of interaction with technology and will be summarised and discussed in Section \ref{sec:trustMeasurement}.

\section{Perceived versus actual trustworthiness}

As we discussed the trustworthiness of AI systems and the underlying models is a growing concern, leading to the need to develop trustworthy AI.
When designing models to be trustworthy some of the main concerns (from trustworthy AI) are model fairness, reliability and safety \cite{Mohseni2021}. Addressing each of these concerns poses challenges in their own right. 
To further complicate the matter of trustworthiness, users may perceive a system's trustworthiness incorrectly and may, therefore, undertrust, or disuse, trustworthy systems and overtrust untrustworthy ones \cite{Schlicker2021}.
To highlight this complication, \citeA{Schlicker2021} distinguish between actual and perceived trustworthiness and introduce the notion of cues, that the system can provide users, to bridge the gap between the two.

In general terms, the actual trustworthiness of an AI system encompasses the system's functionalities and capabilities, the underlying motivations and objectives of its developers, the conscientiousness of the development, design, and validation process, and the societal value ascribed to them \cite{hlegAI}.
The perceived trustworthiness on the other hand is based on the user's assessment of the systems actual trustworthiness. 
A system is perceived as trustworthy only in respect to its ability to perform a specific task within a specific context and at a specific point of time \cite{Schlicker2021}.
The user bases their assessment on their cognitive and affective evaluation of the system, which can include the fairness or ability of the system or the purpose for which it was developed \cite{Baer2018, lee2004trust, Madsen2000, Schlicker2021}.

With the term cues \citeA{Schlicker2021} group any observable piece of information that provides the user with insight on the actual trustworthiness of the system. The cues form an interface between the user's perceived trustworthiness of the system and its actual trustworthiness, where the degree of agreement between these two reflect the accuracy of the user's assessment of the system \cite{Schlicker2021}. The accuracy of the assessment of the actual trustworthiness depends on the ecological validity of the system cues and how the cues are applied by the user \cite{Schlicker2021}. 
It is not possible to observe the actual trustworthiness of an AI system directly, therefore users utilize the different cues to form their perceived trustworthiness \cite{Schlicker2021}.
For example, it is not possible to test and observe most models for all situations they might encounter.
If the user is able to perfectly assess the actual trustworthiness, they neither over- nor underestimate the system's capabilities \cite{Schlicker2021}.
When interacting with an AI system users are constantly searching for, using, and interpreting cues to assess its trustworthiness, which results in perceived trustworthiness \cite{Schlicker2021}.
As a result, users must continuously evaluate the system's actual trustworthiness to determine their own perceived trustworthiness of the system \cite{Schlicker2021}. 

\citeA{Chen2023} highlight the risk of overreliance on untrustworthy AI. They show a number of studies that suggest that explanations can increase users' reliance on AI predictions even if the AI system's predictions are wrong. 
This highlights the difficulties the user has in correctly calibrating between perceived trustworthiness, actual trustworthiness, and reliance on the system.
This is even more of a problem as the current understanding of user trust is limited. A common impulse, for example, is to nudge people into trusting the AI system (i.e. to find an explanation plausible and to accept it) rather than to distrust it \cite{Ehsan2020}. 
The importance of accurate assessment of actual trustworthiness and appropriate reliance is highlighted by the risks associated with the misleading of user with XAI. Several studies investigate or observe the misleading effects that explanations can have \cite{Eiband2019, Lakkaraju2020, Dimanov2020, Suresh2020}. 
This makes an exclusive focus on trust, rather than healthy scepticism and distrust, potentially dangerous, as a mismatch between actual and perceived trustworthiness can result in overtrust on the system.

\section{Human-centric evaluation of XAI}
\label{sec:human-centric-evaluation}

As we described, the development and application of explainability methods is one way to make AI systems more transparent and trustworthy.
The application of XAI can serve a number of purposes, namely it can be used to justify, to control and manage, to evaluate and improve, to discover and learn, or to calibrate trust \cite{adadi2018peeking, Meske2020, vanderWaa2021, Tintarev2012}.
XAI systems allow users to assess how reliable the system is and, therefore, to correctly calibrate their perception of the system's accuracy \cite{Mohseni2021} or, in other words, to calibrate their perceived trustworthiness of the system.
\cite{Schmidt2019} note that providing explanations to users more than doubled their productivity in annotation tasks, which indicates that they do not only serve to calibrate reliance but can also benefit other outcomes, such as productivity.

When developing explainability methods two main criteria need to be met. Firstly, we need to determine whether the developed method is mathematically and computationally sound and that it correctly represents the underlying model (i.e. does not come up with random or meaningless explanations that are not in line with the model). Secondly, we need to consider whether the method is having the originally intended effect on the user (e.g. providing them with meaningful insight into why the model came up with some output).

As explainability is an inherently human-centric property \cite{Lopes2022}, \cite{Mohseni2021} and \cite{Liao2021} believe that HCI and human-centric evaluation can contribute to solving XAI algorithms and applications' limitations.
\cite{Lopes2022} point out that most XAI methods are developed in a more technically, computationally, focused environment and, as a result, potentially useful contributions/insights from other fields, such as HCI, are often ignored. They note that the lack of a multidisciplinary approach is a potential pitfall for both the development and evaluation of XAI methods.
From their survey \cite{adadi2018peeking} found that of the works related to the development of interpretable ML only 5\% considered the evaluation of these methods and quantification of their relevance.

Another problem is that while many measurements exist for the computational evaluation of XAI (e.g. completeness, soundness, and fidelity of explanations) \cite{Zhou2021, Nauta2022, Mohseni2021, Lopes2022}, there is a clear lack of well-defined and validated measurements for the user-centric evaluation of XAI \cite{hoffman2018metrics}.
Given these facts, there is a need for properly defined metrics in order to be able meaningfully compare how explainable a model is \cite{Arrieta2020}. 

Common human-centric evaluation criteria that are used in the context of XAI are related to understanding, user's mental model, trust and reliance, satisfaction, usefulness and usability, performance, fairness \cite{Mueller2021, Mohseni2021, Lai2021}.
\cite{Vilone2020} distinguish two types of human-centric evaluation studies: qualitative and quantitative. Qualitative studies make use of open-ended questions aimed at obtaining deeper insights, while quantitative studies make use of close-ended questions that can be analysed in a statistical manner.
Similarly, \cite{Zhou2021} distinguish between subjective and objective metrics for human-centric evaluation of XAI.

The quality criteria of validity and reliability from the social sciences provide standards for scientifically sound user-centric evaluation of XAI \cite{vanderWaa2021}.
The validity of a method refers to its ability to accurately measure what it sets out to measure \cite{Field.2012}, which may be harmed by poor design, ill defined constructs, or arbitrarily selected measurements \cite{vanderWaa2021}.
Reliability refers to the extend whether a method's interpretation is consistent across different situations \cite{Field.2012},  
which may be harmed by a lack of documentation, application in an unsuitable use case, or noisy measurements \cite{vanderWaa2021}.
In order to be able to properly compare the results of different studies and experiments it is necessary that a user evaluation is both reliable and valid \cite{Joppe2000}. 
\cite{vanderWaa2021} state that this can be (partially) achieved by developing different types of measurements for common constructs, such as for example using self-reported subjective and behavioral measurements to measure task performance or trust.

In the development and usage of XAI systems several potential user types and various other stakeholders exist. Stakeholder that are not directly interacting with the AI system itself, still might have an impact on the design of the systems themselves. 
Some of the stakeholder are AI regulators or individuals that may be affected by the decision made by or based upon AI \cite{Meske2020}.
Different users have different objectives and levels of expertise and might, therefore, be impacted differently by different types of applications \cite{Liao2021}.
Which means that, the choice of a particular XAI method should be guided by the needs of the specific type of user \cite{Lopes2022, Liao2021}.
Already many ways exist to group and identify different user types and stakeholder in the context of XAI \cite{Mohseni2021, Liao2021, Meske2020}.
Still, one of the issues in existing research is that diverse and dynamic user objectives are often not explicitly considered when developing XAI algorithms \cite{Liao2021}.
Algorithms are often developed based on the intuitions of the AI researchers about what constitutes a good explanation rather than on the needs of the intended users \cite{Liao2021}.
While in our survey the user studies focus primarily on direct users of AI system, it is important to note that passive and indirect stakeholders should also be considered a type of user, something which is mostly ignored by current research on trust in AI \cite{Lukyanenko2022}. 
These indirect stakeholders are affected by the outcomes of AI and it may, therefore, be important to also study their perceptions of and views on AI systems.

In our survey we focus on the stakeholders, or user groups, that directly interact with an AI system across the different stages of the ML-lifecycle (from design and development to being used by actual end-users). This includes ML researchers, engineers and developers, domain experts, and end- or lay-users. We categorize the user types according to their objectives, in what stage of the ML development life-cycle they interact with the system (which informs their objectives), and their level of expertise. This led to the following grouping: 

\begin{itemize}
    \item Interacting stakeholders
    \begin{itemize}
        \item (X)AI Developers, Designers, ML experts
        \item Domain experts
        \item Lay-users
    \end{itemize}
    \item Non-user/Other stakeholders:
    \begin{itemize}
        \item Regulators/regulatory bodies
        \item Business owners or administators
        \item Impacted groups
    \end{itemize}
\end{itemize}

\section{Studying the effects of (X)AI on (user) trust}

In empirical research the calibration of user trust and distrust has two primary objectives: achieving appropriate reliance and improving task performance. Appropriate reliance in this case should increase task performance by causing people to neither overtrust nor disuse an AI system's outputs.
When studying the effects that (X)AI systems have on its users and their trust a number of different components need to be considered. 
Relevant questions to consider are: 
\begin{itemize} 
    \item How does the user interface, or interact, with the system?
    \item What factors can influence user trust?
    \item Which information can be provided to the user to affect their trust in the system?
    \item What ML and XAI methods can be used to provide such information?
    \item How to measure user trust (and other important evaluation criteria)?
    \item What constitutes a good outcome and what are the main objectives, e.g. aiming for maximizing user trust versus calibrating trust and distrust for fostering appropriate reliance? 
\end{itemize}

In order to give a comprehensive overview of the evaluation of the effects of XAI on user trust, we have performed a survey of both theoretical and empirical research on this subject.
For this we have developed a taxonomy of components and considerations that are important when studying the effects of (X)AI systems on user trust and provide an overview of existing work that has been done for each of these elements. Additionally, based on the survey of existing work we determine what the gaps and open questions are within the existing literature and provide a number of recommendations for each component.
With our survey we try to give a comprehensive overview of existing work that studies the effects that (X)AI methods can have on user trust. 
For the survey we collected for each of the paper the following information:
\begin{itemize}
    \item \textbf{Application domain \& type} the application domain(s) in which the experiments were performed (see Section~\ref{sec:contextual_factors_trust})
    \item \textbf{Main aims:} a summary of the main aims and objectives of the researchers.
    \item \textbf{(X)AI system} (description of relevant features of the AI system used in the experiments): 
    \begin{itemize}
        \item \textbf{ML methods:} the machine learning methods that were used in the studies (see Section~\ref{sec:model_descriptors})
        \item \textbf{ML model modulation:} if applicable, how the ML models were varied in order to observe that effects that had on users (e.g. variety of models with different overall accuracy/performance) (see Section~\ref{sec:model_descriptors})
        \item \textbf{(Model) descriptors:} the descriptors used in the experiments (e.g. showing uncertainty estimate or example-based local explanation to the user) (see Section \ref{sec:xai_system} and \ref{sec:model_descriptors} for more details)
        \item \textbf{Descriptors modulation:} if applicable, how the descriptors were applied or varied in the experiments (e.g. showing versus not showing an uncertainty estimate or local output explanation) (see Section~\ref{sec:model_descriptors})
        \item \textbf{Interface modality \& modulation:} the modality of the interface between the system and the user (e.g. image-based outputs and explanations, textual explanations, or the researchers providing feedback and explanations) (see Section~\ref{sec:ai_system_interface})
    \end{itemize}
    \item \textbf{Experimental setup} (description of relevant information about the experiments besides the XAI system used):
    \begin{itemize}
        \item \textbf{Methodology:} a summary description of the methodology for the empirical studies and experimental setups
        \item \textbf{Targeted users:} which types of users were considered (e.g. domain experts, lay-users, or users with varying of levels of AI expertise) (see Section~\ref{sec:user_factors_trust})
        \item \textbf{Evaluation criteria:} evaluation criteria that were considered besides trust/reliance (e.g. understanding, user's mental models, etc.). (see Section~\ref{sec:trustMeasurement})
        \item \textbf{Trust measurement type:} the specific trust measurements that were used (see Section~\ref{sec:trustMeasurement})
        \item \textbf{Trust outcomes/effects:} summary of the reported outcomes of the experiments with respect to the measurement and evaluation of the effects on user trust (see Section~\ref{sec:trustMeasurement})
    \end{itemize}
    \item \textbf{Conclusions:} summary of the conclusions of the paper and/or description of the main outcomes
    \item \textbf{Limitations:} summary of the main limitations observed by the authors
\end{itemize}

In the following sections we first describe the factors that can influence user trust. We connect these factors to the ML life-cycle and (X)AI system that a user can interact with. Finally we give a detailed description of types of trust measurement methods and evaluation criteria that exist and are used in the surveyed papers.

\subsection{Factors of trust}
\label{sec:trustFactors}
When designing human-centric evaluation of user trust in AI it is important to keep in mind which factors can influence trust. Both those factors that are directly influenced by the human-AI interaction and confounding factors which might not be influenced by the AI system and its design directly, but that may impact whether and to what degree a user (dis)trusts the system.
There are factors that are directly influenced within the human-AI interaction and other, potentially confounding, factors that are not influenced by the AI system and its design directly, but that may impact whether and to what degree a user (dis)trusts the system. 

Many different models to classify and categorize the factors that influence trust (i.e. its antecedents) exist \cite{Toreini2020, Siau2018, hoff2015trust, Schaefer2016, Riegelsberger2005, Kee1970, Kaplan2023, Yang2022, Mayer1995, Dietz2006}. 
The first of these models on trust \cite{Kee1970, Mayer1995, Dietz2006} stem from social sciences, including organisational science and psychology, and are concerned with interpersonal trust. These earlier models have later been adapted to the context of human-technology, human-automation, and human-AI relations \cite{Toreini2020, Siau2018, hoff2015trust, Schaefer2016, Riegelsberger2005, Kaplan2023, Yang2022}. 

\citeA{Yang2022} give a comprehensive survey of the different models and provide conceptual framework detailing the components of trust in relation to AI. Similarly, \citeA{Siau2018} and \citeA{Toreini2020} provide a comparison/overview between different earlier models and the adaptation into the context of AI. Many of these works derive (partial) inspiration from the work by \citeA{hoff2015trust} in which the authors give a detailed of account of the factors that influence trust in automation.
\citeA{Toreini2020} establish a conceptualization that makes the connection between and integrates the factors of trustworthiness (Ability, Benevolence, Integrity) factors with the human, environmental, and technological categorization of trust factors and connects these to the (Trustworthy AI) objectives important to regulators and policy makers e.g. FAES (fairness, explainability, auditability, safety) and FATE (Fairness, Accountability, Transparency \& Ethics).
\citeA{Koerber2018} provide a model of trust in automation with factors related to perceived trustworthiness based on the dimensions proposed in the \citeA{Mayer1995} and \citeA{lee2004trust} models, thereby placing the ABI notions of trustworthiness within the context of automation.

In this paper we use a categorization of the factors influencing trust similar to the one used by \citeA{Kaplan2023}. 
These factors are related to the human, the AI, and the context \cite{Kaplan2023}.
When looking at the human-centric evaluation of the effect of (X)AI on user trust we want to focus on the factors that may influence the user, differentiating between the factors that the AI system designers and developers (e.g. researchers) can control and those that they have no direct control of (e.g. demographic or cultural aspects of the user, or societal views on AI in general). 
Of the ones that have been studied there is a relatively robust picture of which individual antecedents are effective at predicting trust and which are not, however there is a lack of empirical research for many of the other factors \cite{Kaplan2023}.
Additionally, the interactive effects between the antecedents of trust have been largely unexplored \cite{Kaplan2023}.

\subsubsection{Contextual factors of trust}
\label{sec:contextual_factors_trust}
As said, there are many factors outside of the XAI system itself that can influence user trust.
These can include cultural, organisational factors (e.g. organisational setting), organisational trust, and institutional factors, i.e. the trust that people have in the institutions that develop or regulate the AI systems (i.e. institution-based trust) \cite{hoff2015trust, Siau2018, Yang2022}, as well as task related ones such as the type of system, system complexity, task complexity (e.g. tasks of varying cognitive load \cite{Wang2022, Jiang2022}), workload, perceived risks (e.g. using AI in a medical application \cite{Bussone2015}), perceived benefits, and framing of task \cite{hoff2015trust, Siau2018, Schaefer2016}.

Research into XAI systems cover a wide variety of application domains \cite{adadi2018peeking, Ferreira2020, Lai2021}, which is supported by our survey.
The surveyed papers are concerned with various domains, ranging from higher risk ones (e.g. medical diagnosis, self-driving cars, finance, or law, e.g. recidivism prediction) to low-medium risk (e.g. entertainment or social media related such as movie recommendation) applications.
Table~\ref{tab:application_domain_categories} shows the most frequently specified application domains. Many studies did not target a specific application domain but rather had users perform a mock or proxy task, such as a generic image classification or some other ML recommendation task. 

Different types of applications are used in the user studies, such as decision support systems \cite{Wanner2020, Bayer2021, Anik2021, Thaler2021, Wang2023, Chen2023}, recommender systems \cite{Guesmi2021, Shin2021, Bansal2021, Eslami2018, Cramer2008, Eiband2019, Gurney2022, Kunkel2019, Berkovsky2017, Zhao2019, Suresh2020, Guesmi2023}, image classification \cite{Yu2018, Yu2017, Yu2019, nourani2019effects, Nourani2020, Leichtmann2023, Yang2020, Yu2016, Suresh2020}, text annotation \cite{Papenmeier2021, Papenmeier2022, Schmidt2019, Linder2021}, or speech recognition \cite{Anik2021}.

\begin{table}
\centering
\caption{Application domain categories}
\label{tab:application_domain_categories}
\begin{tabular}{lp{7cm}}
\toprule
              Application domain &                                                                                                                                                                                                                                      Papers \\
\midrule
      Entertainment &                                                                                                                              \citeA{kulesza2013too, Ehsan2020, Cramer2008, Kunkel2019, Berkovsky2017, Bayer2021, Schmidt2020, Schmidt2019} \\
          Law/Legal &                                                                                                                                                                            \citeA{Wang2021, Wang2022, Bansal2021, Lakkaraju2020, Anik2021} \\
Medicine/Healthcare &                                                                                                                                                                        \citeA{Bussone2015, Leffrang2021, Jiang2022, Alam2021} \\
  Social Media/News &                                                                                                                                                                             \citeA{Papenmeier2022, Papenmeier2021, Linder2021, Eslami2018} \\
          Education &                                                                                                                                                                                                  \citeA{Kizilcec2016, Cheng2019, Anik2021} \\
     Transportation &                                                                                                                                                                                                            \citeA{Omeiza2021, Koerber2018} \\
            Finance &                                                                                                                                                                                                                           \citeA{Wang2023} \\
Scientific Research &                                                                                                                                                                                                                         \citeA{Thaler2021, Guesmi2023} \\
\bottomrule
\end{tabular}
\end{table}

\subsubsection{Factors related to the user}
\label{sec:user_factors_trust}
The user itself can also have an impact on the human-AI system interaction and how user trust develops.
Common terms that relate to users that are frequently employed to analyze and evaluate trust, include user knowledge, technical proficiency, familiarity, confidence, beliefs, faith, emotions, and personal attachments \cite{Mohseni2021}.
A person's personality (disposition to trust) \cite{Siau2018, hoff2015trust} and ability are of concern \cite{Siau2018}.
Similarly, a user's self-confidence, mood, and emotional state can also have an impact of level of trust \cite{Schaefer2016, hoff2015trust}. 
Additionally, demographics factors such as gender, age, or culture \cite{hoff2015trust} and various cultural aspects such as individualism and power relations within a culture can impact a person's propensity to trust an AI system \cite{Chien2015, Chien2016}. 
Data experts, similar to AI novices, benefit from interpretability in order to assess model uncertainty and trustworthiness \cite{Mohseni2021}. 

Another important factor that can influence the effect of explanations on users is their domain knowledge \cite{Wang2023}.
Domain experts are capable of dynamically adjusting the perceived trustworthiness of an AI model by using its explanations \cite{Nourani2020}.
Furthermore, \cite{Nourani2020} observe that novice (non-expert) users suffer from over-reliance due to their lack of knowledge which results in their inability to properly detect errors.
However, while domain expertise does impact user trust \cite{Ooge2021} observe that as expectations and personal experiences play a significant role, domain expertise alone cannot fully predict people's trust in a model.

In our survey we find that most papers do not target a specific type of user (31 papers). Often generic, mixed, participants are recruited via crowdsource platforms, such as Amazon Mechanical Turk or Prolific.
Other authors select specifically one user type such as non-experts \cite{Yu2017, Yu2019, Leffrang2021, Nourani2020, Jiang2022, Cheng2019, Yang2020, MingYin2019, Ehsan2020, Eslami2018, Rechkemmer2022, Koerber2018} or domain experts \cite{Lakkaraju2020, Wanner2020, Zhou2019, drozdal2020trust, Kunkel2019, Berkovsky2017}.
In most studies a variety of demographic information about the user is collected (e.g. education level \cite{Omeiza2021}), even when the authors do not target any group specifically.
Moreover, a number of studies look specifically at a variety of groups with varying attributes, such as varying levels of AI \cite{nourani2019effects, Anik2021, Suresh2020, Wang2023} or domain \cite{Bayer2021, Suresh2020, Wang2023} expertise.

While many studies evaluate user trust as a static property, it is essential, when interacting with complex AI systems, to take into account the evolution of users' experience and learning over time \cite{Mohseni2021, Lopes2022}. 
The long term evaluation of XAI systems can help in the estimation of valuable user experience factors such as over- and undertrust \cite{Mohseni2021}.
Prior knowledge and belief are important factors in shaping a person's initial trust \cite{Mohseni2021}.
Users of ML systems are in a constant learning state, therefore, even without model updates, their mental models and trust depend on their knowledge and familiarity with the system \cite{Lopes2022}.

\subsection{(X)AI system}
\label{sec:xai_system}
The factors of trust related to the AI are often categorized according to performance, process, and purpose factors \cite{lee2004trust, Siau2018, hoff2015trust}.
Many of the factors are not related to model performance, but instead depend on its design, appearance or usability \cite{Kaplan2023, hoff2015trust}.
The type of technology (e.g. using a DNN (deep neural network) blackbox model \cite{Bansal2021} or decision tree \cite{Zhang2020}) can also have an impact on trust \cite{Schaefer2016}.
When defining and constructing an XAI system a distinction between two key components should be considered, namely the (model) descriptors and the (explanation) interface. \citeA{Nauta2022} define an explanation as "a presentation of (aspects of) the reasoning, functioning and/or behavior of a machine learning model in human-understandable terms". 
Similarly, the concept of cues introduced by \citeA{Schlicker2021} contains both the notion of the presentation (e.g. the aesthetics of the interface \cite{Schlicker2021}) and content of explanations (e.g. descriptors such as inputs or outputs of the system \cite{Schlicker2021}).
So, from the perspective of studying human-evaluation of XAI an explanation consists of both its content (e.g. the descriptive information about the reasoning, functioning, and behavior of the system and underlying ML model) of the explanation and the way in which its represented (e.g. the user interface, its modality, or a researcher describing how the system is supposed to work or what its limitations are).

The explanation content can be any descriptive information about the model and its outputs (e.g. local output explanation or descriptive statistics of the training data), which we refer to as the (model/system) descriptors that can be used. 
While the presentation is the way in which this information is presented to the user, such as for example the design of the user interface and the modality in which the explanation is presented (e.g. visual or textual).
Important factors for descriptors and interface of the XAI system are the relevance, availability, detection and utilization of the cues that this provides the users \cite{Schlicker2021}.
In the following sections we describe both the different model descriptors (Section~\ref{sec:model_descriptors}) that can be used as the content and the presentation of the interface (Section~\ref{sec:ai_system_interface}).

\subsubsection{Model descriptors}
\label{sec:model_descriptors}

When asking what information to provide the user we need to consider what questions users may ask about or of the system. \citeA{Liao2021, Lim2009} provide a list of such questions. These questions range from asking what inputs the model used to asking how the system came to its output and what reasoning it applied. 
\citeA{Mohseni2021, Anik2021, Liao2021} give an overview of different explanation approaches, or types, that are already used to answer such questions. 
Several authors \cite{Schlicker2021, Lai2021, Vilone2020, Liao2021} have focused on the types of information or feedback users can get from an AI system. 
As we mentioned, \citeA{Schlicker2021} describe the cues that users can make use of. Similarly, \citeA{Lai2021} describe AI assistance elements by which they mean additional information such as information about the output (e.g. explanations or uncertainty estimates), the model or training data (e.g. feature importances, model performance, or what type of algorithm that is used), or other AI system elements that affect a user's agency or experience, that can be provided to the user besides simply giving them the model's prediction. \citeA{Vilone2020, Vilone2021} mention the use of explanators which is what the end-user will interact with, this is similar to \citeA{Schlicker2021} and \citeA{Lai2021}'s concepts of cues and assistance elements. 
These similar approaches highlight an overlapping sense of what information users would require from the system and how this information might be provided by explanations.
We summarize the available descriptive information that can be used in the AI system's interface and accompanying explanations as the system or model descriptors. 
The used descriptors provide the content of the explanations that the user is given about the system.

From this we can see that an explanation does not have to be confined to the outputs generated by explainability methods, but can be any descriptive information about the model and its outputs. 
Based on the observations from our survey and the related literature, which we described in the previous paragraph, we summarize the descriptors that have already been of interest in user studies concerned with user trust. Therefore, to help the identification of existing insights on descriptors and effects of different explanatory content, we provide the following list:

\begin{itemize}
    \item Input sample (e.g. (tabular) feature values, text, image)
    \item Output prediction
    \item Explanations of outputs
    \begin{itemize}
        \item Uncertainty estimates
        \item Local explanation (e.g. example based, counterfactuals, salience maps)
        \item Local feature importances / input influences
    \end{itemize}
    \item Underlying model
    \begin{itemize}
        \item Training data (e.g. data distribution, features used, modality of data)
        \item Algorithm type / training procedure
        \item Inherently interpretable model information (e.g. Decision Tree)
        \item Global feature ranking / importance
        \item Model performance metrics (e.g. model accuracy, precision/recall scores)
    \end{itemize}
\end{itemize}

In our survey we find that the most commonly used descriptors are different types of local explanations (depending on input data modality and application) or uncertainty estimates. Many types of local explanations are used, namely example based \cite{Yang2020, Chen2023, Alam2021, Leichtmann2023, kulesza2013too}, contrastive \cite{Omeiza2021}, (pixel) attribution \cite{nourani2019effects, Nourani2020, Leichtmann2023, Bussone2015}, counterfactual or 'What if' \cite{Guesmi2023, Wang2022}, nearest neighbor \cite{Wang2022} explanations or describing the model's reasoning or logical rationale for the output. 
As we detail later in this section, it is common that a mock model is used and not an actual underlying ML model, therefore, in these cases, the specific local explanation method is not always described in detail but rather that the users receive some sort of local explanation of which a high-level description is given.
Other studies make use of feature importance or feature contributions \cite{Cheng2019, Wang2022, Bucinca2021, Zhou2019, Lu2021, Chen2023}
or display or describe the input features of the model and the output sample \cite{Cheng2019, Lim2009a, Zhou2019, drozdal2020trust, Zhao2019, Rechkemmer2022, Lu2021, Wang2023, Chen2023}.

User are also often provided with some form of model performance metrics, such as the accuracy \cite{Papenmeier2022, Papenmeier2021, MingYin2019, Lim2009a, Rechkemmer2022, Lu2021, Thaler2021, Suresh2020, drozdal2020trust}, confusion matrices \cite{drozdal2020trust}, or class-based error rates \cite{Thaler2021}.
Some less frequently used descriptors are providing the users with (introductory) information about the system and underlying model \cite{Koerber2018, Chen2023}, such as the model architecture \cite{Suresh2020}, which algorithm was used, system limitations \cite{Koerber2018}, what the explanatory process is and how it works \cite{Zhao2019}, or information about the training data, data distributions, and process of feature engineering \cite{Anik2021, drozdal2020trust}.

In order to test the impact of different descriptors they are manipulate in various ways. This can be done either in a within-subject or a between-subject design.
The most common strategy is to either show or not show a specific descriptor, which is usually applied for various local explanations and uncertainty estimates.
Another strategy is to supply the user with explanations of various levels of detail (e.g. only showing an uncertainty estimate, combining it with a pixel attribution) or various combinations of different explanations (e.g. combining uncertainty estimates with an example based explanations).
The modality of the explanation and the interface can also vary or be combined (e.g. showing a text based explanation along with an input image). We discuss the design and modality of the user interface in the following section (Section~\ref{sec:ai_system_interface}).

Additionally, authors may also try to determine the effects of the trustworthiness of the explanations and underlying model.
In most of the papers the actual underlying model that is used is not clearly specified. 
Often this is because they use a mock or proxy task with a simulated model. This type of setup is referred to as a Wizard of Oz setup \cite{Lai2021}. The benefit of this is that it allows the researchers full control over the outputs that the user is seeing and allows them to specifically tune the (actual) trustworthiness of the system to determine to what extend this affects the users \cite{Lai2021}.
For example, it allows the researchers to control many elements of the explanations interface such the usage of different descriptor classes (rather than actual XAI methods) and interface designs.
The drawback of this is of course that this does not tell us anything about a specific XAI method or algorithm. Additionally, given the complexity of model behavior, it can be challenging to design realistic studies using this method. While a lack of realism can hinder the validity and generalizability of study results \cite{Lai2021}.

The papers that do state a specific ML method make use of a range of different models of varying complexity and interpretability from decision trees \cite{Wanner2020, Gurney2022}, SVMs \cite{Rechkemmer2022}, Logistic regression \cite{Wang2022}, and Naive Bayes \cite{Rechkemmer2022} models to Random Forest \cite{Rechkemmer2022}, CNN (e.g. ResNet) \cite{Papenmeier2021, Wanner2020, Zhou2019, Leichtmann2023}, language (e.g. LSTM) \cite{Wanner2020, Bansal2021}, time series \cite{Leffrang2021}, and Reinforcement Learning \cite{Pynadath2018} models.
Most researchers study the effects of different models whether they are proxy or mock models (i.e. Wizard of Oz setups) or actual underlying models. The most common approach is to use models of varying accuracy \cite{Yu2018, Yu2017, Yu2019, Bansal2021, Honeycutt2020, MingYin2019, Zhou2018, Zhou2019, nourani2019effects} and to observe whether this has an effect on the user or whether or not users can accurately detect the change and adjust their perceived trustworthiness of the model and reliance accordingly.
\citeA{Wang2023} observe what the effect of model changes are. 
\citeA{Miller2016} use two models with different levels of automation capability.
\citeA{Yu2017, Yu2019} study the effects of varying accuracy over time.
\citeA{Zhang2020} use models that are based on different sets of features, where one uses all available features and another uses only part of the feature set.

\subsubsection{AI interface}
\label{sec:ai_system_interface}
Along with the explanations' content (i.e. used descriptors) the presentation of this information also has an impact on the user \cite{hoff2015trust}.
\citeA{hoff2015trust} describe several design features related to trust in automation and the development of trustworthy automation, such as communication style, ease-of-use, transparency and feedback, level of control, and appearance. For the design of which they provide a number of recommendation.
\citeA{Nauta2022} describe three quality properties, or criteria, related to the presentation of the explanation: compactness (i.e. the size of the explanation), compositionality (i.e. formatting and organization of the information), and confidence (i.e. presence and accuracy of probability information).
In our survey we find different interface modalities being used, where textual interfaces, such as showing a worded explanation of the output, are most frequently used (45 papers), followed by visual or graphical descriptions (28 papers), such as image based methods (e.g. salience maps or displaying the input image) \cite{Yu2018, Yu2017, Yu2019, nourani2019effects, Nourani2020, Leichtmann2023, Thaler2021, Wang2021, Wang2022, Yang2020, Honeycutt2020, Wang2023}, using graphs \cite{Leffrang2021, Cheng2019, Bansal2021, MingYin2019, Linder2021, Wanner2020, Bucinca2021, Zhou2019, drozdal2020trust, Gurney2022, Pynadath2018, Ooge2021, Rechkemmer2022, Lu2021, Chen2023, Guesmi2023}, or video \cite{Suresh2020}.

Many studies make use of feature based display methods \cite{kulesza2013too, Bussone2015, Zhang2020, Guesmi2021, Jiang2022, Cheng2019, Wang2022, MingYin2019, Linder2021, Lim2009a, Wanner2020, Bucinca2021, Zhou2019, drozdal2020trust, Lu2021}, such as input features (e.g. showing the values of the input features for the sample) and output explanation (e.g. local feature importance), or global feature importance to describe the model.
Some papers perform visual highlighting of different elements of the interface \cite{Papenmeier2022, Papenmeier2021, Bansal2021, Linder2021, Schmidt2019, Wang2023, Chen2023, Guesmi2023}, such as highlighting the words with the highest feature importances, to make them stand out more compared to the other parts of the interface and/or explanations.
\citeA{Miller2016} provide users with both visual and voice cues about navigation in an automated driving system.
\citeA{Guesmi2023} make use of an interactive interface that the user can use to explore the problem domain, model, outputs, and output explanations, in order to the determine the effects that different levels of interactivity can have on users.

In our survey we find that several studies attempt to discern the effects of different kinds of interface design rather than the descriptor content.
\citeA{Guesmi2021, Bansal2021, Bayer2021, Alam2021, Linder2021, Miller2016, Omeiza2021, Gurney2022} make use of varying presentation formats with different combinations of explanations in different types of modalities.
\citeA{Yang2020} investigate the effect of different kinds of spatial layouts for displaying graphical example-based explanations. Comparing a grid, tree, and graph based structure, as well as using the input image versus a rose chart of features.
While \citeA{Kulms2019} investigate the effects of different kinds of antropomorphism.

\subsection{Evaluation \& trust measurement}
\label{sec:trustMeasurement}

\begin{figure}
  \centering
  \includegraphics[width=\textwidth]{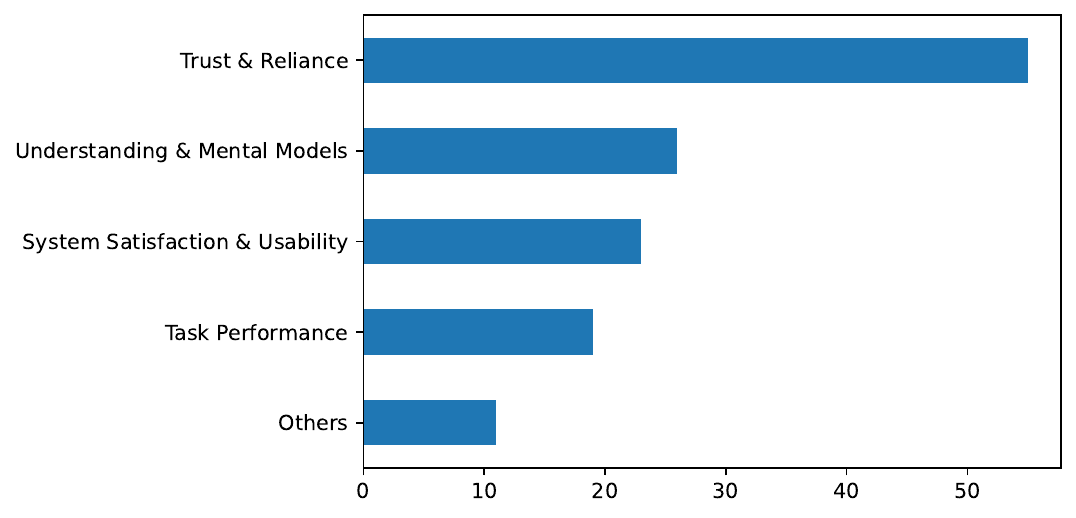}
  \caption{Number of papers that make use of one or more evaluation criteria per category.}
  \label{fig:evaluation_criteria_categories}
\end{figure}

In Section~\ref{sec:human-centric-evaluation} we mentioned the evaluation criteria which are often used in the context of the human-centric evaluation of XAI. For each of the surveyed paper we have collected the evaluation criteria that are applied alongside trust or reliance related criteria. The criteria found in the surveyed papers fall in four main categories: Trust and Reliance, Understanding and Mental Models, System Satisfaction and Usability, and Task Performance. Some miscellaneous criteria we classify in the 'Other' category. 
Figure~\ref{fig:evaluation_criteria_categories} shows for each of the categories how many studies use criteria related to that category. In Table~\ref{tab:evaluation_criteria_per_category} we provide a full overview of the specific criteria that are applied for each category and which papers use of that criteria. 

\begingroup
\setlength{\tabcolsep}{6pt} 
\renewcommand{\arraystretch}{1.5} 
\begin{longtable}{l|p{9.5cm}}
\caption{Evaluation criteria per category}
\label{tab:evaluation_criteria_per_category}\\
\toprule
                                                                                Category &                                                                                                                                                                                                                                                                                                                                                                                                                                                                                                                                                                                                                                                                                                                                                                                                                                                                                                                                                                                                                                                                                                                                                                                                                                                                                                                                                                                                                                                                                                                                                                                                                                       Criteria \\
\midrule
\endfirsthead
\caption[]{Evaluation criteria per category} \\
\toprule
                                                                                Category &                                                                                                                                                                                                                                                                                                                                                                                                                                                                                                                                                                                                                                                                                                                                                                                                                                                                                                                                                                                                                                                                                                                                                                                                                                                                                                                                                                                                                                                                                                                                                                                                                                       Criteria \\
\midrule
\endhead
\midrule
\multicolumn{2}{r}{{Continued on next page}} \\
\midrule
\endfoot

\bottomrule
\endlastfoot
                                                                          Trust \& Reliance & 
                                                                          
                                                                          trust (ABI) \cite{Guesmi2021, Berkovsky2017}, trust change \cite{Yu2017, Nourani2020, Wang2023}, trust in automation \cite{Papenmeier2022}, trust score \cite{kulesza2013too}, propensity to trust \cite{Papenmeier2022, Yang2020, Schmidt2020}, reliance \cite{Bussone2015, Yu2018, Yu2017, Wang2021, Wang2022, Miller2016, Yu2016, Bucinca2021, Koerber2018, Lu2021, Schmidt2019, Chen2023}, willingness to accept \cite{Wanner2020}, willingness to follow predictions \cite{Leffrang2021, Yang2020, Wanner2020, Suresh2020}, ability to form appropriate trust (control factor) \cite{Kulms2019}, perceived user trust \cite{Omeiza2021}, effect demographic characteristic of users on trust \cite{Papenmeier2022}, acceptance \cite{Jiang2022, Bansal2021, Cramer2008, Wanner2020}, switch rate \cite{Leffrang2021, Chen2023}, advice adoption \cite{Jiang2022, Bansal2021, Wang2023}, agreement rate \cite{Nourani2020, Wang2021, Wang2022, Chen2023}, disagreement rate \cite{Leffrang2021}, advice seeking \cite{Kulms2019} \\
                                                             Understanding \& Mental Models &                                                                                                                                                                                                                                                                                                                                                                                                                                                                                                                                                                                                                                                                                                                 understanding \cite{kulesza2013too, Papenmeier2022, Omeiza2021, Cheng2019, Wang2021, Wang2022, Yang2020, Linder2021, Alam2021, Eslami2018, Lim2009a, Cramer2008, Eiband2019, drozdal2020trust, Wang2023}, self-reported understanding \cite{Cheng2019}, objective understanding \cite{Cheng2019}, perceived understanding \cite{Papenmeier2021, Eiband2019, Zhao2019}, comprehension \cite{Leichtmann2023, Cheng2019, Thaler2021}, perception of accuracy \cite{Alam2021}, user's mental model fidelity \cite{kulesza2013too}, uncertainty awareness \cite{Wang2021, Wang2022}, perceived system ability \cite{Holliday2016}, perceived system accuracy \cite{nourani2019effects, Nourani2020, Honeycutt2020, Alam2021, Rechkemmer2022}, perceived system performance \cite{Yu2019, Shin2021}, perceived task performance \cite{Yu2019}, perceived model accuracy \cite{Wang2023}, perceived model change \cite{Honeycutt2020} \\
                                                                        Task Performance &                                                                                                                                                                                                                                                                                                                                                                                                                                                                                                                                                                                                                                                                                                                                                                                                                                                                                                                                                                                                              task performance \cite{Omeiza2021, Linder2021, Lim2009a, Bucinca2021, Suresh2020, Schmidt2019, Wang2023, Chen2023}, combined accuracy \cite{Zhang2020, Schmidt2020}, combined performance \cite{Zhang2020, Leichtmann2023, Yang2020, Bansal2021, Thaler2021}, complementary performance \cite{Bansal2021}, correctness and response time as measures of human performance \cite{Linder2021}, decision performance / combined performance \cite{Zhou2018}, task execution speed \cite{Schmidt2019}, take-over performance \cite{Koerber2018}, team performance \cite{Bansal2021, Kulms2019, Gurney2022, Pynadath2018, Koerber2018} \\
System Satisfaction \& Usability &                                                                                                                                                                                                                                                                                                                                           usability \cite{Yang2020}, usefulness \cite{Bansal2021, Alam2021, Anik2021}, satisfaction \cite{Guesmi2021, Alam2021, Wang2023, Guesmi2023}, user satisfaction \cite{Ehsan2020}, user confidence \cite{Zhou2018, Wanner2020, Thaler2021, Guesmi2023}, perceived ease of use \cite{Cramer2008, Guesmi2023}, perceived usefulness \cite{Cramer2008}, recommendation quality \cite{Kunkel2019}, perceived control \cite{Holliday2016, Guesmi2023}, mental demand \cite{Bucinca2021}, perceived competence \cite{Cramer2008, Berkovsky2017, Bayer2021}, persuasiveness \cite{Guesmi2021}, quality of interpretability \cite{Schmidt2019}, explanation quality \cite{Kunkel2019}, explanation type preference \cite{Chen2023}, perceived accountability \cite{Shin2021}, perceived consistency of AI explanations with prior knowledge \cite{Wang2023}, perceived quality of explanation \cite{Guesmi2023}, perceived system complexity \cite{Bucinca2021}, perceived transparency \cite{Shin2021, Cramer2008, Holliday2016}, perceived trustworthiness \cite{Kulms2019, Bruzzese2020, Berkovsky2017, Bayer2021}, perceived fairness \cite{Shin2021, Anik2021}, perceived predictability \cite{Holliday2016}, perceived need for explanation \cite{Cramer2008}, expectation violation \cite{Kizilcec2016} \\
                                                                                  Others &                                                                                                                                                                                                                                                                                                                                                                                                                                                                                                                                                                                                                                                                                                                                                                                                                                                                                                                                                                                                                                                                                                                                                                                                                                        plausibility \cite{Ehsan2020}, completeness \cite{Alam2021}, effectiveness \cite{Guesmi2021}, efficiency \cite{Guesmi2021}, knowledge gaps \cite{kulesza2013too}, self-confidence in decision \cite{Yang2020, Wanner2020}, situation awareness \cite{Pynadath2018}, sufficiency \cite{Alam2021}, transparency \cite{Guesmi2021, Berkovsky2017, Zhao2019, Guesmi2023}, familiarity \cite{Papenmeier2022} \\
\end{longtable}

\endgroup

In \citeA{Kohn.2021} the authors give a review of different measurements of trust in automation. They describe three categories in which these methods can be placed: self-report, physiological, and behavioral.
\textbf{Self-report} measures are measures where respondents report on their own behaviors, beliefs, attitudes, or intentions by receiving a question or prompt and selecting or detailing a response \cite{Kohn.2021}. Self-report measures are typically set up as surveys or questionnaires. \citeA{Kohn.2021} describe 16 different types of self-report methods used in trust in automation.
\textbf{Behavioral} methods use the observation of participants' behavioral processes or tendencies \cite{Kohn.2021}. Examples of behavioral measures include the tracking of combined human-machine performance, agreement rate, decision or response time, delegation or reliance, and economic trust games.
\textbf{Physiological} methods are based on the measurement of biological responses from the user, such as muscle movements, heart rate, or neural activation \cite{Kohn.2021}.

With respect to the use of self-report and behavioral measures, \citeA{scharowski2023distrust} clarify that trust as an attitude should be measured via questionnaires whereas behavioural measures are suitable to assess reliance. 
The outcome of questionnaires is influenced by the participants' ability to reflect their attitude towards a system, which can be difficult for some participants \cite{Papenmeier2022}. Such questionnaires measure the participants' perception of their trust \cite{Miller2022} that do not always correspond to their actions.
Across multiple studies in our overview a discrepancy between self-report and behavioural measures is observed \cite<e.g.,>[]{Papenmeier2022, Papenmeier2021, Wang2023}. 

In practice most often self-reported trust is employed. Typically, this entails a short questionnaire, in which the participants rate their agreement to different statements (items) on a likert-scale. 
Frequently, custom self-report scales are used for assessing trust in automation \cite{Kohn.2021}. A procedure that is also observed in the evaluation of XAI methods \cite{Lopes2022} or in the broad context of trust in information systems \cite{Rusk.2018}. 
\citeA{Kohn.2021} highlight that researchers either state that they developed their own way of measuring trust, or they do not cite any source for their method. 
These custom scales can even take the form single item measurements \cite{Spain.2008}, which, in essence simply ask "Did you trust our system?" \cite{Kohn.2021}. Despite the advantage in efficiency of single item measures, they are usually less reliable than multiple items and narrow down the complex concept of trust to a single (dis)agreement.

The customization of self-report items has the further problem that their validity has not been tested. This hinders the interpretation of results because one can not be certain if they are actually measuring trust. 
Moreover, this contributes to a lack of generalization of trust measurement, because results obtained with different self-report items can not be easily compared. A difference in outcomes could be due to both the method or manipulation that is investigated or the measurement applied. 
Thus, without standardizing measurements the comparison between results and methods is obstructed. Consistent and standardized items would allow researchers to develop a better understanding of trust and distrust \cite{Rusk.2018}.

As discussed in section \ref{sec:distrustSeparate} the distinction of trust and distrust may be of interest in the context of appropriate reliance. Some examples identifying the merit of considering trust and distrust as separate dimensions, which can be found across different subfields of human-technology interaction \cite{Kohn.2021, HarrisonMcKnight.2001, McKnight.2004, Benamati.2006, Ou.2010, Fang.2015, Thielsch.2018}.
A difference between dispositional trust and dispositional distrust was observed in the context of online expert advice \cite{McKnight.2004}, and trust and distrust co-existed as distinct construct in the context of online banking \cite{Benamati.2006} and online shopping \cite{Ou.2010}. A study on website design showed that trust and distrust are affected by different antecedents, and the performance of a trust-aware recommender system was improved by predicting not only trust but also distrust \cite{Fang.2015}. \citeA{Thielsch.2018} investigated work-related information systems and also identified trust and distrust as related yet separate influences on different outcome variables. 

\citeA{Kohn.2021} note, however, that notwithstanding the evidence for the two-dimensional conceptualisation, uni-dimen\-sional scales are the common form for assessing trust in automation \cite{Kohn.2021}.
Of these, the Checklist for Trust between People and Automation \cite{Jian.2000} is the most frequently used \cite{Kohn.2021}. 
This checklist measures trust and distrust as polar opposites along a single dimension. Five of the 12 items (statements rated by the user) measure distrust. In practice, these items are often reverse-scored and summed with the trust items to form one trust score, which was also suggested by the original authors of the scale \cite{Spain.2008}. In a critical validation attempt of this scale by \citeA{Spain.2008}, a one-factor model (indicating the polar opposites along a single dimension) and a two-factor model were compared. 
This factor analysis provided evidence for the conceptualization of trust and distrust as separate, yet related constructs \cite{Spain.2008}. Reverse scoring distrust items to then sum with the trust items entails a problematic entanglement of the two factors identified by \citeA{Spain.2008} and disregards the incremental insight by measuring trust and distrust individually.

In his dissertation \citeA{Rusk.2018} sets out to close a research gap and introduces a scale that measures trust and distrust separately and provides a first validation. The scale is developed for the context of information systems and might be applicable to the (X)AI context as well. Importantly, the author points out that the results need to be independently re-evaluated in a different study. In that regard we could not find any work with that aim and highlight this as promising work for future research. 

Referring back to the clarification of the underlying aim of appropriate trust (Section \ref{sec:desideratumApproprTrust}) the work of \citeA{Wang2021, Wang2022} should be highlighted. When assessing behavioural trust they distinguished between appropriate trust, undertrust and overtrust, depending on the participants usage of the model and the correctness of the model. When a ground truth is known, this is an useful approach to measure effects of an introduced method on the user in a clarified manner. For instance in one study, this approach allowed to observe that feature importance and feature contribution slightly increased appropriate trust, decreased their undertrust, and that, for feature importance, this came at the cost of a slight increase in overtrust \cite{Wang2022}.

\begin{figure}
  \centering
  \includegraphics[width=0.9\textwidth]{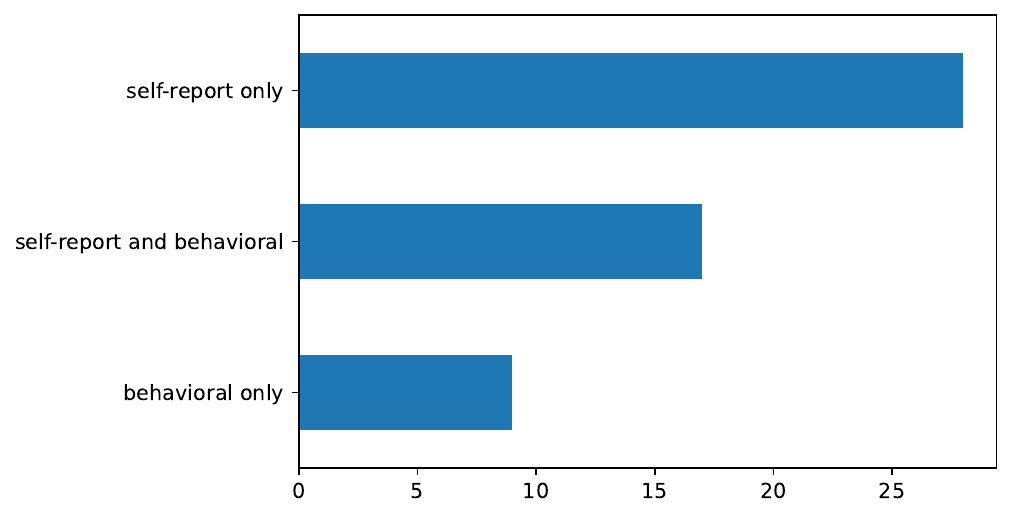}
  \caption{Number of papers that use self-report and/or behavioral trust measurement strategies.}
  \label{fig:measure_type}
\end{figure}

\begin{figure}
  \centering
  \includegraphics[width=0.9\textwidth]{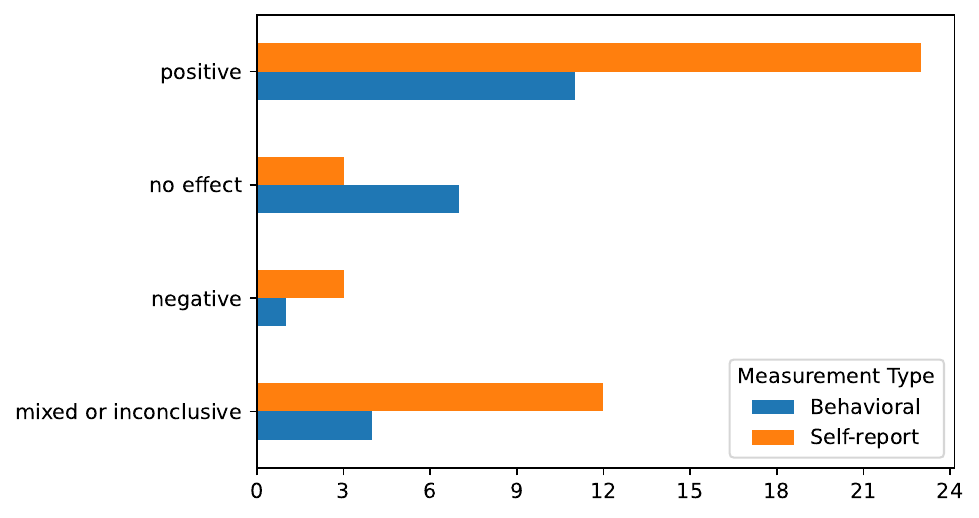}
  \caption{Stated effects of methods on trust (counted across all studies). If a paper includes multiple experiments or uses both self-report and behavioral it may be counted multiple times.}
  \label{fig:outcomes_trust_effects}
\end{figure}

Figure~\ref{fig:measure_type} show the number of surveyed papers that use some self-report, behavioral, or a combination of both trust measurements.
For each of the experiments in the papers we have summarized the outcomes related to the effects for the different measurements (e.g. whether the intervention (changing AI system) had a positive measured effect on behavioral trust). In Figure~\ref{fig:outcomes_trust_effects} we provide a barplot summarizing the effects for self-report and behavioral trust. From this we can see that mostly there is a (moderately) positive effect on user trust from the application of explanations, while a clear negative effect is not as common especially for behavioral trust measures. Additionally, we observe that many experiments result in mixed or inconclusive outcomes.
In Section~\ref{appendix:trust_outcomes} in the appendix we provide an overview of the outcomes for each of the experiments in the papers.
\citeA{Zhou2018} is the only study in our survey in which a physiological trust measurement is applied, therefore we did not include it in the strategies shown in Figure~\ref{fig:measure_type}.

\section{Open questions \& recommendations}

From our survey we find that while there is growing interest in the empirical evaluation of trust several open questions and gaps in the field remain from which we can describe a number of recommendations.
\citeA{vanderWaa2021} provides a number of recommendations for the improvement of user-centric evaluation of XAI, such as on the constructs, use cases and experimental context, and measurements.
\citeA{Miller2022} lay out four requirements for the proper evaluation of warranted and unwarranted trust and distrust, namely the need to take into account and measure task performance, the need for risk, the need to allow users to choose to rely or not, and the ability to manipulate the trustworthiness of the AI system.

We find that from the possible types of explanations, the descriptors used, and the ways in which they are presented only a limited scope has already been investigated. This means that there are ample opportunities for researchers left. Not least by integrating the evaluation of user trust and reliance in existing user studies/workflows related to (X)AI. Many different setups for various application domains, types, and users remain un(der)investigated for many of the potential antecedent and confounding factors of trust.

When comparing the descriptors, modulation strategies, and (XAI) methods used with those user questions and needs, types of explanations, descriptors, and XAI methods that are described in the various overviews we find that there is still a number of areas and methods that have been either not investigated yet or those for which results are currently inconclusive. 

For researchers it is important to note that they can control certain factors that influence trust while not other.
These factors that lie beyond the control of the developer should be considered as potential confounding variables in user studies. 
Potential confounders can be addressed in multiple ways. They can be controlled for by keeping them on a constant level, by balancing them between conditions, or by randomisation \cite{Sedlmeier.2013}. They can also be treated as further independent variables and be manipulated as well by, for example recruiting lay-users and domain experts as two experimental groups that can be compared.
Most importantly, one should be conscious of the potential confounders and address them within the limits of the planned experiment. Considering which of the known confounders maybe be present in one’s study and addressing the relevant ones improves the quality of results and improves their interpretation and the potential comparability of study outcomes.    

Given that a number of the surveyed papers observe discrepancies between self-report and behavioural measurements, a combination of measurement types would be beneficial. This allows to study potential differences of the reported attitude and the actual reliance. For example, in a low risk scenario a person who reports trust could also rely. Yet, under higher risk the person may report trust but could be more hesitant and would not act upon it. With combining measurement types such hypothetical patterns are possible to observe and investigate.     

To at least partially overcome the mixed and inconclusive results in empirical research on the explainability-trust hypothesis, we make three suggestions.
First, make explicit what you mean by trust by indicating which definition of trust you are following. Studies on trust (and distrust) in the XAI context need to continue to draw from the established work on (dis)trust without conflating it with common-sense reasoning on trust and distrust. Furthermore, state which perspective on trust is of interest for your study. There is an important difference between studying whether a XAI method affects the user’s trust in a single interaction and whether it affects the general attitude in AI, which should not be mixed.
Second, standardized and validated scales should be used for self-report questionnaires to facilitate comparability across studies and generalization of observed results. 

Third, consider that the distinction of trust and distrust may be of interest in the XAI context. Because explanations help identify both correct and wrong outputs, XAI methods may affect both the user’s trust and distrust.
As a starting point to advance the consideration and evaluation of trust and distrust, we suggest the scale developed by \citeA{Rusk.2018}, which requires further independent validation as recommended by the author. This would provide further evidence for the two-dimensional concept of trust and distrust and make it easier to consider and evaluated trust and distrust separately.

In general, one should be aware of the context of one’s study. As the varying results of the surveyed studies show the application domain, the targeted users and their level of expertise are relevant to the effects of a method on the user’s trust and other outcomes in general.
The application domain and the entailed risk may inform which reliance problem (disuse or overtrust) is more important. Depending on the risk, unwarranted trust or unwarranted distrust can be problematic in different ways.

\section{Conclusion}

We have shown that there is growing interest in user trust in (X)AI and the development of methods that can foster appropriate reliance in trustworthy AI methods. 
We aimed to give an overview of the current theoretical view on trust in (X)AI and its related concepts. We provided a survey of existing empirical studies that perform human-centric evaluation of the effects that (X)AI methods have on its users and which factors are most important and the extend to which they can affect user trust and distrust.
We hope that this gives researchers and ML practitioners insights into concepts such as trust in AI, appropriate reliance, and trustworthiness of AI are and support their ability to construct valuable studies on the effects of ML methods on user trust and its evaluation.

Everything discussed here is not necessarily relevant for each study or context.
Yet, this contribution seeks to stimulate awareness of one's application domain, user types and needs, the involved risk, the chosen type measurement, and the implications all this encompasses. 
We find that work needs to be done both on the theoretical underpinnings of the constructs of trust and distrust and how they can be properly measured and evaluated in the context of (X)AI.

\newpage
\appendix
\section{Appendix}

\subsection{Description of surveyed papers on empirical evaluation of the effects of ML methods on user trust}

\begin{longtable}{lp{1cm}p{1cm}p{1cm}|p{1cm}p{1cm}p{1cm}p{1cm}}
\caption{Survey papers.}
\label{tab:full_table}\\
\toprule
                   & \multicolumn{3}{c|}{Trust} &\multicolumn{4}{|c}{Other criteria} \\ \hline
                     Paper & \rotatebox{90}{Self-report} & \rotatebox{90}{Behavioral} & \rotatebox{90}{Physiological} & \rotatebox{90}{Understanding \& Mental models} & \rotatebox{90}{Task Performance} & \rotatebox{90}{System satisfaction \& usability} & \rotatebox{90}{Others} \\
\midrule
\endfirsthead
\caption[]{Survey papers.} \\
\toprule
                     Paper & \rotatebox{90}{Self-report} & \rotatebox{90}{Behavioral} & \rotatebox{90}{Physiological} & \rotatebox{90}{Understanding \& Mental models} & \rotatebox{90}{Task Performance} & \rotatebox{90}{System satisfaction \& usability} & \rotatebox{90}{Others} \\
\midrule
\endhead
\midrule
\multicolumn{8}{r}{{Continued on next page}} \\
\midrule
\endfoot

\bottomrule
\endlastfoot
    \citeA{kulesza2013too} &                      \cmark &                            &                               &                                         \cmark &                                  &                                                  &                 \cmark \\
      \citeA{Kizilcec2016} &                      \cmark &                            &                               &                                                &                                  &                                           \cmark &                        \\
       \citeA{Bussone2015} &                      \cmark &                     \cmark &                               &                                                &                                  &                                                  &                        \\
    \citeA{Papenmeier2022} &                      \cmark &                     \cmark &                               &                                         \cmark &                                  &                                                  &                 \cmark \\
          \citeA{Zhou2018} &                             &                     \cmark &                        \cmark &                                                &                           \cmark &                                           \cmark &                        \\
            \citeA{Yu2018} &                      \cmark &                     \cmark &                               &                                                &                                  &                                                  &                        \\
            \citeA{Yu2017} &                      \cmark &                     \cmark &                               &                                                &                                  &                                                  &                        \\
            \citeA{Yu2019} &                      \cmark &                     \cmark &                               &                                         \cmark &                                  &                                                  &                        \\
      \citeA{Leffrang2021} &                             &                            &                               &                                                &                                  &                                                  &                        \\
         \citeA{Zhang2020} &                             &                     \cmark &                               &                                                &                           \cmark &                                                  &                        \\
\citeA{nourani2019effects} &                             &                            &                               &                                         \cmark &                                  &                                                  &                        \\
        \citeA{Omeiza2021} &                      \cmark &                            &                               &                                         \cmark &                           \cmark &                                                  &                        \\
       \citeA{Nourani2020} &                      \cmark &                            &                               &                                         \cmark &                                  &                                                  &                        \\
        \citeA{Guesmi2021} &                      \cmark &                            &                               &                                                &                                  &                                           \cmark &                 \cmark \\
         \citeA{Jiang2022} &                             &                            &                               &                                                &                                  &                                                  &                        \\
          \citeA{Shin2021} &                      \cmark &                            &                               &                                         \cmark &                                  &                                           \cmark &                        \\
    \citeA{Leichtmann2023} &                      \cmark &                            &                               &                                         \cmark &                           \cmark &                                                  &                        \\
    \citeA{Papenmeier2021} &                      \cmark &                     \cmark &                               &                                         \cmark &                                  &                                                  &                        \\
         \citeA{Cheng2019} &                      \cmark &                            &                               &                                         \cmark &                                  &                                                  &                        \\
          \citeA{Wang2021} &                             &                     \cmark &                               &                                         \cmark &                                  &                                                  &                        \\
          \citeA{Wang2022} &                             &                     \cmark &                               &                                         \cmark &                                  &                                                  &                        \\
          \citeA{Yang2020} &                      \cmark &                     \cmark &                               &                                         \cmark &                           \cmark &                                           \cmark &                 \cmark \\
        \citeA{Bansal2021} &                             &                     \cmark &                               &                                                &                           \cmark &                                           \cmark &                        \\
     \citeA{Honeycutt2020} &                             &                     \cmark &                               &                                         \cmark &                                  &                                                  &                        \\
       \citeA{MingYin2019} &                      \cmark &                     \cmark &                               &                                                &                                  &                                                  &                        \\
        \citeA{Linder2021} &                             &                            &                               &                                         \cmark &                           \cmark &                                                  &                        \\
        \citeA{Miller2016} &                      \cmark &                     \cmark &                               &                                                &                                  &                                                  &                        \\
     \citeA{Lakkaraju2020} &                      \cmark &                            &                               &                                                &                                  &                                                  &                        \\
         \citeA{Ehsan2020} &                             &                            &                               &                                                &                                  &                                           \cmark &                 \cmark \\
          \citeA{Alam2021} &                      \cmark &                            &                               &                                         \cmark &                                  &                                           \cmark &                 \cmark \\
        \citeA{Eslami2018} &                      \cmark &                            &                               &                                         \cmark &                                  &                                                  &                        \\
         \citeA{Kulms2019} &                      \cmark &                     \cmark &                               &                                                &                           \cmark &                                           \cmark &                        \\
            \citeA{Yu2016} &                      \cmark &                     \cmark &                               &                                                &                                  &                                                  &                        \\
          \citeA{Lim2009a} &                      \cmark &                            &                               &                                         \cmark &                           \cmark &                                                  &                        \\
        \citeA{Cramer2008} &                      \cmark &                            &                               &                                         \cmark &                                  &                                           \cmark &                        \\
        \citeA{Eiband2019} &                      \cmark &                            &                               &                                         \cmark &                                  &                                                  &                        \\
      \citeA{Bruzzese2020} &                      \cmark &                            &                               &                                                &                                  &                                           \cmark &                        \\
        \citeA{Wanner2020} &                      \cmark &                            &                               &                                                &                                  &                                           \cmark &                 \cmark \\
       \citeA{Bucinca2021} &                      \cmark &                            &                               &                                                &                           \cmark &                                           \cmark &                        \\
          \citeA{Zhou2019} &                      \cmark &                            &                               &                                                &                                  &                                                  &                        \\
        \citeA{Thaler2021} &                      \cmark &                            &                               &                                         \cmark &                           \cmark &                                           \cmark &                        \\
  \citeA{drozdal2020trust} &                      \cmark &                            &                               &                                         \cmark &                                  &                                                  &                        \\
        \citeA{Gurney2022} &                      \cmark &                     \cmark &                               &                                                &                           \cmark &                                                  &                        \\
        \citeA{Kunkel2019} &                      \cmark &                            &                               &                                                &                                  &                                           \cmark &                        \\
      \citeA{Pynadath2018} &                      \cmark &                            &                               &                                                &                           \cmark &                                                  &                 \cmark \\
     \citeA{Berkovsky2017} &                      \cmark &                            &                               &                                                &                                  &                                           \cmark &                 \cmark \\
         \citeA{Bayer2021} &                      \cmark &                     \cmark &                               &                                                &                                  &                                           \cmark &                        \\
          \citeA{Zhao2019} &                      \cmark &                            &                               &                                         \cmark &                                  &                                                  &                 \cmark \\
       \citeA{Schmidt2020} &                             &                     \cmark &                               &                                                &                           \cmark &                                                  &                        \\
          \citeA{Ooge2021} &                      \cmark &                            &                               &                                                &                                  &                                                  &                        \\
    \citeA{Rechkemmer2022} &                      \cmark &                     \cmark &                               &                                         \cmark &                                  &                                                  &                        \\
       \citeA{Koerber2018} &                      \cmark &                     \cmark &                               &                                                &                           \cmark &                                                  &                        \\
          \citeA{Anik2021} &                      \cmark &                            &                               &                                                &                                  &                                           \cmark &                        \\
        \citeA{Suresh2020} &                             &                     \cmark &                               &                                                &                           \cmark &                                                  &                        \\
            \citeA{Lu2021} &                      \cmark &                     \cmark &                               &                                                &                                  &                                                  &                        \\
       \citeA{Schmidt2019} &                             &                     \cmark &                               &                                                &                           \cmark &                                           \cmark &                        \\
      \citeA{Holliday2016} &                      \cmark &                            &                               &                                         \cmark &                                  &                                           \cmark &                        \\
          \citeA{Wang2023} &                      \cmark &                     \cmark &                               &                                         \cmark &                           \cmark &                                           \cmark &                        \\
          \citeA{Chen2023} &                             &                     \cmark &                               &                                                &                           \cmark &                                           \cmark &                        \\
        \citeA{Guesmi2023} &                      \cmark &                            &                               &                                                &                                  &                                           \cmark &                 \cmark \\
\end{longtable}

\subsection{Trust measurements outcomes}
\label{appendix:trust_outcomes}

\begin{longtable}{l|c|c}
\caption{Measurement outcomes}
\label{tab:measurement_outcomes}\\
\toprule
                     Paper & Measurement Type &                  Measured Effect \\
\midrule
\endfirsthead
\caption[]{Measurement outcomes} \\
\toprule
                     Paper & Measurement Type &                  Measured Effect \\
\midrule
\endhead
\midrule
\multicolumn{3}{r}{{Continued on next page}} \\
\midrule
\endfoot

\bottomrule
\endlastfoot
    \citeA{kulesza2013too} &      self-report &                         positive \\
      \citeA{Kizilcec2016} &      self-report &                            mixed \\
       \citeA{Bussone2015} &      self-report &                         positive \\
                           &       behavioral &                            mixed \\
    \citeA{Papenmeier2022} &      self-report &                            mixed \\
                           &       behavioral &                        no effect \\
            \citeA{Yu2018} &      self-report &                         positive \\
                           &       behavioral &                        no effect \\
            \citeA{Yu2017} &      self-report &                         positive \\
                           &       behavioral &                         positive \\
            \citeA{Yu2019} &      self-report &                         positive \\
                           &       behavioral &                         positive \\
      \citeA{Leffrang2021} &       behavioral &                        no effect \\
         \citeA{Zhang2020} &       behavioral &                         positive \\
\citeA{nourani2019effects} &      self-report &                         positive \\
        \citeA{Omeiza2021} &      self-report &                         negative \\
       \citeA{Nourani2020} &      self-report &                         negative \\
        \citeA{Guesmi2021} &      self-report &                            mixed \\
         \citeA{Jiang2022} &      self-report &                            mixed \\
          \citeA{Shin2021} &      self-report &                         positive \\
    \citeA{Leichtmann2023} &      self-report &                         negative \\
    \citeA{Papenmeier2021} &      self-report &                            mixed \\
                           &       behavioral &                         positive \\
         \citeA{Cheng2019} &      self-report &                        no effect \\
          \citeA{Wang2021} &       behavioral &                            mixed \\
          \citeA{Wang2022} &       behavioral &                            mixed \\
          \citeA{Yang2020} &       behavioral &                         positive \\
        \citeA{Bansal2021} &       behavioral &                            mixed \\
       \citeA{MingYin2019} &      self-report &                         positive \\
                           &       behavioral &                         positive \\
        \citeA{Linder2021} &       behavioral &                        no effect \\
        \citeA{Miller2016} &      self-report &                        no effect \\
                           &       behavioral &                        no effect \\
     \citeA{Lakkaraju2020} &      self-report &                         positive \\
          \citeA{Alam2021} &      self-report &                         positive \\
        \citeA{Eslami2018} &      self-report &                     inconclusive \\
         \citeA{Kulms2019} &      self-report &                         positive \\
                           &       behavioral &                        no effect \\
            \citeA{Yu2016} &      self-report &                         positive \\
          \citeA{Lim2009a} &      self-report &                         positive \\
        \citeA{Cramer2008} &      self-report &                        no effect \\
        \citeA{Eiband2019} &      self-report & positive (but not fully desired) \\
      \citeA{Bruzzese2020} &      self-report &                            mixed \\
       \citeA{Bucinca2021} &      self-report &                     inconclusive \\
          \citeA{Zhou2019} &      self-report &                         positive \\
        \citeA{Thaler2021} &      self-report &                         positive \\
  \citeA{drozdal2020trust} &      self-report &                         positive \\
        \citeA{Kunkel2019} &      self-report &                         positive \\
      \citeA{Pynadath2018} &      self-report &                         positive \\
     \citeA{Berkovsky2017} &      self-report &                            mixed \\
         \citeA{Bayer2021} &      self-report &                            mixed \\
       \citeA{Schmidt2020} &       behavioral &                         negative \\
          \citeA{Ooge2021} &      self-report &                            mixed \\
    \citeA{Rechkemmer2022} &      self-report &                         positive \\
                           &       behavioral &                         positive \\
       \citeA{Koerber2018} &       behavioral &                         positive \\
          \citeA{Anik2021} &      self-report &                         positive \\
        \citeA{Suresh2020} &       behavioral &                         positive \\
       \citeA{Schmidt2019} &       behavioral &                         positive \\
      \citeA{Holliday2016} &      self-report &                            mixed \\
          \citeA{Wang2023} &      self-report &                         positive \\
                           &       behavioral &                        no effect \\
          \citeA{Chen2023} &       behavioral &                         positive \\
        \citeA{Guesmi2023} &      self-report &                         positive \\
\end{longtable}

\bibliographystyle{apacite}
\bibliography{bibliography}

\begin{thebibliography}{}

\bibitem [\protect \citeauthoryear {%
Adadi%
\ \BBA {} Berrada%
}{%
Adadi%
\ \BBA {} Berrada%
}{%
{\protect \APACyear {2018}}%
}]{%
adadi2018peeking}
\APACinsertmetastar {%
adadi2018peeking}%
\begin{APACrefauthors}%
Adadi, A.%
\BCBT {}\ \BBA {} Berrada, M.%
\end{APACrefauthors}%
\unskip\
\newblock
\APACrefYearMonthDay{2018}{}{}.
\newblock
{\BBOQ}\APACrefatitle {Peeking inside the black-box: a survey on explainable
  artificial intelligence ({XAI})} {Peeking inside the black-box: a survey on
  explainable artificial intelligence ({XAI})}.{\BBCQ}
\newblock
\APACjournalVolNumPages{IEEE access}{6}{}{52138--52160}.
\newblock
\begin{APACrefDOI} \doi{10.1109/ACCESS.2018.2870052} \end{APACrefDOI}
\PrintBackRefs{\CurrentBib}

\bibitem [\protect \citeauthoryear {%
Alam%
\ \BBA {} Mueller%
}{%
Alam%
\ \BBA {} Mueller%
}{%
{\protect \APACyear {2021}}%
}]{%
Alam2021}
\APACinsertmetastar {%
Alam2021}%
\begin{APACrefauthors}%
Alam, L.%
\BCBT {}\ \BBA {} Mueller, S.%
\end{APACrefauthors}%
\unskip\
\newblock
\APACrefYearMonthDay{2021}{}{}.
\newblock
{\BBOQ}\APACrefatitle {Examining the effect of explanation on satisfaction and
  trust in AI diagnostic systems} {Examining the effect of explanation on
  satisfaction and trust in ai diagnostic systems}.{\BBCQ}
\newblock
\APACjournalVolNumPages{BMC medical informatics and decision
  making}{21}{1}{178}.
\newblock
\begin{APACrefURL}
  \url{https://bmcmedinformdecismak.biomedcentral.com/articles/10.1186/s12911-021-01542-6}
  \end{APACrefURL}
\PrintBackRefs{\CurrentBib}

\bibitem [\protect \citeauthoryear {%
Anik%
\ \BBA {} Bunt%
}{%
Anik%
\ \BBA {} Bunt%
}{%
{\protect \APACyear {2021}}%
}]{%
Anik2021}
\APACinsertmetastar {%
Anik2021}%
\begin{APACrefauthors}%
Anik, A\BPBI I.%
\BCBT {}\ \BBA {} Bunt, A.%
\end{APACrefauthors}%
\unskip\
\newblock
\APACrefYearMonthDay{2021}{}{}.
\newblock
{\BBOQ}\APACrefatitle {Data-Centric Explanations: Explaining Training Data of
  Machine Learning Systems to Promote Transparency} {Data-centric explanations:
  Explaining training data of machine learning systems to promote
  transparency}.{\BBCQ}
\newblock
\BIn{} \APACrefbtitle {Proceedings of the 2021 CHI Conference on Human Factors
  in Computing Systems.} {Proceedings of the 2021 chi conference on human
  factors in computing systems.}
\newblock
\APACaddressPublisher{New York, NY, USA}{Association for Computing Machinery}.
\newblock
\begin{APACrefURL} \url{https://doi.org/10.1145/3411764.3445736}
  \end{APACrefURL}
\newblock
\begin{APACrefDOI} \doi{10.1145/3411764.3445736} \end{APACrefDOI}
\PrintBackRefs{\CurrentBib}

\bibitem [\protect \citeauthoryear {%
Arrieta%
\ \protect \BOthers {.}}{%
Arrieta%
\ \protect \BOthers {.}}{%
{\protect \APACyear {2020}}%
}]{%
Arrieta2020}
\APACinsertmetastar {%
Arrieta2020}%
\begin{APACrefauthors}%
Arrieta, A.%
, Díaz-Rodríguez, N.%
, Ser, J.%
, Bennetot, A.%
, Tabik, S.%
, Barbado, A.%
\BDBL {}Herrera, F.%
\end{APACrefauthors}%
\unskip\
\newblock
\APACrefYearMonthDay{2020}{12}{30}.
\newblock
{\BBOQ}\APACrefatitle {Explainable Artificial Intelligence ({XAI}): Concepts,
  Taxonomies, Opportunities and Challenges toward Responsible AI} {Explainable
  artificial intelligence ({XAI}): Concepts, taxonomies, opportunities and
  challenges toward responsible ai}.{\BBCQ}.
\newblock
\begin{APACrefURL}
  \url{https://www.sciencedirect.com/science/article/abs/pii/S1566253519308103}
  \end{APACrefURL}
\PrintBackRefs{\CurrentBib}

\bibitem [\protect \citeauthoryear {%
Baer%
\ \BBA {} Colquitt%
}{%
Baer%
\ \BBA {} Colquitt%
}{%
{\protect \APACyear {2018}}%
}]{%
Baer2018}
\APACinsertmetastar {%
Baer2018}%
\begin{APACrefauthors}%
Baer, M.%
\BCBT {}\ \BBA {} Colquitt, J\BPBI A.%
\end{APACrefauthors}%
\unskip\
\newblock
\APACrefYearMonthDay{2018}{}{}.
\newblock
{\BBOQ}\APACrefatitle {Moving toward a more comprehensive consideration of the
  antecedents of trust} {Moving toward a more comprehensive consideration of
  the antecedents of trust}.{\BBCQ}
\newblock
\APACjournalVolNumPages{Routledge companion to trust}{}{}{163--182}.
\PrintBackRefs{\CurrentBib}

\bibitem [\protect \citeauthoryear {%
Bansal%
\ \protect \BOthers {.}}{%
Bansal%
\ \protect \BOthers {.}}{%
{\protect \APACyear {2021}}%
}]{%
Bansal2021}
\APACinsertmetastar {%
Bansal2021}%
\begin{APACrefauthors}%
Bansal, G.%
, Wu, T.%
, Zhou, J.%
, Fok, R.%
, Nushi, B.%
, Kamar, E.%
\BDBL {}Weld, D.%
\end{APACrefauthors}%
\unskip\
\newblock
\APACrefYearMonthDay{2021}{}{}.
\newblock
{\BBOQ}\APACrefatitle {Does the Whole Exceed Its Parts? The Effect of AI
  Explanations on Complementary Team Performance} {Does the whole exceed its
  parts? the effect of ai explanations on complementary team
  performance}.{\BBCQ}
\newblock
\BIn{} \APACrefbtitle {Proceedings of the 2021 CHI Conference on Human Factors
  in Computing Systems.} {Proceedings of the 2021 chi conference on human
  factors in computing systems.}
\newblock
\APACaddressPublisher{New York, NY, USA}{Association for Computing Machinery}.
\newblock
\begin{APACrefURL} \url{https://doi.org/10.1145/3411764.3445717}
  \end{APACrefURL}
\newblock
\begin{APACrefDOI} \doi{10.1145/3411764.3445717} \end{APACrefDOI}
\PrintBackRefs{\CurrentBib}

\bibitem [\protect \citeauthoryear {%
{Barredo Arrieta}%
\ \protect \BOthers {.}}{%
{Barredo Arrieta}%
\ \protect \BOthers {.}}{%
{\protect \APACyear {2020}}%
}]{%
BarredoArrieta.2020}
\APACinsertmetastar {%
BarredoArrieta.2020}%
\begin{APACrefauthors}%
{Barredo Arrieta}, A.%
, D{\'i}az-Rodr{\'i}guez, N.%
, {Del Ser}, J.%
, Bennetot, A.%
, Tabik, S.%
, Barbado, A.%
\BDBL {}Herrera, F.%
\end{APACrefauthors}%
\unskip\
\newblock
\APACrefYearMonthDay{2020}{}{}.
\newblock
{\BBOQ}\APACrefatitle {Explainable Artificial Intelligence (XAI): Concepts,
  taxonomies, opportunities and challenges toward responsible AI} {Explainable
  artificial intelligence (xai): Concepts, taxonomies, opportunities and
  challenges toward responsible ai}.{\BBCQ}
\newblock
\APACjournalVolNumPages{Information Fusion}{58}{}{82--115}.
\newblock
\begin{APACrefDOI} \doi{10.1016/j.inffus.2019.12.012} \end{APACrefDOI}
\PrintBackRefs{\CurrentBib}

\bibitem [\protect \citeauthoryear {%
Benamati%
, Serva%
\BCBL {}\ \BBA {} Fuller%
}{%
Benamati%
\ \protect \BOthers {.}}{%
{\protect \APACyear {2006}}%
}]{%
Benamati.2006}
\APACinsertmetastar {%
Benamati.2006}%
\begin{APACrefauthors}%
Benamati, J.%
, Serva, M\BPBI A.%
\BCBL {}\ \BBA {} Fuller, M\BPBI A.%
\end{APACrefauthors}%
\unskip\
\newblock
\APACrefYearMonthDay{2006}{}{}.
\newblock
{\BBOQ}\APACrefatitle {Are Trust and Distrust Distinct Constructs? An Empirical
  Study of the Effects of Trust and Distrust among Online Banking Users} {Are
  trust and distrust distinct constructs? an empirical study of the effects of
  trust and distrust among online banking users}.{\BBCQ}
\newblock
\BIn{} \APACrefbtitle {Proceedings of the 39th Annual Hawaii International
  Conference on System Sciences (HICSS'06).} {Proceedings of the 39th annual
  hawaii international conference on system sciences (hicss'06).}
\newblock
\APACaddressPublisher{}{IEEE}.
\newblock
\begin{APACrefDOI} \doi{10.1109/hicss.2006.63} \end{APACrefDOI}
\PrintBackRefs{\CurrentBib}

\bibitem [\protect \citeauthoryear {%
Berkovsky%
, Taib%
\BCBL {}\ \BBA {} Conway%
}{%
Berkovsky%
\ \protect \BOthers {.}}{%
{\protect \APACyear {2017}}%
}]{%
Berkovsky2017}
\APACinsertmetastar {%
Berkovsky2017}%
\begin{APACrefauthors}%
Berkovsky, S.%
, Taib, R.%
\BCBL {}\ \BBA {} Conway, D.%
\end{APACrefauthors}%
\unskip\
\newblock
\APACrefYearMonthDay{2017}{}{}.
\newblock
{\BBOQ}\APACrefatitle {How to Recommend? User Trust Factors in Movie
  Recommender Systems} {How to recommend? user trust factors in movie
  recommender systems}.{\BBCQ}
\newblock
\BIn{} \APACrefbtitle {Proceedings of the 22nd International Conference on
  Intelligent User Interfaces} {Proceedings of the 22nd international
  conference on intelligent user interfaces}\ (\BPG~287–300).
\newblock
\APACaddressPublisher{New York, NY, USA}{Association for Computing Machinery}.
\newblock
\begin{APACrefURL} \url{https://doi.org/10.1145/3025171.3025209}
  \end{APACrefURL}
\newblock
\begin{APACrefDOI} \doi{10.1145/3025171.3025209} \end{APACrefDOI}
\PrintBackRefs{\CurrentBib}

\bibitem [\protect \citeauthoryear {%
Bruzzese%
, Gao%
, Dietz%
, Ding%
\BCBL {}\ \BBA {} Romanos%
}{%
Bruzzese%
\ \protect \BOthers {.}}{%
{\protect \APACyear {2020}}%
}]{%
Bruzzese2020}
\APACinsertmetastar {%
Bruzzese2020}%
\begin{APACrefauthors}%
Bruzzese, T.%
, Gao, I.%
, Dietz, G.%
, Ding, C.%
\BCBL {}\ \BBA {} Romanos, A.%
\end{APACrefauthors}%
\unskip\
\newblock
\APACrefYearMonthDay{2020}{}{}.
\newblock
{\BBOQ}\APACrefatitle {Effect of Confidence Indicators on Trust in AI-Generated
  Profiles} {Effect of confidence indicators on trust in ai-generated
  profiles}.{\BBCQ}
\newblock
\BIn{} \APACrefbtitle {Extended Abstracts of the 2020 CHI Conference on Human
  Factors in Computing Systems} {Extended abstracts of the 2020 chi conference
  on human factors in computing systems}\ (\BPG~1–8).
\newblock
\APACaddressPublisher{New York, NY, USA}{Association for Computing Machinery}.
\newblock
\begin{APACrefURL} \url{https://doi.org/10.1145/3334480.3382842}
  \end{APACrefURL}
\newblock
\begin{APACrefDOI} \doi{10.1145/3334480.3382842} \end{APACrefDOI}
\PrintBackRefs{\CurrentBib}

\bibitem [\protect \citeauthoryear {%
Bu\c{c}inca%
, Malaya%
\BCBL {}\ \BBA {} Gajos%
}{%
Bu\c{c}inca%
\ \protect \BOthers {.}}{%
{\protect \APACyear {2021}}%
}]{%
Bucinca2021}
\APACinsertmetastar {%
Bucinca2021}%
\begin{APACrefauthors}%
Bu\c{c}inca, Z.%
, Malaya, M\BPBI B.%
\BCBL {}\ \BBA {} Gajos, K\BPBI Z.%
\end{APACrefauthors}%
\unskip\
\newblock
\APACrefYearMonthDay{2021}{apr}{}.
\newblock
{\BBOQ}\APACrefatitle {To Trust or to Think: Cognitive Forcing Functions Can
  Reduce Overreliance on AI in AI-Assisted Decision-Making} {To trust or to
  think: Cognitive forcing functions can reduce overreliance on ai in
  ai-assisted decision-making}.{\BBCQ}
\newblock
\APACjournalVolNumPages{Proc. ACM Hum.-Comput. Interact.}{5}{CSCW1}{}.
\newblock
\begin{APACrefURL} \url{https://doi.org/10.1145/3449287} \end{APACrefURL}
\newblock
\begin{APACrefDOI} \doi{10.1145/3449287} \end{APACrefDOI}
\PrintBackRefs{\CurrentBib}

\bibitem [\protect \citeauthoryear {%
Bussone%
, Stumpf%
\BCBL {}\ \BBA {} O'Sullivan%
}{%
Bussone%
\ \protect \BOthers {.}}{%
{\protect \APACyear {2015}}%
}]{%
Bussone2015}
\APACinsertmetastar {%
Bussone2015}%
\begin{APACrefauthors}%
Bussone, A.%
, Stumpf, S.%
\BCBL {}\ \BBA {} O'Sullivan, D.%
\end{APACrefauthors}%
\unskip\
\newblock
\APACrefYearMonthDay{2015}{}{}.
\newblock
{\BBOQ}\APACrefatitle {The Role of Explanations on Trust and Reliance in
  Clinical Decision Support Systems} {The role of explanations on trust and
  reliance in clinical decision support systems}.{\BBCQ}
\newblock
\BIn{} \APACrefbtitle {2015 International Conference on Healthcare Informatics}
  {2015 international conference on healthcare informatics}\ (\BPGS\ 160--169).
\newblock
\begin{APACrefDOI} \doi{10.1109/ICHI.2015.26} \end{APACrefDOI}
\PrintBackRefs{\CurrentBib}

\bibitem [\protect \citeauthoryear {%
Chen%
, Liao%
, Vaughan%
\BCBL {}\ \BBA {} Bansal%
}{%
Chen%
\ \protect \BOthers {.}}{%
{\protect \APACyear {2023}}%
}]{%
Chen2023}
\APACinsertmetastar {%
Chen2023}%
\begin{APACrefauthors}%
Chen, V.%
, Liao, Q\BPBI V.%
, Vaughan, J\BPBI W.%
\BCBL {}\ \BBA {} Bansal, G.%
\end{APACrefauthors}%
\unskip\
\newblock
\APACrefYearMonthDay{2023}{}{}.
\newblock
{\BBOQ}\APACrefatitle {Understanding the Role of Human Intuition on Reliance in
  Human-AI Decision-Making with Explanations} {Understanding the role of human
  intuition on reliance in human-ai decision-making with explanations}.{\BBCQ}
\newblock
\APACjournalVolNumPages{ArXiv}{abs/2301.07255}{}{}.
\PrintBackRefs{\CurrentBib}

\bibitem [\protect \citeauthoryear {%
Cheng%
\ \protect \BOthers {.}}{%
Cheng%
\ \protect \BOthers {.}}{%
{\protect \APACyear {2019}}%
}]{%
Cheng2019}
\APACinsertmetastar {%
Cheng2019}%
\begin{APACrefauthors}%
Cheng, H\BHBI F.%
, Wang, R.%
, Zhang, Z.%
, O'Connell, F.%
, Gray, T.%
, Harper, F\BPBI M.%
\BCBL {}\ \BBA {} Zhu, H.%
\end{APACrefauthors}%
\unskip\
\newblock
\APACrefYearMonthDay{2019}{}{}.
\newblock
{\BBOQ}\APACrefatitle {Explaining decision-making algorithms through UI:
  Strategies to help non-expert stakeholders} {Explaining decision-making
  algorithms through ui: Strategies to help non-expert stakeholders}.{\BBCQ}
\newblock
\BIn{} \APACrefbtitle {Proceedings of the 2019 chi conference on human factors
  in computing systems} {Proceedings of the 2019 chi conference on human
  factors in computing systems}\ (\BPGS\ 1--12).
\PrintBackRefs{\CurrentBib}

\bibitem [\protect \citeauthoryear {%
Chien%
, Lewis%
, Hergeth%
, Semnani-Azad%
\BCBL {}\ \BBA {} Sycara%
}{%
Chien%
\ \protect \BOthers {.}}{%
{\protect \APACyear {2015}}%
}]{%
Chien2015}
\APACinsertmetastar {%
Chien2015}%
\begin{APACrefauthors}%
Chien, S\BHBI Y.%
, Lewis, M.%
, Hergeth, S.%
, Semnani-Azad, Z.%
\BCBL {}\ \BBA {} Sycara, K.%
\end{APACrefauthors}%
\unskip\
\newblock
\APACrefYearMonthDay{2015}{sep}{}.
\newblock
{\BBOQ}\APACrefatitle {Cross-Country Validation of a Cultural Scale in
  Measuring Trust in Automation} {Cross-country validation of a cultural scale
  in measuring trust in automation}.{\BBCQ}
\newblock
\APACjournalVolNumPages{Proceedings of the Human Factors and Ergonomics Society
  Annual Meeting}{59}{1}{686--690}.
\newblock
\begin{APACrefDOI} \doi{10.1177/1541931215591149} \end{APACrefDOI}
\PrintBackRefs{\CurrentBib}

\bibitem [\protect \citeauthoryear {%
Chien%
, Lewis%
\BCBL {}\ \BBA {} Sycara%
}{%
Chien%
\ \protect \BOthers {.}}{%
{\protect \APACyear {2016}}%
}]{%
Chien2016}
\APACinsertmetastar {%
Chien2016}%
\begin{APACrefauthors}%
Chien, S\BHBI Y.%
, Lewis, M.%
\BCBL {}\ \BBA {} Sycara, K.%
\end{APACrefauthors}%
\unskip\
\newblock
\APACrefYearMonthDay{2016}{}{}.
\newblock
{\BBOQ}\APACrefatitle {Influence of Cultural Factors in Dynamic Trust in
  Automation} {Influence of cultural factors in dynamic trust in
  automation}.{\BBCQ}
\newblock
\BIn{} \APACrefbtitle {2016 IEEE International Conference on Systems, Man, and
  Cybernetics' SMC 20161 October 9-12,2016' Budapest, Hungary.} {2016 ieee
  international conference on systems, man, and cybernetics' smc 20161 october
  9-12,2016' budapest, hungary.}
\newblock
\APACaddressPublisher{}{IEEE}.
\newblock
\begin{APACrefURL} \url{https://ieeexplore.ieee.org/abstract/document/7844677}
  \end{APACrefURL}
\PrintBackRefs{\CurrentBib}

\bibitem [\protect \citeauthoryear {%
Cho%
}{%
Cho%
}{%
{\protect \APACyear {2006}}%
}]{%
Cho.2006}
\APACinsertmetastar {%
Cho.2006}%
\begin{APACrefauthors}%
Cho, J.%
\end{APACrefauthors}%
\unskip\
\newblock
\APACrefYearMonthDay{2006}{}{}.
\newblock
{\BBOQ}\APACrefatitle {The mechanism of trust and distrust formation and their
  relational outcomes} {The mechanism of trust and distrust formation and their
  relational outcomes}.{\BBCQ}
\newblock
\APACjournalVolNumPages{Journal of Retailing}{82}{1}{25--35}.
\newblock
\begin{APACrefDOI} \doi{10.1016/j. jretai.2005.11.002} \end{APACrefDOI}
\PrintBackRefs{\CurrentBib}

\bibitem [\protect \citeauthoryear {%
Cramer%
\ \protect \BOthers {.}}{%
Cramer%
\ \protect \BOthers {.}}{%
{\protect \APACyear {2008}}%
}]{%
Cramer2008}
\APACinsertmetastar {%
Cramer2008}%
\begin{APACrefauthors}%
Cramer, H.%
, Evers, V.%
, Ramlal, S.%
, Van~Someren, M.%
, Rutledge, L.%
, Stash, N.%
\BDBL {}Wielinga, B.%
\end{APACrefauthors}%
\unskip\
\newblock
\APACrefYearMonthDay{2008}{}{}.
\newblock
{\BBOQ}\APACrefatitle {The effects of transparency on trust in and acceptance
  of a content-based art recommender} {The effects of transparency on trust in
  and acceptance of a content-based art recommender}.{\BBCQ}
\newblock
\APACjournalVolNumPages{User Modeling and User-adapted
  interaction}{18}{5}{455--496}.
\newblock
\begin{APACrefURL}
  \url{https://link.springer.com/article/10.1007/s11257-008-9051-3}
  \end{APACrefURL}
\PrintBackRefs{\CurrentBib}

\bibitem [\protect \citeauthoryear {%
de Visser%
\ \protect \BOthers {.}}{%
de Visser%
\ \protect \BOthers {.}}{%
{\protect \APACyear {2020}}%
}]{%
Visser.2020}
\APACinsertmetastar {%
Visser.2020}%
\begin{APACrefauthors}%
de Visser, E\BPBI J.%
, Peeters, M\BPBI M\BPBI M.%
, Jung, M\BPBI F.%
, Kohn, S.%
, Shaw, T\BPBI H.%
, Pak, R.%
\BCBL {}\ \BBA {} Neerincx, M\BPBI A.%
\end{APACrefauthors}%
\unskip\
\newblock
\APACrefYearMonthDay{2020}{}{}.
\newblock
{\BBOQ}\APACrefatitle {Towards a Theory of Longitudinal Trust Calibration in
  Human--Robot Teams} {Towards a theory of longitudinal trust calibration in
  human--robot teams}.{\BBCQ}
\newblock
\APACjournalVolNumPages{International Journal of Social
  Robotics}{12}{2}{459--478}.
\newblock
\begin{APACrefDOI} \doi{10.1007/s12369-019-00596-x} \end{APACrefDOI}
\PrintBackRefs{\CurrentBib}

\bibitem [\protect \citeauthoryear {%
Dietz%
\ \BBA {} Hartog%
}{%
Dietz%
\ \BBA {} Hartog%
}{%
{\protect \APACyear {2006}}%
}]{%
Dietz2006}
\APACinsertmetastar {%
Dietz2006}%
\begin{APACrefauthors}%
Dietz, G.%
\BCBT {}\ \BBA {} Hartog, D\BPBI N\BPBI D.%
\end{APACrefauthors}%
\unskip\
\newblock
\APACrefYearMonthDay{2006}{sep}{}.
\newblock
{\BBOQ}\APACrefatitle {Measuring trust inside organisations} {Measuring trust
  inside organisations}.{\BBCQ}
\newblock
\APACjournalVolNumPages{Personnel Review}{35}{5}{557--588}.
\newblock
\begin{APACrefDOI} \doi{10.1108/00483480610682299} \end{APACrefDOI}
\PrintBackRefs{\CurrentBib}

\bibitem [\protect \citeauthoryear {%
Dimanov%
, Bhatt%
, Jamnik%
\BCBL {}\ \BBA {} Weller%
}{%
Dimanov%
\ \protect \BOthers {.}}{%
{\protect \APACyear {2020}}%
}]{%
Dimanov2020}
\APACinsertmetastar {%
Dimanov2020}%
\begin{APACrefauthors}%
Dimanov, B.%
, Bhatt, U.%
, Jamnik, M.%
\BCBL {}\ \BBA {} Weller, A.%
\end{APACrefauthors}%
\unskip\
\newblock
\APACrefYearMonthDay{2020}{}{}.
\newblock
{\BBOQ}\APACrefatitle {You Shouldn't Trust Me: Learning Models Which Conceal
  Unfairness From Multiple Explanation Methods} {You shouldn't trust me:
  Learning models which conceal unfairness from multiple explanation
  methods}.{\BBCQ}.
\PrintBackRefs{\CurrentBib}

\bibitem [\protect \citeauthoryear {%
Donald R.~Honeycutt%
}{%
Donald R.~Honeycutt%
}{%
{\protect \APACyear {2020}}%
}]{%
Honeycutt2020}
\APACinsertmetastar {%
Honeycutt2020}%
\begin{APACrefauthors}%
Donald R.~Honeycutt, E\BPBI D\BPBI R., Mahsan~Nourani.%
\end{APACrefauthors}%
\unskip\
\newblock
\APACrefYearMonthDay{2020}{}{}.
\newblock
{\BBOQ}\APACrefatitle {Soliciting Human-in-the-Loop User Feedback for
  Interactive Machine Learning Reduces User Trust and Impressions of Model
  Accuracy} {Soliciting human-in-the-loop user feedback for interactive machine
  learning reduces user trust and impressions of model accuracy}.{\BBCQ}
\newblock
\BIn{} \APACrefbtitle {Proceedings of the Eighth AAAI Conference on Human
  Computation and Crowdsourcing (HCOMP-20).} {Proceedings of the eighth aaai
  conference on human computation and crowdsourcing (hcomp-20).}
\PrintBackRefs{\CurrentBib}

\bibitem [\protect \citeauthoryear {%
Drozdal%
\ \protect \BOthers {.}}{%
Drozdal%
\ \protect \BOthers {.}}{%
{\protect \APACyear {2020}}%
}]{%
drozdal2020trust}
\APACinsertmetastar {%
drozdal2020trust}%
\begin{APACrefauthors}%
Drozdal, J.%
, Weisz, J.%
, Wang, D.%
, Dass, G.%
, Yao, B.%
, Zhao, C.%
\BDBL {}Su, H.%
\end{APACrefauthors}%
\unskip\
\newblock
\APACrefYearMonthDay{2020}{}{}.
\newblock
{\BBOQ}\APACrefatitle {Trust in AutoML: Exploring Information Needs for
  Establishing Trust in Automated Machine Learning Systems} {Trust in automl:
  Exploring information needs for establishing trust in automated machine
  learning systems}.{\BBCQ}
\newblock
\BIn{} \APACrefbtitle {Proceedings of the 25th International Conference on
  Intelligent User Interfaces} {Proceedings of the 25th international
  conference on intelligent user interfaces}\ (\BPGS\ 297--307).
\newblock
\begin{APACrefDOI} \doi{.org/10.1145/3377325.3377501} \end{APACrefDOI}
\PrintBackRefs{\CurrentBib}

\bibitem [\protect \citeauthoryear {%
Ehsan%
\ \BBA {} Riedl%
}{%
Ehsan%
\ \BBA {} Riedl%
}{%
{\protect \APACyear {2020}}%
}]{%
Ehsan2020}
\APACinsertmetastar {%
Ehsan2020}%
\begin{APACrefauthors}%
Ehsan, U.%
\BCBT {}\ \BBA {} Riedl, M\BPBI O.%
\end{APACrefauthors}%
\unskip\
\newblock
\APACrefYearMonthDay{2020}{}{}.
\newblock
{\BBOQ}\APACrefatitle {Human-Centered Explainable AI: Towards a Reflective
  Sociotechnical Approach} {Human-centered explainable ai: Towards a reflective
  sociotechnical approach}.{\BBCQ}
\newblock
\BIn{} \APACrefbtitle {International Conference on Human-Computer Interaction}
  {International conference on human-computer interaction}\ (\BPGS\ 449--466).
\newblock
\APACaddressPublisher{}{{Springer, Cham}}.
\newblock
\begin{APACrefDOI} \doi{10.1007/978-3-030-60117-1\_33} \end{APACrefDOI}
\PrintBackRefs{\CurrentBib}

\bibitem [\protect \citeauthoryear {%
Eiband%
, Buschek%
, Kremer%
\BCBL {}\ \BBA {} Hussmann%
}{%
Eiband%
\ \protect \BOthers {.}}{%
{\protect \APACyear {2019}}%
}]{%
Eiband2019}
\APACinsertmetastar {%
Eiband2019}%
\begin{APACrefauthors}%
Eiband, M.%
, Buschek, D.%
, Kremer, A.%
\BCBL {}\ \BBA {} Hussmann, H.%
\end{APACrefauthors}%
\unskip\
\newblock
\APACrefYearMonthDay{2019}{}{}.
\newblock
{\BBOQ}\APACrefatitle {The Impact of Placebic Explanations on Trust in
  Intelligent Systems} {The impact of placebic explanations on trust in
  intelligent systems}.{\BBCQ}
\newblock
\BIn{} \APACrefbtitle {Extended Abstracts of the 2019 CHI Conference on Human
  Factors in Computing Systems} {Extended abstracts of the 2019 chi conference
  on human factors in computing systems}\ (\BPG~1–6).
\newblock
\APACaddressPublisher{New York, NY, USA}{Association for Computing Machinery}.
\newblock
\begin{APACrefURL} \url{https://doi.org/10.1145/3290607.3312787}
  \end{APACrefURL}
\newblock
\begin{APACrefDOI} \doi{10.1145/3290607.3312787} \end{APACrefDOI}
\PrintBackRefs{\CurrentBib}

\bibitem [\protect \citeauthoryear {%
Eslami%
, Krishna~Kumaran%
, Sandvig%
\BCBL {}\ \BBA {} Karahalios%
}{%
Eslami%
\ \protect \BOthers {.}}{%
{\protect \APACyear {2018}}%
}]{%
Eslami2018}
\APACinsertmetastar {%
Eslami2018}%
\begin{APACrefauthors}%
Eslami, M.%
, Krishna~Kumaran, S\BPBI R.%
, Sandvig, C.%
\BCBL {}\ \BBA {} Karahalios, K.%
\end{APACrefauthors}%
\unskip\
\newblock
\APACrefYearMonthDay{2018}{}{}.
\newblock
{\BBOQ}\APACrefatitle {Communicating Algorithmic Process in Online Behavioral
  Advertising} {Communicating algorithmic process in online behavioral
  advertising}.{\BBCQ}
\newblock
\BIn{} \APACrefbtitle {Proceedings of the 2018 CHI Conference on Human Factors
  in Computing Systems} {Proceedings of the 2018 chi conference on human
  factors in computing systems}\ (\BPG~1–13).
\newblock
\APACaddressPublisher{New York, NY, USA}{Association for Computing Machinery}.
\newblock
\begin{APACrefURL} \url{https://doi.org/10.1145/3173574.3174006}
  \end{APACrefURL}
\newblock
\begin{APACrefDOI} \doi{10.1145/3173574.3174006} \end{APACrefDOI}
\PrintBackRefs{\CurrentBib}

\bibitem [\protect \citeauthoryear {%
Fang%
, Guo%
\BCBL {}\ \BBA {} Zhang%
}{%
Fang%
\ \protect \BOthers {.}}{%
{\protect \APACyear {2015}}%
}]{%
Fang.2015}
\APACinsertmetastar {%
Fang.2015}%
\begin{APACrefauthors}%
Fang, H.%
, Guo, G.%
\BCBL {}\ \BBA {} Zhang, J.%
\end{APACrefauthors}%
\unskip\
\newblock
\APACrefYearMonthDay{2015}{}{}.
\newblock
{\BBOQ}\APACrefatitle {Multi-faceted trust and distrust prediction for
  recommender systems} {Multi-faceted trust and distrust prediction for
  recommender systems}.{\BBCQ}
\newblock
\APACjournalVolNumPages{Decision Support Systems}{71}{}{37--47}.
\newblock
\begin{APACrefDOI} \doi{10.1016/j. dss.2015.01.005} \end{APACrefDOI}
\PrintBackRefs{\CurrentBib}

\bibitem [\protect \citeauthoryear {%
Fein%
}{%
Fein%
}{%
{\protect \APACyear {1996}}%
}]{%
Fein.1996}
\APACinsertmetastar {%
Fein.1996}%
\begin{APACrefauthors}%
Fein, S.%
\end{APACrefauthors}%
\unskip\
\newblock
\APACrefYearMonthDay{1996}{}{}.
\newblock
{\BBOQ}\APACrefatitle {Effects of suspicion on attributional thinking and the
  correspondence bias} {Effects of suspicion on attributional thinking and the
  correspondence bias}.{\BBCQ}
\newblock
\APACjournalVolNumPages{Journal of Personality and Social
  Psychology}{70}{6}{1164--1184}.
\newblock
\begin{APACrefDOI} \doi{10.1037/0022-3514.70.6.1164} \end{APACrefDOI}
\PrintBackRefs{\CurrentBib}

\bibitem [\protect \citeauthoryear {%
Ferrario%
\ \BBA {} Loi%
}{%
Ferrario%
\ \BBA {} Loi%
}{%
{\protect \APACyear {2022}}%
}]{%
ferrario2022}
\APACinsertmetastar {%
ferrario2022}%
\begin{APACrefauthors}%
Ferrario, A.%
\BCBT {}\ \BBA {} Loi, M.%
\end{APACrefauthors}%
\unskip\
\newblock
\APACrefYearMonthDay{2022}{}{}.
\newblock
{\BBOQ}\APACrefatitle {How explainability contributes to trust in AI} {How
  explainability contributes to trust in ai}.{\BBCQ}
\newblock
\BIn{} \APACrefbtitle {2022 ACM Conference on Fairness, Accountability, and
  Transparency} {2022 acm conference on fairness, accountability, and
  transparency}\ (\BPGS\ 1457--1466).
\PrintBackRefs{\CurrentBib}

\bibitem [\protect \citeauthoryear {%
Ferreira%
\ \BBA {} Monteiro%
}{%
Ferreira%
\ \BBA {} Monteiro%
}{%
{\protect \APACyear {2020}}%
}]{%
Ferreira2020}
\APACinsertmetastar {%
Ferreira2020}%
\begin{APACrefauthors}%
Ferreira, J\BPBI J.%
\BCBT {}\ \BBA {} Monteiro, M\BPBI S.%
\end{APACrefauthors}%
\unskip\
\newblock
\APACrefYearMonthDay{2020}{}{}.
\newblock
{\BBOQ}\APACrefatitle {What are people doing about {XAI} user experience? A
  survey on AI explainability research and practice} {What are people doing
  about {XAI} user experience? a survey on ai explainability research and
  practice}.{\BBCQ}
\newblock
\BIn{} \APACrefbtitle {International Conference on Human-Computer Interaction}
  {International conference on human-computer interaction}\ (\BPGS\ 56--73).
\newblock
\begin{APACrefURL}
  \url{https://link.springer.com/chapter/10.1007/978-3-030-49760-6\_4}
  \end{APACrefURL}
\PrintBackRefs{\CurrentBib}

\bibitem [\protect \citeauthoryear {%
Field%
, Miles%
\BCBL {}\ \BBA {} Field%
}{%
Field%
\ \protect \BOthers {.}}{%
{\protect \APACyear {2012}}%
}]{%
Field.2012}
\APACinsertmetastar {%
Field.2012}%
\begin{APACrefauthors}%
Field, Z.%
, Miles, J.%
\BCBL {}\ \BBA {} Field, A.%
\end{APACrefauthors}%
\unskip\
\newblock
\APACrefYear{2012}.
\newblock
\APACrefbtitle {Discovering statistics using {R}} {Discovering statistics using
  {R}}.
\newblock
\APACaddressPublisher{}{Sage}.
\PrintBackRefs{\CurrentBib}

\bibitem [\protect \citeauthoryear {%
Frison%
\ \protect \BOthers {.}}{%
Frison%
\ \protect \BOthers {.}}{%
{\protect \APACyear {2019}}%
}]{%
Frison.2019}
\APACinsertmetastar {%
Frison.2019}%
\begin{APACrefauthors}%
Frison, A\BHBI K.%
, Wintersberger, P.%
, Riener, A.%
, Schartm{\"u}ller, C.%
, Boyle, L\BPBI N.%
, Miller, E.%
\BCBL {}\ \BBA {} Weigl, K.%
\end{APACrefauthors}%
\unskip\
\newblock
\APACrefYearMonthDay{2019}{}{}.
\newblock
{\BBOQ}\APACrefatitle {In UX We Trust} {In ux we trust}.{\BBCQ}
\newblock
\BIn{} S.~Brewster\ (\BED), \APACrefbtitle {Proceedings of the 2019 CHI
  Conference on Human Factors in Computing Systems} {Proceedings of the 2019
  chi conference on human factors in computing systems}\ (\BPGS\ 1--13).
\newblock
\APACaddressPublisher{New York,NY,United States}{{Association for Computing
  Machinery}}.
\newblock
\begin{APACrefDOI} \doi{10.1145/3290605.3300374} \end{APACrefDOI}
\PrintBackRefs{\CurrentBib}

\bibitem [\protect \citeauthoryear {%
Gaube%
\ \protect \BOthers {.}}{%
Gaube%
\ \protect \BOthers {.}}{%
{\protect \APACyear {2021}}%
}]{%
Gaube.2021}
\APACinsertmetastar {%
Gaube.2021}%
\begin{APACrefauthors}%
Gaube, S.%
, Suresh, H.%
, Raue, M.%
, Merritt, A.%
, Berkowitz, S\BPBI J.%
, Lermer, E.%
\BDBL {}Ghassemi, M.%
\end{APACrefauthors}%
\unskip\
\newblock
\APACrefYearMonthDay{2021}{}{}.
\newblock
{\BBOQ}\APACrefatitle {Do as AI say: susceptibility in deployment of clinical
  decision-aids} {Do as ai say: susceptibility in deployment of clinical
  decision-aids}.{\BBCQ}
\newblock
\APACjournalVolNumPages{NPJ digital medicine}{4}{1}{31}.
\PrintBackRefs{\CurrentBib}

\bibitem [\protect \citeauthoryear {%
Glikson%
\ \BBA {} Woolley%
}{%
Glikson%
\ \BBA {} Woolley%
}{%
{\protect \APACyear {2020}}%
}]{%
Glikson.2020}
\APACinsertmetastar {%
Glikson.2020}%
\begin{APACrefauthors}%
Glikson, E.%
\BCBT {}\ \BBA {} Woolley, A\BPBI W.%
\end{APACrefauthors}%
\unskip\
\newblock
\APACrefYearMonthDay{2020}{}{}.
\newblock
{\BBOQ}\APACrefatitle {Human Trust in Artificial Intelligence: Review of
  Empirical Research} {Human trust in artificial intelligence: Review of
  empirical research}.{\BBCQ}
\newblock
\APACjournalVolNumPages{Academy of Management Annals}{14}{2}{627--660}.
\newblock
\begin{APACrefDOI} \doi{10.5465/annals.2018.0057} \end{APACrefDOI}
\PrintBackRefs{\CurrentBib}

\bibitem [\protect \citeauthoryear {%
Guesmi%
\ \protect \BOthers {.}}{%
Guesmi%
\ \protect \BOthers {.}}{%
{\protect \APACyear {2023}}%
}]{%
Guesmi2023}
\APACinsertmetastar {%
Guesmi2023}%
\begin{APACrefauthors}%
Guesmi, M.%
, Chatti, M\BPBI A.%
, Joarder, S.%
, Ain, Q\BPBI U.%
, Alatrash, R.%
, Siepmann, C.%
\BCBL {}\ \BBA {} Vahidi, T.%
\end{APACrefauthors}%
\unskip\
\newblock
\APACrefYearMonthDay{2023}{}{}.
\newblock
{\BBOQ}\APACrefatitle {Interactive Explanation with Varying Level of Details in
  an Explainable Scientific Literature Recommender System} {Interactive
  explanation with varying level of details in an explainable scientific
  literature recommender system}.{\BBCQ}
\newblock
\APACjournalVolNumPages{ArXiv}{abs/2306.05809}{}{}.
\PrintBackRefs{\CurrentBib}

\bibitem [\protect \citeauthoryear {%
Guesmi%
\ \protect \BOthers {.}}{%
Guesmi%
\ \protect \BOthers {.}}{%
{\protect \APACyear {2021}}%
}]{%
Guesmi2021}
\APACinsertmetastar {%
Guesmi2021}%
\begin{APACrefauthors}%
Guesmi, M.%
, Chatti, M\BPBI A.%
, Vorgerd, L.%
, Joarder, S\BPBI A.%
, Ain, Q\BPBI U.%
, Ngo, T.%
\BDBL {}Muslim, A.%
\end{APACrefauthors}%
\unskip\
\newblock
\APACrefYearMonthDay{2021}{}{}.
\newblock
{\BBOQ}\APACrefatitle {Input or Output: Effects of Explanation Focus on the
  Perception of Explainable Recommendation with Varying Level of Details.}
  {Input or output: Effects of explanation focus on the perception of
  explainable recommendation with varying level of details.}{\BBCQ}
\newblock
\BIn{} \APACrefbtitle {IntRS@ RecSys} {Intrs@ recsys}\ (\BPGS\ 55--72).
\PrintBackRefs{\CurrentBib}

\bibitem [\protect \citeauthoryear {%
Guidotti%
\ \protect \BOthers {.}}{%
Guidotti%
\ \protect \BOthers {.}}{%
{\protect \APACyear {2019}}%
}]{%
Guidotti2019}
\APACinsertmetastar {%
Guidotti2019}%
\begin{APACrefauthors}%
Guidotti, R.%
, Monreale, A.%
, Ruggieri, S.%
, Turini, F.%
, Giannotti, F.%
\BCBL {}\ \BBA {} Pedreschi, D.%
\end{APACrefauthors}%
\unskip\
\newblock
\APACrefYearMonthDay{2019}{sep}{}.
\newblock
{\BBOQ}\APACrefatitle {A Survey of Methods for Explaining Black Box Models} {A
  survey of methods for explaining black box models}.{\BBCQ}
\newblock
\APACjournalVolNumPages{{ACM} Computing Surveys}{51}{5}{1--42}.
\newblock
\begin{APACrefDOI} \doi{10.1145/3236009} \end{APACrefDOI}
\PrintBackRefs{\CurrentBib}

\bibitem [\protect \citeauthoryear {%
Gunning%
\ \BBA {} Aha%
}{%
Gunning%
\ \BBA {} Aha%
}{%
{\protect \APACyear {2019}}%
}]{%
Gunning.2019}
\APACinsertmetastar {%
Gunning.2019}%
\begin{APACrefauthors}%
Gunning, D.%
\BCBT {}\ \BBA {} Aha, D.%
\end{APACrefauthors}%
\unskip\
\newblock
\APACrefYearMonthDay{2019}{}{}.
\newblock
{\BBOQ}\APACrefatitle {DARPA's Explainable Artificial Intelligence (XAI)
  Program} {Darpa's explainable artificial intelligence (xai) program}.{\BBCQ}
\newblock
\APACjournalVolNumPages{AI Magazine}{40}{2}{44--58}.
\newblock
\begin{APACrefDOI} \doi{10.1609/aimag.v40i2.2850} \end{APACrefDOI}
\PrintBackRefs{\CurrentBib}

\bibitem [\protect \citeauthoryear {%
S\BHBI L.~Guo%
, Lumineau%
\BCBL {}\ \BBA {} Lewicki%
}{%
S\BHBI L.~Guo%
\ \protect \BOthers {.}}{%
{\protect \APACyear {2017}}%
}]{%
Guo.2017}
\APACinsertmetastar {%
Guo.2017}%
\begin{APACrefauthors}%
Guo, S\BHBI L.%
, Lumineau, F.%
\BCBL {}\ \BBA {} Lewicki, R\BPBI J.%
\end{APACrefauthors}%
\unskip\
\newblock
\APACrefYearMonthDay{2017}{}{}.
\newblock
{\BBOQ}\APACrefatitle {Revisiting the Foundations of Organizational Distrust}
  {Revisiting the foundations of organizational distrust}.{\BBCQ}
\newblock
\APACjournalVolNumPages{Foundations and Trends{\circledR} in
  Management}{1}{1}{1--88}.
\newblock
\begin{APACrefDOI} \doi{10.1561/3400000001} \end{APACrefDOI}
\PrintBackRefs{\CurrentBib}

\bibitem [\protect \citeauthoryear {%
Y.~Guo%
\ \BBA {} Yang%
}{%
Y.~Guo%
\ \BBA {} Yang%
}{%
{\protect \APACyear {2021}}%
}]{%
guo2021modeling}
\APACinsertmetastar {%
guo2021modeling}%
\begin{APACrefauthors}%
Guo, Y.%
\BCBT {}\ \BBA {} Yang, X\BPBI J.%
\end{APACrefauthors}%
\unskip\
\newblock
\APACrefYearMonthDay{2021}{}{}.
\newblock
{\BBOQ}\APACrefatitle {Modeling and Predicting Trust Dynamics in Human--Robot
  Teaming: A Bayesian Inference Approach} {Modeling and predicting trust
  dynamics in human--robot teaming: A bayesian inference approach}.{\BBCQ}
\newblock
\APACjournalVolNumPages{International Journal of Social
  Robotics}{13}{8}{1899--1909}.
\newblock
\begin{APACrefURL}
  \url{https://link.springer.com/article/10.1007/s12369-020-00703-3}
  \end{APACrefURL}
\newblock
\begin{APACrefDOI} \doi{10.1007/s12369-020-00703-3} \end{APACrefDOI}
\PrintBackRefs{\CurrentBib}

\bibitem [\protect \citeauthoryear {%
Gurney%
, Pynadath%
\BCBL {}\ \BBA {} Wang%
}{%
Gurney%
\ \protect \BOthers {.}}{%
{\protect \APACyear {2022}}%
}]{%
Gurney2022}
\APACinsertmetastar {%
Gurney2022}%
\begin{APACrefauthors}%
Gurney, N.%
, Pynadath, D\BPBI V.%
\BCBL {}\ \BBA {} Wang, N.%
\end{APACrefauthors}%
\unskip\
\newblock
\APACrefYearMonthDay{2022}{}{}.
\newblock
{\BBOQ}\APACrefatitle {Measuring and~Predicting Human Trust in~Recommendations
  from~an~{AI} Teammate} {Measuring and~predicting human trust
  in~recommendations from~an~{AI} teammate}.{\BBCQ}
\newblock
\BIn{} ©c The~Author(s), under exclusive license~to Springer Nature
  Switzerland AG 2022 H.~Degen\BCBL {}\ \BBA {} S.~Ntoa\ (\BEDS),
  \APACrefbtitle {Artificial Intelligence in {HCI}} {Artificial intelligence in
  {HCI}}\ (\BVOL\ 13336, \BPGS\ 22--34).
\newblock
\APACaddressPublisher{}{Springer International Publishing}.
\newblock
\begin{APACrefDOI} \doi{10.1007/978-3-031-05643-7\_2} \end{APACrefDOI}
\PrintBackRefs{\CurrentBib}

\bibitem [\protect \citeauthoryear {%
Han%
\ \BBA {} Schulz%
}{%
Han%
\ \BBA {} Schulz%
}{%
{\protect \APACyear {2020}}%
}]{%
Han2020}
\APACinsertmetastar {%
Han2020}%
\begin{APACrefauthors}%
Han, W.%
\BCBT {}\ \BBA {} Schulz, H\BHBI J.%
\end{APACrefauthors}%
\unskip\
\newblock
\APACrefYearMonthDay{2020}{}{}.
\newblock
{\BBOQ}\APACrefatitle {Beyond Trust Building — Calibrating Trust in Visual
  Analytics} {Beyond trust building — calibrating trust in visual
  analytics}.{\BBCQ}
\newblock
\BIn{} \APACrefbtitle {2020 IEEE Workshop on TRust and EXpertise in Visual
  Analytics (TREX)} {2020 ieee workshop on trust and expertise in visual
  analytics (trex)}\ (\BPGS\ 9--15).
\newblock
\begin{APACrefDOI} \doi{10.1109/TREX51495.2020.00006} \end{APACrefDOI}
\PrintBackRefs{\CurrentBib}

\bibitem [\protect \citeauthoryear {%
{Harrison McKnight}%
\ \BBA {} Chervany%
}{%
{Harrison McKnight}%
\ \BBA {} Chervany%
}{%
{\protect \APACyear {2001}}%
}]{%
HarrisonMcKnight.2001}
\APACinsertmetastar {%
HarrisonMcKnight.2001}%
\begin{APACrefauthors}%
{Harrison McKnight}, D.%
\BCBT {}\ \BBA {} Chervany, N\BPBI L.%
\end{APACrefauthors}%
\unskip\
\newblock
\APACrefYearMonthDay{2001}{}{}.
\newblock
{\BBOQ}\APACrefatitle {Trust and Distrust Definitions: One Bite at a Time}
  {Trust and distrust definitions: One bite at a time}.{\BBCQ}
\newblock
\BIn{} \APACrefbtitle {Trust in Cyber-societies} {Trust in cyber-societies}\
  (\BPGS\ 27--54).
\newblock
\APACaddressPublisher{}{{Springer, Berlin, Heidelberg}}.
\newblock
\begin{APACrefDOI} \doi{10.1007/3-540-45547-7\_3} \end{APACrefDOI}
\PrintBackRefs{\CurrentBib}

\bibitem [\protect \citeauthoryear {%
HLEG%
}{%
HLEG%
}{%
{\protect \APACyear {2019}}%
}]{%
hlegAI}
\APACinsertmetastar {%
hlegAI}%
\begin{APACrefauthors}%
HLEG, A.%
\end{APACrefauthors}%
\unskip\
\newblock
\APACrefYearMonthDay{2019}{}{}.
\newblock
\APACrefbtitle {Ethics guidelines for trustworthy AI.} {Ethics guidelines for
  trustworthy ai.}
\newblock
\begin{APACrefURL}
  [{22.03.2023}]\url{https://digital-strategy.ec.europa.eu/en/library/ethics-guidelines-trustworthy-ai}
  \end{APACrefURL}
\PrintBackRefs{\CurrentBib}

\bibitem [\protect \citeauthoryear {%
Hoff%
\ \BBA {} Bashir%
}{%
Hoff%
\ \BBA {} Bashir%
}{%
{\protect \APACyear {2015}}%
}]{%
hoff2015trust}
\APACinsertmetastar {%
hoff2015trust}%
\begin{APACrefauthors}%
Hoff, K\BPBI A.%
\BCBT {}\ \BBA {} Bashir, M.%
\end{APACrefauthors}%
\unskip\
\newblock
\APACrefYearMonthDay{2015}{}{}.
\newblock
{\BBOQ}\APACrefatitle {Trust in automation: Integrating empirical evidence on
  factors that influence trust} {Trust in automation: Integrating empirical
  evidence on factors that influence trust}.{\BBCQ}
\newblock
\APACjournalVolNumPages{Human factors}{57}{3}{407--434}.
\newblock
\begin{APACrefDOI} \doi{10.1177/0018720814547570} \end{APACrefDOI}
\PrintBackRefs{\CurrentBib}

\bibitem [\protect \citeauthoryear {%
R.~Hoffman%
, Mueller%
, Klein%
\BCBL {}\ \BBA {} Litman%
}{%
R.~Hoffman%
\ \protect \BOthers {.}}{%
{\protect \APACyear {2018}}%
}]{%
Hoffman.2018}
\APACinsertmetastar {%
Hoffman.2018}%
\begin{APACrefauthors}%
Hoffman, R.%
, Mueller, S\BPBI T.%
, Klein, G.%
\BCBL {}\ \BBA {} Litman, J.%
\end{APACrefauthors}%
\unskip\
\newblock
\APACrefYear{2018}.
\newblock
\APACrefbtitle {Measuring Trust in the XAI Context: Technical Report, DARPA
  Explainable AI Program.} {Measuring trust in the xai context: Technical
  report, darpa explainable ai program.}
\newblock
\begin{APACrefDOI} \doi{10.31234/osf.io/e3kv9} \end{APACrefDOI}
\PrintBackRefs{\CurrentBib}

\bibitem [\protect \citeauthoryear {%
R\BPBI R.~Hoffman%
\ \protect \BOthers {.}}{%
R\BPBI R.~Hoffman%
\ \protect \BOthers {.}}{%
{\protect \APACyear {2009}}%
}]{%
hoffman2009dynamics}
\APACinsertmetastar {%
hoffman2009dynamics}%
\begin{APACrefauthors}%
Hoffman, R\BPBI R.%
, Lee, J\BPBI D.%
, Woods, D\BPBI D.%
, Shadbolt, N.%
, Miller, J.%
\BCBL {}\ \BBA {} Bradshaw, J\BPBI M.%
\end{APACrefauthors}%
\unskip\
\newblock
\APACrefYearMonthDay{2009}{}{}.
\newblock
{\BBOQ}\APACrefatitle {The dynamics of trust in cyberdomains} {The dynamics of
  trust in cyberdomains}.{\BBCQ}
\newblock
\APACjournalVolNumPages{IEEE Intelligent Systems}{24}{6}{5--11}.
\PrintBackRefs{\CurrentBib}

\bibitem [\protect \citeauthoryear {%
R\BPBI R.~Hoffman%
, Mueller%
, Klein%
\BCBL {}\ \BBA {} Litman%
}{%
R\BPBI R.~Hoffman%
\ \protect \BOthers {.}}{%
{\protect \APACyear {2018}}%
}]{%
hoffman2018metrics}
\APACinsertmetastar {%
hoffman2018metrics}%
\begin{APACrefauthors}%
Hoffman, R\BPBI R.%
, Mueller, S\BPBI T.%
, Klein, G.%
\BCBL {}\ \BBA {} Litman, J.%
\end{APACrefauthors}%
\unskip\
\newblock
\APACrefYearMonthDay{2018}{}{}.
\newblock
{\BBOQ}\APACrefatitle {Metrics for explainable AI: Challenges and prospects}
  {Metrics for explainable ai: Challenges and prospects}.{\BBCQ}
\newblock
\APACjournalVolNumPages{arXiv preprint arXiv:1812.04608}{}{}{}.
\newblock
\begin{APACrefURL} \url{https://arxiv.org/abs/1812.04608} \end{APACrefURL}
\PrintBackRefs{\CurrentBib}

\bibitem [\protect \citeauthoryear {%
Holliday%
, Wilson%
\BCBL {}\ \BBA {} Stumpf%
}{%
Holliday%
\ \protect \BOthers {.}}{%
{\protect \APACyear {2016}}%
}]{%
Holliday2016}
\APACinsertmetastar {%
Holliday2016}%
\begin{APACrefauthors}%
Holliday, D.%
, Wilson, S.%
\BCBL {}\ \BBA {} Stumpf, S.%
\end{APACrefauthors}%
\unskip\
\newblock
\APACrefYearMonthDay{2016}{}{}.
\newblock
{\BBOQ}\APACrefatitle {User Trust in Intelligent Systems: A Journey Over Time}
  {User trust in intelligent systems: A journey over time}.{\BBCQ}
\newblock
\BIn{} \APACrefbtitle {Proceedings of the 21st International Conference on
  Intelligent User Interfaces} {Proceedings of the 21st international
  conference on intelligent user interfaces}\ (\BPG~164–168).
\newblock
\APACaddressPublisher{New York, NY, USA}{Association for Computing Machinery}.
\newblock
\begin{APACrefURL} \url{https://doi.org/10.1145/2856767.2856811}
  \end{APACrefURL}
\newblock
\begin{APACrefDOI} \doi{10.1145/2856767.2856811} \end{APACrefDOI}
\PrintBackRefs{\CurrentBib}

\bibitem [\protect \citeauthoryear {%
Jacovi%
, Marasovi{\'c}%
, Miller%
\BCBL {}\ \BBA {} Goldberg%
}{%
Jacovi%
\ \protect \BOthers {.}}{%
{\protect \APACyear {2021}}%
}]{%
Jacovi.2021}
\APACinsertmetastar {%
Jacovi.2021}%
\begin{APACrefauthors}%
Jacovi, A.%
, Marasovi{\'c}, A.%
, Miller, T.%
\BCBL {}\ \BBA {} Goldberg, Y.%
\end{APACrefauthors}%
\unskip\
\newblock
\APACrefYearMonthDay{2021}{}{}.
\newblock
{\BBOQ}\APACrefatitle {Formalizing Trust in Artificial Intelligence}
  {Formalizing trust in artificial intelligence}.{\BBCQ}
\newblock
\BIn{} \APACrefbtitle {Proceedings of the 2021 ACM Conference on Fairness,
  Accountability, and Transparency} {Proceedings of the 2021 acm conference on
  fairness, accountability, and transparency}\ (\BPGS\ 624--635).
\newblock
\APACaddressPublisher{New York,NY,United States}{{Association for Computing
  Machinery}}.
\newblock
\begin{APACrefDOI} \doi{10.1145/3442188.3445923} \end{APACrefDOI}
\PrintBackRefs{\CurrentBib}

\bibitem [\protect \citeauthoryear {%
Jian%
, Bisantz%
\BCBL {}\ \BBA {} Drury%
}{%
Jian%
\ \protect \BOthers {.}}{%
{\protect \APACyear {2000}}%
}]{%
Jian.2000}
\APACinsertmetastar {%
Jian.2000}%
\begin{APACrefauthors}%
Jian, J\BHBI Y.%
, Bisantz, A\BPBI M.%
\BCBL {}\ \BBA {} Drury, C\BPBI G.%
\end{APACrefauthors}%
\unskip\
\newblock
\APACrefYearMonthDay{2000}{}{}.
\newblock
{\BBOQ}\APACrefatitle {Foundations for an Empirically Determined Scale of Trust
  in Automated Systems} {Foundations for an empirically determined scale of
  trust in automated systems}.{\BBCQ}
\newblock
\APACjournalVolNumPages{International Journal of Cognitive
  Ergonomics}{4}{1}{53--71}.
\newblock
\begin{APACrefDOI} \doi{10.1207/S15327566 IJCE0401{\textunderscore }04}
  \end{APACrefDOI}
\PrintBackRefs{\CurrentBib}

\bibitem [\protect \citeauthoryear {%
Jiang%
, Kahai%
\BCBL {}\ \BBA {} Yang%
}{%
Jiang%
\ \protect \BOthers {.}}{%
{\protect \APACyear {2022}}%
}]{%
Jiang2022}
\APACinsertmetastar {%
Jiang2022}%
\begin{APACrefauthors}%
Jiang, J.%
, Kahai, S.%
\BCBL {}\ \BBA {} Yang, M.%
\end{APACrefauthors}%
\unskip\
\newblock
\APACrefYearMonthDay{2022}{}{}.
\newblock
{\BBOQ}\APACrefatitle {Who needs explanation and when? Juggling explainable AI
  and user epistemic uncertainty} {Who needs explanation and when? juggling
  explainable ai and user epistemic uncertainty}.{\BBCQ}
\newblock
\APACjournalVolNumPages{International Journal of Human-Computer
  Studies}{165}{}{102839}.
\newblock
\begin{APACrefURL}
  \url{https://www.sciencedirect.com/science/article/pii/S1071581922000660}
  \end{APACrefURL}
\newblock
\begin{APACrefDOI} \doi{10.1016/j.ijhcs.2022.102839} \end{APACrefDOI}
\PrintBackRefs{\CurrentBib}

\bibitem [\protect \citeauthoryear {%
Joppe%
}{%
Joppe%
}{%
{\protect \APACyear {2000}}%
}]{%
Joppe2000}
\APACinsertmetastar {%
Joppe2000}%
\begin{APACrefauthors}%
Joppe, M.%
\end{APACrefauthors}%
\unskip\
\newblock
\APACrefYearMonthDay{2000}{}{}.
\newblock
\APACrefbtitle {The Research Process. Retrieved February 25, 1998.} {The
  research process. retrieved february 25, 1998.}
\PrintBackRefs{\CurrentBib}

\bibitem [\protect \citeauthoryear {%
Kaplan%
, Kessler%
, Brill%
\BCBL {}\ \BBA {} Hancock%
}{%
Kaplan%
\ \protect \BOthers {.}}{%
{\protect \APACyear {2023}}%
}]{%
Kaplan2023}
\APACinsertmetastar {%
Kaplan2023}%
\begin{APACrefauthors}%
Kaplan, A\BPBI D.%
, Kessler, T\BPBI T.%
, Brill, J\BPBI C.%
\BCBL {}\ \BBA {} Hancock, P\BPBI A.%
\end{APACrefauthors}%
\unskip\
\newblock
\APACrefYearMonthDay{2023}{}{}.
\newblock
{\BBOQ}\APACrefatitle {Trust in Artificial Intelligence: Meta-Analytic
  Findings} {Trust in artificial intelligence: Meta-analytic findings}.{\BBCQ}
\newblock
\APACjournalVolNumPages{Human Factors}{65}{2}{337--359}.
\newblock
\begin{APACrefURL} \url{https://doi.org/10.1177/00187208211013988}
  \end{APACrefURL}
\newblock
\APACrefnote{PMID: 34048287}
\newblock
\begin{APACrefDOI} \doi{10.1177/00187208211013988} \end{APACrefDOI}
\PrintBackRefs{\CurrentBib}

\bibitem [\protect \citeauthoryear {%
Kastner%
\ \protect \BOthers {.}}{%
Kastner%
\ \protect \BOthers {.}}{%
{\protect \APACyear {2021}}%
}]{%
Kastner.2021}
\APACinsertmetastar {%
Kastner.2021}%
\begin{APACrefauthors}%
Kastner, L.%
, Langer, M.%
, Lazar, V.%
, Schomacker, A.%
, Speith, T.%
\BCBL {}\ \BBA {} Sterz, S.%
\end{APACrefauthors}%
\unskip\
\newblock
\APACrefYearMonthDay{2021}{}{}.
\newblock
{\BBOQ}\APACrefatitle {On the Relation of Trust and Explainability: Why to
  Engineer for Trustworthiness} {On the relation of trust and explainability:
  Why to engineer for trustworthiness}.{\BBCQ}
\newblock
\BIn{} \APACrefbtitle {Proceedings, 29th IEEE International Requirements
  Engineering Conference Workshops : REW 2021 : September 20-24 2021, online
  event} {Proceedings, 29th ieee international requirements engineering
  conference workshops : Rew 2021 : September 20-24 2021, online event}\
  (\BPGS\ 169--175).
\newblock
\APACaddressPublisher{Los Alamitos, California}{{IEEE Computer Society,
  Conference Publishing Services}}.
\newblock
\begin{APACrefDOI} \doi{10.1109/REW53955.2021.00031} \end{APACrefDOI}
\PrintBackRefs{\CurrentBib}

\bibitem [\protect \citeauthoryear {%
Kee%
\ \BBA {} Knox%
}{%
Kee%
\ \BBA {} Knox%
}{%
{\protect \APACyear {1970}}%
}]{%
Kee1970}
\APACinsertmetastar {%
Kee1970}%
\begin{APACrefauthors}%
Kee, H\BPBI W.%
\BCBT {}\ \BBA {} Knox, R\BPBI E.%
\end{APACrefauthors}%
\unskip\
\newblock
\APACrefYearMonthDay{1970}{}{}.
\newblock
{\BBOQ}\APACrefatitle {Conceptual and methodological considerations in the
  study of trust and suspicion} {Conceptual and methodological considerations
  in the study of trust and suspicion}.{\BBCQ}
\newblock
\APACjournalVolNumPages{Journal of conflict resolution}{14}{3}{357--366}.
\PrintBackRefs{\CurrentBib}

\bibitem [\protect \citeauthoryear {%
Kizilcec%
}{%
Kizilcec%
}{%
{\protect \APACyear {2016}}%
}]{%
Kizilcec2016}
\APACinsertmetastar {%
Kizilcec2016}%
\begin{APACrefauthors}%
Kizilcec, R\BPBI F.%
\end{APACrefauthors}%
\unskip\
\newblock
\APACrefYearMonthDay{2016}{may}{}.
\newblock
{\BBOQ}\APACrefatitle {How Much Information? Effects of Transparency on Trust
  in an Algorithmic Interface} {How much information? effects of transparency
  on trust in an algorithmic interface}.{\BBCQ}
\newblock
\BIn{} \APACrefbtitle {Proceedings of the 2016 {CHI} Conference on Human
  Factors in Computing Systems.} {Proceedings of the 2016 {CHI} conference on
  human factors in computing systems.}
\newblock
\APACaddressPublisher{}{{ACM}}.
\newblock
\begin{APACrefDOI} \doi{10.1145/2858036.2858402} \end{APACrefDOI}
\PrintBackRefs{\CurrentBib}

\bibitem [\protect \citeauthoryear {%
Kohn%
, de Visser%
, Wiese%
, Lee%
\BCBL {}\ \BBA {} Shaw%
}{%
Kohn%
\ \protect \BOthers {.}}{%
{\protect \APACyear {2021}}%
}]{%
Kohn.2021}
\APACinsertmetastar {%
Kohn.2021}%
\begin{APACrefauthors}%
Kohn, S\BPBI C.%
, de Visser, E\BPBI J.%
, Wiese, E.%
, Lee, Y\BHBI C.%
\BCBL {}\ \BBA {} Shaw, T\BPBI H.%
\end{APACrefauthors}%
\unskip\
\newblock
\APACrefYearMonthDay{2021}{}{}.
\newblock
{\BBOQ}\APACrefatitle {Measurement of Trust in Automation: A Narrative Review
  and Reference Guide} {Measurement of trust in automation: A narrative review
  and reference guide}.{\BBCQ}
\newblock
\APACjournalVolNumPages{Frontiers in psychology}{12}{}{604977}.
\newblock
\begin{APACrefDOI} \doi{10.3389/fpsyg.2021.604977} \end{APACrefDOI}
\PrintBackRefs{\CurrentBib}

\bibitem [\protect \citeauthoryear {%
Kulesza%
\ \protect \BOthers {.}}{%
Kulesza%
\ \protect \BOthers {.}}{%
{\protect \APACyear {2013}}%
}]{%
kulesza2013too}
\APACinsertmetastar {%
kulesza2013too}%
\begin{APACrefauthors}%
Kulesza, T.%
, Stumpf, S.%
, Burnett, M.%
, Yang, S.%
, Kwan, I.%
\BCBL {}\ \BBA {} Wong, W\BHBI K.%
\end{APACrefauthors}%
\unskip\
\newblock
\APACrefYearMonthDay{2013}{}{}.
\newblock
{\BBOQ}\APACrefatitle {Too much, too little, or just right? Ways explanations
  impact end users' mental models} {Too much, too little, or just right? ways
  explanations impact end users' mental models}.{\BBCQ}
\newblock
\BIn{} \APACrefbtitle {2013 IEEE Symposium on visual languages and human
  centric computing} {2013 ieee symposium on visual languages and human centric
  computing}\ (\BPGS\ 3--10).
\newblock
\begin{APACrefDOI} \doi{.org/10.1109/VLHCC.2013.6645235} \end{APACrefDOI}
\PrintBackRefs{\CurrentBib}

\bibitem [\protect \citeauthoryear {%
Kulms%
\ \BBA {} Kopp%
}{%
Kulms%
\ \BBA {} Kopp%
}{%
{\protect \APACyear {2019}}%
}]{%
Kulms2019}
\APACinsertmetastar {%
Kulms2019}%
\begin{APACrefauthors}%
Kulms, P.%
\BCBT {}\ \BBA {} Kopp, S.%
\end{APACrefauthors}%
\unskip\
\newblock
\APACrefYearMonthDay{2019}{}{}.
\newblock
{\BBOQ}\APACrefatitle {More Human-Likeness, More Trust? The Effect of
  Anthropomorphism on Self-Reported and Behavioral Trust in Continued and
  Interdependent Human-Agent Cooperation} {More human-likeness, more trust? the
  effect of anthropomorphism on self-reported and behavioral trust in continued
  and interdependent human-agent cooperation}.{\BBCQ}
\newblock
\BIn{} \APACrefbtitle {Proceedings of Mensch Und Computer 2019} {Proceedings of
  mensch und computer 2019}\ (\BPG~31–42).
\newblock
\APACaddressPublisher{New York, NY, USA}{Association for Computing Machinery}.
\newblock
\begin{APACrefURL} \url{https://doi.org/10.1145/3340764.3340793}
  \end{APACrefURL}
\newblock
\begin{APACrefDOI} \doi{10.1145/3340764.3340793} \end{APACrefDOI}
\PrintBackRefs{\CurrentBib}

\bibitem [\protect \citeauthoryear {%
Kunkel%
, Donkers%
, Michael%
, Barbu%
\BCBL {}\ \BBA {} Ziegler%
}{%
Kunkel%
\ \protect \BOthers {.}}{%
{\protect \APACyear {2019}}%
}]{%
Kunkel2019}
\APACinsertmetastar {%
Kunkel2019}%
\begin{APACrefauthors}%
Kunkel, J.%
, Donkers, T.%
, Michael, L.%
, Barbu, C\BHBI M.%
\BCBL {}\ \BBA {} Ziegler, J.%
\end{APACrefauthors}%
\unskip\
\newblock
\APACrefYearMonthDay{2019}{}{}.
\newblock
{\BBOQ}\APACrefatitle {Let Me Explain: Impact of Personal and Impersonal
  Explanations on Trust in Recommender Systems} {Let me explain: Impact of
  personal and impersonal explanations on trust in recommender systems}.{\BBCQ}
\newblock
\BIn{} \APACrefbtitle {Proceedings of the 2019 CHI Conference on Human Factors
  in Computing Systems} {Proceedings of the 2019 chi conference on human
  factors in computing systems}\ (\BPG~1–12).
\newblock
\APACaddressPublisher{New York, NY, USA}{Association for Computing Machinery}.
\newblock
\begin{APACrefURL} \url{https://doi.org/10.1145/3290605.3300717}
  \end{APACrefURL}
\newblock
\begin{APACrefDOI} \doi{10.1145/3290605.3300717} \end{APACrefDOI}
\PrintBackRefs{\CurrentBib}

\bibitem [\protect \citeauthoryear {%
Körber%
, Baseler%
\BCBL {}\ \BBA {} Bengler%
}{%
Körber%
\ \protect \BOthers {.}}{%
{\protect \APACyear {2018}}%
}]{%
Koerber2018}
\APACinsertmetastar {%
Koerber2018}%
\begin{APACrefauthors}%
Körber, M.%
, Baseler, E.%
\BCBL {}\ \BBA {} Bengler, K.%
\end{APACrefauthors}%
\unskip\
\newblock
\APACrefYearMonthDay{2018}{jan}{}.
\newblock
{\BBOQ}\APACrefatitle {Introduction matters: Manipulating trust in automation
  and reliance in automated driving} {Introduction matters: Manipulating trust
  in automation and reliance in automated driving}.{\BBCQ}
\newblock
\APACjournalVolNumPages{Applied Ergonomics}{66}{Munich.}{18--31}.
\newblock
\begin{APACrefDOI} \doi{10.1016/j.apergo.2017.07.006} \end{APACrefDOI}
\PrintBackRefs{\CurrentBib}

\bibitem [\protect \citeauthoryear {%
Lai%
, Chen%
, Liao%
, Smith-Renner%
\BCBL {}\ \BBA {} Tan%
}{%
Lai%
\ \protect \BOthers {.}}{%
{\protect \APACyear {2021}}%
}]{%
Lai2021}
\APACinsertmetastar {%
Lai2021}%
\begin{APACrefauthors}%
Lai, V.%
, Chen, C.%
, Liao, Q\BPBI V.%
, Smith-Renner, A.%
\BCBL {}\ \BBA {} Tan, C.%
\end{APACrefauthors}%
\unskip\
\newblock
\APACrefYearMonthDay{2021}{}{}.
\newblock
\APACrefbtitle {Towards a Science of Human-AI Decision Making: A Survey of
  Empirical Studies.} {Towards a science of human-ai decision making: A survey
  of empirical studies.}
\PrintBackRefs{\CurrentBib}

\bibitem [\protect \citeauthoryear {%
Lakkaraju%
\ \BBA {} Bastani%
}{%
Lakkaraju%
\ \BBA {} Bastani%
}{%
{\protect \APACyear {2020}}%
}]{%
Lakkaraju2020}
\APACinsertmetastar {%
Lakkaraju2020}%
\begin{APACrefauthors}%
Lakkaraju, H.%
\BCBT {}\ \BBA {} Bastani, O.%
\end{APACrefauthors}%
\unskip\
\newblock
\APACrefYearMonthDay{2020}{feb}{}.
\newblock
{\BBOQ}\APACrefatitle {"How do I fool you?" Manipulating User Trust via
  Misleading Black Box Explanations} {"how do i fool you?" manipulating user
  trust via misleading black box explanations}.{\BBCQ}
\newblock
\BIn{} \APACrefbtitle {Proceedings of the AAAI/ACM Conference on AI, Ethics,
  and Society} {Proceedings of the aaai/acm conference on ai, ethics, and
  society}\ (\BPGS\ 79--85).
\newblock
\APACaddressPublisher{}{{ACM}}.
\newblock
\begin{APACrefDOI} \doi{10.1145/3375627.3375833} \end{APACrefDOI}
\PrintBackRefs{\CurrentBib}

\bibitem [\protect \citeauthoryear {%
Lee%
\ \BBA {} Moray%
}{%
Lee%
\ \BBA {} Moray%
}{%
{\protect \APACyear {1994}}%
}]{%
Lee1994}
\APACinsertmetastar {%
Lee1994}%
\begin{APACrefauthors}%
Lee, J\BPBI D.%
\BCBT {}\ \BBA {} Moray, N.%
\end{APACrefauthors}%
\unskip\
\newblock
\APACrefYearMonthDay{1994}{}{}.
\newblock
{\BBOQ}\APACrefatitle {Trust, self-confidence, and operators' adaptation to
  automation} {Trust, self-confidence, and operators' adaptation to
  automation}.{\BBCQ}
\newblock
\APACjournalVolNumPages{International Journal of Human-Computer
  Studies}{40}{1}{153--184}.
\newblock
\begin{APACrefDOI} \doi{https://doi.org/10.1006/ijhc.1994.1007}
  \end{APACrefDOI}
\PrintBackRefs{\CurrentBib}

\bibitem [\protect \citeauthoryear {%
Lee%
\ \BBA {} See%
}{%
Lee%
\ \BBA {} See%
}{%
{\protect \APACyear {2004}}%
}]{%
lee2004trust}
\APACinsertmetastar {%
lee2004trust}%
\begin{APACrefauthors}%
Lee, J\BPBI D.%
\BCBT {}\ \BBA {} See, K\BPBI A.%
\end{APACrefauthors}%
\unskip\
\newblock
\APACrefYearMonthDay{2004}{}{}.
\newblock
{\BBOQ}\APACrefatitle {Trust in automation: Designing for appropriate reliance}
  {Trust in automation: Designing for appropriate reliance}.{\BBCQ}
\newblock
\APACjournalVolNumPages{Human factors}{46}{1}{50--80}.
\newblock
\begin{APACrefURL}
  \url{https://journals.sagepub.com/doi/abs/10.1518/hfes.46.1.50\_30392}
  \end{APACrefURL}
\PrintBackRefs{\CurrentBib}

\bibitem [\protect \citeauthoryear {%
Leffrang%
\ \BBA {} Müller%
}{%
Leffrang%
\ \BBA {} Müller%
}{%
{\protect \APACyear {2021}}%
}]{%
Leffrang2021}
\APACinsertmetastar {%
Leffrang2021}%
\begin{APACrefauthors}%
Leffrang, D.%
\BCBT {}\ \BBA {} Müller, O.%
\end{APACrefauthors}%
\unskip\
\newblock
\APACrefYearMonthDay{2021}{}{}.
\newblock
{\BBOQ}\APACrefatitle {Should I Follow this Model? The Effect of Uncertainty
  Visualization on the Acceptance of Time Series Forecasts} {Should i follow
  this model? the effect of uncertainty visualization on the acceptance of time
  series forecasts}.{\BBCQ}
\newblock
\BIn{} \APACrefbtitle {2021 IEEE Workshop on TRust and EXpertise in Visual
  Analytics (TREX)} {2021 ieee workshop on trust and expertise in visual
  analytics (trex)}\ (\BPGS\ 20--26).
\newblock
\begin{APACrefDOI} \doi{10.1109/TREX53765.2021.00009} \end{APACrefDOI}
\PrintBackRefs{\CurrentBib}

\bibitem [\protect \citeauthoryear {%
Leichtmann%
, Humer%
, Hinterreiter%
, Streit%
\BCBL {}\ \BBA {} Mara%
}{%
Leichtmann%
\ \protect \BOthers {.}}{%
{\protect \APACyear {2023}}%
}]{%
Leichtmann2023}
\APACinsertmetastar {%
Leichtmann2023}%
\begin{APACrefauthors}%
Leichtmann, B.%
, Humer, C.%
, Hinterreiter, A.%
, Streit, M.%
\BCBL {}\ \BBA {} Mara, M.%
\end{APACrefauthors}%
\unskip\
\newblock
\APACrefYearMonthDay{2023}{}{}.
\newblock
{\BBOQ}\APACrefatitle {Effects of Explainable Artificial Intelligence on trust
  and human behavior in a high-risk decision task} {Effects of explainable
  artificial intelligence on trust and human behavior in a high-risk decision
  task}.{\BBCQ}
\newblock
\APACjournalVolNumPages{Computers in Human Behavior}{139}{}{107539}.
\newblock
\begin{APACrefURL}
  \url{https://www.sciencedirect.com/science/article/pii/S0747563222003594}
  \end{APACrefURL}
\newblock
\begin{APACrefDOI} \doi{https://doi.org/10.1016/j.chb.2022.107539}
  \end{APACrefDOI}
\PrintBackRefs{\CurrentBib}

\bibitem [\protect \citeauthoryear {%
Lewicki%
, McAllister%
\BCBL {}\ \BBA {} Bies%
}{%
Lewicki%
\ \protect \BOthers {.}}{%
{\protect \APACyear {1998}}%
}]{%
Lewicki.1998}
\APACinsertmetastar {%
Lewicki.1998}%
\begin{APACrefauthors}%
Lewicki, R\BPBI J.%
, McAllister, D\BPBI J.%
\BCBL {}\ \BBA {} Bies, R\BPBI J.%
\end{APACrefauthors}%
\unskip\
\newblock
\APACrefYearMonthDay{1998}{}{}.
\newblock
{\BBOQ}\APACrefatitle {Trust And Distrust: New Relationships and Realities}
  {Trust and distrust: New relationships and realities}.{\BBCQ}
\newblock
\APACjournalVolNumPages{Academy of Management Review}{23}{3}{438--458}.
\newblock
\begin{APACrefDOI} \doi{10.5465/amr.1998.926620} \end{APACrefDOI}
\PrintBackRefs{\CurrentBib}

\bibitem [\protect \citeauthoryear {%
Liao%
\ \BBA {} Varshney%
}{%
Liao%
\ \BBA {} Varshney%
}{%
{\protect \APACyear {2021}}%
}]{%
Liao2021}
\APACinsertmetastar {%
Liao2021}%
\begin{APACrefauthors}%
Liao, Q\BPBI V.%
\BCBT {}\ \BBA {} Varshney, K\BPBI R.%
\end{APACrefauthors}%
\unskip\
\newblock
\APACrefYearMonthDay{2021}{}{}.
\newblock
{\BBOQ}\APACrefatitle {Human-centered explainable ai ({XAI}): From algorithms
  to user experiences} {Human-centered explainable ai ({XAI}): From algorithms
  to user experiences}.{\BBCQ}
\newblock
\APACjournalVolNumPages{arXiv preprint arXiv:2110.10790}{}{}{}.
\PrintBackRefs{\CurrentBib}

\bibitem [\protect \citeauthoryear {%
Lim%
\ \BBA {} Dey%
}{%
Lim%
\ \BBA {} Dey%
}{%
{\protect \APACyear {2009}}%
}]{%
Lim2009}
\APACinsertmetastar {%
Lim2009}%
\begin{APACrefauthors}%
Lim, B\BPBI Y.%
\BCBT {}\ \BBA {} Dey, A\BPBI K.%
\end{APACrefauthors}%
\unskip\
\newblock
\APACrefYearMonthDay{2009}{}{}.
\newblock
{\BBOQ}\APACrefatitle {Assessing Demand for Intelligibility in Context-Aware
  Applications} {Assessing demand for intelligibility in context-aware
  applications}.{\BBCQ}
\newblock
\BIn{} \APACrefbtitle {Proceedings of the 11th International Conference on
  Ubiquitous Computing} {Proceedings of the 11th international conference on
  ubiquitous computing}\ (\BPG~195–204).
\newblock
\APACaddressPublisher{New York, NY, USA}{Association for Computing Machinery}.
\newblock
\begin{APACrefURL} \url{https://doi.org/10.1145/1620545.1620576}
  \end{APACrefURL}
\newblock
\begin{APACrefDOI} \doi{10.1145/1620545.1620576} \end{APACrefDOI}
\PrintBackRefs{\CurrentBib}

\bibitem [\protect \citeauthoryear {%
Lim%
, Dey%
\BCBL {}\ \BBA {} Avrahami%
}{%
Lim%
\ \protect \BOthers {.}}{%
{\protect \APACyear {2009}}%
}]{%
Lim2009a}
\APACinsertmetastar {%
Lim2009a}%
\begin{APACrefauthors}%
Lim, B\BPBI Y.%
, Dey, A\BPBI K.%
\BCBL {}\ \BBA {} Avrahami, D.%
\end{APACrefauthors}%
\unskip\
\newblock
\APACrefYearMonthDay{2009}{}{}.
\newblock
{\BBOQ}\APACrefatitle {Why and Why Not Explanations Improve the Intelligibility
  of Context-Aware Intelligent Systems} {Why and why not explanations improve
  the intelligibility of context-aware intelligent systems}.{\BBCQ}
\newblock
\BIn{} \APACrefbtitle {Proceedings of the SIGCHI Conference on Human Factors in
  Computing Systems} {Proceedings of the sigchi conference on human factors in
  computing systems}\ (\BPG~2119–2128).
\newblock
\APACaddressPublisher{New York, NY, USA}{Association for Computing Machinery}.
\newblock
\begin{APACrefURL} \url{https://doi.org/10.1145/1518701.1519023}
  \end{APACrefURL}
\newblock
\begin{APACrefDOI} \doi{10.1145/1518701.1519023} \end{APACrefDOI}
\PrintBackRefs{\CurrentBib}

\bibitem [\protect \citeauthoryear {%
Linder%
\ \protect \BOthers {.}}{%
Linder%
\ \protect \BOthers {.}}{%
{\protect \APACyear {2021}}%
}]{%
Linder2021}
\APACinsertmetastar {%
Linder2021}%
\begin{APACrefauthors}%
Linder, R.%
, Mohseni, S.%
, Yang, F.%
, Pentyala, S\BPBI K.%
, Ragan, E\BPBI D.%
\BCBL {}\ \BBA {} Hu, X\BPBI B.%
\end{APACrefauthors}%
\unskip\
\newblock
\APACrefYearMonthDay{2021}{}{}.
\newblock
{\BBOQ}\APACrefatitle {How level of explanation detail affects human
  performance in interpretable intelligent systems: A study on explainable fact
  checking} {How level of explanation detail affects human performance in
  interpretable intelligent systems: A study on explainable fact
  checking}.{\BBCQ}
\newblock
\APACjournalVolNumPages{Applied AI Letters}{2}{4}{e49}.
\newblock
\begin{APACrefURL}
  \url{https://onlinelibrary.wiley.com/doi/abs/10.1002/ail2.49}
  \end{APACrefURL}
\newblock
\begin{APACrefDOI} \doi{https://doi.org/10.1002/ail2.49} \end{APACrefDOI}
\PrintBackRefs{\CurrentBib}

\bibitem [\protect \citeauthoryear {%
Lopes%
, Silva%
, Braga%
, Oliveira%
\BCBL {}\ \BBA {} Rosado%
}{%
Lopes%
\ \protect \BOthers {.}}{%
{\protect \APACyear {2022}}%
}]{%
Lopes2022}
\APACinsertmetastar {%
Lopes2022}%
\begin{APACrefauthors}%
Lopes, P.%
, Silva, E.%
, Braga, C.%
, Oliveira, T.%
\BCBL {}\ \BBA {} Rosado, L.%
\end{APACrefauthors}%
\unskip\
\newblock
\APACrefYearMonthDay{2022}{}{}.
\newblock
{\BBOQ}\APACrefatitle {{XAI} Systems Evaluation: A Review of Human and
  Computer-Centred Methods} {{XAI} systems evaluation: A review of human and
  computer-centred methods}.{\BBCQ}
\newblock
\APACjournalVolNumPages{Applied Sciences}{12}{19}{}.
\newblock
\begin{APACrefURL} \url{https://www.mdpi.com/2076-3417/12/19/9423}
  \end{APACrefURL}
\newblock
\begin{APACrefDOI} \doi{10.3390/app12199423} \end{APACrefDOI}
\PrintBackRefs{\CurrentBib}

\bibitem [\protect \citeauthoryear {%
Lu%
\ \BBA {} Yin%
}{%
Lu%
\ \BBA {} Yin%
}{%
{\protect \APACyear {2021}}%
}]{%
Lu2021}
\APACinsertmetastar {%
Lu2021}%
\begin{APACrefauthors}%
Lu, Z.%
\BCBT {}\ \BBA {} Yin, M.%
\end{APACrefauthors}%
\unskip\
\newblock
\APACrefYearMonthDay{2021}{}{}.
\newblock
{\BBOQ}\APACrefatitle {Human Reliance on Machine Learning Models When
  Performance Feedback is Limited: Heuristics and Risks} {Human reliance on
  machine learning models when performance feedback is limited: Heuristics and
  risks}.{\BBCQ}
\newblock
\BIn{} \APACrefbtitle {Proceedings of the 2021 CHI Conference on Human Factors
  in Computing Systems.} {Proceedings of the 2021 chi conference on human
  factors in computing systems.}
\newblock
\APACaddressPublisher{New York, NY, USA}{Association for Computing Machinery}.
\newblock
\begin{APACrefURL} \url{https://doi.org/10.1145/3411764.3445562}
  \end{APACrefURL}
\newblock
\begin{APACrefDOI} \doi{10.1145/3411764.3445562} \end{APACrefDOI}
\PrintBackRefs{\CurrentBib}

\bibitem [\protect \citeauthoryear {%
Luhmann%
}{%
Luhmann%
}{%
{\protect \APACyear {2009}}%
}]{%
luhmann2009a}
\APACinsertmetastar {%
luhmann2009a}%
\begin{APACrefauthors}%
Luhmann, N.%
\end{APACrefauthors}%
\unskip\
\newblock
\APACrefYear{2009}.
\newblock
\APACrefbtitle {Vertrauen : ein Mechanismus der Reduktion sozialer
  Komplexität} {Vertrauen : ein mechanismus der reduktion sozialer
  komplexität}\ (\PrintOrdinal{4. Aufl., Nachdr.}\ \BEd).
\newblock
\APACaddressPublisher{}{Stuttgart : Lucius \& Lucius}.
\PrintBackRefs{\CurrentBib}

\bibitem [\protect \citeauthoryear {%
Lukyanenko%
, Maass%
\BCBL {}\ \BBA {} Storey%
}{%
Lukyanenko%
\ \protect \BOthers {.}}{%
{\protect \APACyear {2022}}%
}]{%
Lukyanenko2022}
\APACinsertmetastar {%
Lukyanenko2022}%
\begin{APACrefauthors}%
Lukyanenko, R.%
, Maass, W.%
\BCBL {}\ \BBA {} Storey, V\BPBI C.%
\end{APACrefauthors}%
\unskip\
\newblock
\APACrefYearMonthDay{2022}{Dec}{01}.
\newblock
{\BBOQ}\APACrefatitle {Trust in artificial intelligence: From a Foundational
  Trust Framework to emerging research opportunities} {Trust in artificial
  intelligence: From a foundational trust framework to emerging research
  opportunities}.{\BBCQ}
\newblock
\APACjournalVolNumPages{Electronic Markets}{32}{4}{1993--2020}.
\newblock
\begin{APACrefURL} \url{https://doi.org/10.1007/s12525-022-00605-4}
  \end{APACrefURL}
\newblock
\begin{APACrefDOI} \doi{10.1007/s12525-022-00605-4} \end{APACrefDOI}
\PrintBackRefs{\CurrentBib}

\bibitem [\protect \citeauthoryear {%
Madsen%
\ \BBA {} Gregor%
}{%
Madsen%
\ \BBA {} Gregor%
}{%
{\protect \APACyear {2000}}%
}]{%
Madsen2000}
\APACinsertmetastar {%
Madsen2000}%
\begin{APACrefauthors}%
Madsen, M.%
\BCBT {}\ \BBA {} Gregor, S.%
\end{APACrefauthors}%
\unskip\
\newblock
\APACrefYearMonthDay{2000}{}{}.
\newblock
{\BBOQ}\APACrefatitle {Measuring human-computer trust} {Measuring
  human-computer trust}.{\BBCQ}
\newblock
\BIn{} \APACrefbtitle {11th australasian conference on information systems}
  {11th australasian conference on information systems}\ (\BVOL~53, \BPGS\
  6--8).
\PrintBackRefs{\CurrentBib}

\bibitem [\protect \citeauthoryear {%
Mahsan~Nourani%
}{%
Mahsan~Nourani%
}{%
{\protect \APACyear {2020}}%
}]{%
Nourani2020}
\APACinsertmetastar {%
Nourani2020}%
\begin{APACrefauthors}%
Mahsan~Nourani, E\BPBI D\BPBI R., Joanie T.~King.%
\end{APACrefauthors}%
\unskip\
\newblock
\APACrefYearMonthDay{2020}{}{}.
\newblock
{\BBOQ}\APACrefatitle {The Role of Domain Expertise in User Trust and the
  Impact of First Impressions with Intelligent Systems} {The role of domain
  expertise in user trust and the impact of first impressions with intelligent
  systems}.{\BBCQ}
\newblock
\BIn{} \APACrefbtitle {Proceedings of the Eighth AAAI Conference on Human
  Computation and Crowdsourcing (HCOMP-20).} {Proceedings of the eighth aaai
  conference on human computation and crowdsourcing (hcomp-20).}
\newblock
\begin{APACrefURL} \url{https://ojs.aaai.org/index.php/HCOMP/article/view/7469}
  \end{APACrefURL}
\PrintBackRefs{\CurrentBib}

\bibitem [\protect \citeauthoryear {%
J.~Mayer%
\ \BBA {} Mussweiler%
}{%
J.~Mayer%
\ \BBA {} Mussweiler%
}{%
{\protect \APACyear {2011}}%
}]{%
Mayer.2011}
\APACinsertmetastar {%
Mayer.2011}%
\begin{APACrefauthors}%
Mayer, J.%
\BCBT {}\ \BBA {} Mussweiler, T.%
\end{APACrefauthors}%
\unskip\
\newblock
\APACrefYearMonthDay{2011}{}{}.
\newblock
{\BBOQ}\APACrefatitle {Suspicious spirits, flexible minds: when distrust
  enhances creativity} {Suspicious spirits, flexible minds: when distrust
  enhances creativity}.{\BBCQ}
\newblock
\APACjournalVolNumPages{Journal of Personality and Social
  Psychology}{101}{6}{1262--1277}.
\newblock
\begin{APACrefDOI} \doi{10.1037/a0024407} \end{APACrefDOI}
\PrintBackRefs{\CurrentBib}

\bibitem [\protect \citeauthoryear {%
R\BPBI C.~Mayer%
, Davis%
\BCBL {}\ \BBA {} Schoorman%
}{%
R\BPBI C.~Mayer%
\ \protect \BOthers {.}}{%
{\protect \APACyear {1995}}%
{\protect \APACexlab {{\protect \BCnt {1}}}}}]{%
Mayer.1995}
\APACinsertmetastar {%
Mayer.1995}%
\begin{APACrefauthors}%
Mayer, R\BPBI C.%
, Davis, J\BPBI H.%
\BCBL {}\ \BBA {} Schoorman, F\BPBI D.%
\end{APACrefauthors}%
\unskip\
\newblock
\APACrefYearMonthDay{1995{\protect \BCnt {1}}}{}{}.
\newblock
{\BBOQ}\APACrefatitle {An Integrative Model Of Organizational Trust} {An
  integrative model of organizational trust}.{\BBCQ}
\newblock
\APACjournalVolNumPages{Academy of Management Review}{20}{3}{709--734}.
\newblock
\begin{APACrefDOI} \doi{10.5465/amr.1995.9508080335} \end{APACrefDOI}
\PrintBackRefs{\CurrentBib}

\bibitem [\protect \citeauthoryear {%
R\BPBI C.~Mayer%
, Davis%
\BCBL {}\ \BBA {} Schoorman%
}{%
R\BPBI C.~Mayer%
\ \protect \BOthers {.}}{%
{\protect \APACyear {1995}}%
{\protect \APACexlab {{\protect \BCnt {2}}}}}]{%
Mayer1995}
\APACinsertmetastar {%
Mayer1995}%
\begin{APACrefauthors}%
Mayer, R\BPBI C.%
, Davis, J\BPBI H.%
\BCBL {}\ \BBA {} Schoorman, F\BPBI D.%
\end{APACrefauthors}%
\unskip\
\newblock
\APACrefYearMonthDay{1995{\protect \BCnt {2}}}{}{}.
\newblock
{\BBOQ}\APACrefatitle {An integrative model of organizational trust} {An
  integrative model of organizational trust}.{\BBCQ}
\newblock
\BIn{} (\BVOL~20, \BPGS\ 709--734).
\newblock
\APACaddressPublisher{}{Academy of Management Briarcliff Manor, NY 10510}.
\PrintBackRefs{\CurrentBib}

\bibitem [\protect \citeauthoryear {%
Mayo%
}{%
Mayo%
}{%
{\protect \APACyear {2015}}%
}]{%
Mayo.2015}
\APACinsertmetastar {%
Mayo.2015}%
\begin{APACrefauthors}%
Mayo, R.%
\end{APACrefauthors}%
\unskip\
\newblock
\APACrefYearMonthDay{2015}{}{}.
\newblock
{\BBOQ}\APACrefatitle {Cognition is a matter of trust: Distrust tunes cognitive
  processes} {Cognition is a matter of trust: Distrust tunes cognitive
  processes}.{\BBCQ}
\newblock
\APACjournalVolNumPages{European Review of Social Psychology}{26}{1}{283--327}.
\newblock
\begin{APACrefDOI} \doi{10.1080/10463283.2015.1117249} \end{APACrefDOI}
\PrintBackRefs{\CurrentBib}

\bibitem [\protect \citeauthoryear {%
McBride%
\ \BBA {} Morgan%
}{%
McBride%
\ \BBA {} Morgan%
}{%
{\protect \APACyear {2010}}%
}]{%
McBride2010}
\APACinsertmetastar {%
McBride2010}%
\begin{APACrefauthors}%
McBride, M.%
\BCBT {}\ \BBA {} Morgan, S.%
\end{APACrefauthors}%
\unskip\
\newblock
\APACrefYearMonthDay{2010}{}{}.
\newblock
{\BBOQ}\APACrefatitle {Trust calibration for automated decision aids} {Trust
  calibration for automated decision aids}.{\BBCQ}
\newblock
\APACjournalVolNumPages{Institute for Homeland Security Solutions}{}{}{1--11}.
\PrintBackRefs{\CurrentBib}

\bibitem [\protect \citeauthoryear {%
McGuirl%
\ \BBA {} Sarter%
}{%
McGuirl%
\ \BBA {} Sarter%
}{%
{\protect \APACyear {2006}}%
}]{%
McGuirl2006}
\APACinsertmetastar {%
McGuirl2006}%
\begin{APACrefauthors}%
McGuirl, J\BPBI M.%
\BCBT {}\ \BBA {} Sarter, N\BPBI B.%
\end{APACrefauthors}%
\unskip\
\newblock
\APACrefYearMonthDay{2006}{}{}.
\newblock
{\BBOQ}\APACrefatitle {Supporting trust calibration and the effective use of
  decision aids by presenting dynamic system confidence information}
  {Supporting trust calibration and the effective use of decision aids by
  presenting dynamic system confidence information}.{\BBCQ}
\newblock
\APACjournalVolNumPages{Human factors}{48}{4}{656--665}.
\PrintBackRefs{\CurrentBib}

\bibitem [\protect \citeauthoryear {%
McKnight%
, Kacmar%
\BCBL {}\ \BBA {} Choudhury%
}{%
McKnight%
\ \protect \BOthers {.}}{%
{\protect \APACyear {2004}}%
}]{%
McKnight.2004}
\APACinsertmetastar {%
McKnight.2004}%
\begin{APACrefauthors}%
McKnight%
, Kacmar%
\BCBL {}\ \BBA {} Choudhury.%
\end{APACrefauthors}%
\unskip\
\newblock
\APACrefYearMonthDay{2004}{}{}.
\newblock
{\BBOQ}\APACrefatitle {Dispositional Trust and Distrust Distinctions in
  Predicting High- and Low-Risk Internet Expert Advice Site Perceptions}
  {Dispositional trust and distrust distinctions in predicting high- and
  low-risk internet expert advice site perceptions}.{\BBCQ}
\newblock
\APACjournalVolNumPages{e-Service Journal}{3}{2}{35}.
\newblock
\begin{APACrefDOI} \doi{10.2979/esj.2004.3.2.35} \end{APACrefDOI}
\PrintBackRefs{\CurrentBib}

\bibitem [\protect \citeauthoryear {%
Meske%
, Bunde%
, Schneider%
\BCBL {}\ \BBA {} Gersch%
}{%
Meske%
\ \protect \BOthers {.}}{%
{\protect \APACyear {2020}}%
}]{%
Meske2020}
\APACinsertmetastar {%
Meske2020}%
\begin{APACrefauthors}%
Meske, C.%
, Bunde, E.%
, Schneider, J.%
\BCBL {}\ \BBA {} Gersch, M.%
\end{APACrefauthors}%
\unskip\
\newblock
\APACrefYearMonthDay{2020}{dec}{}.
\newblock
{\BBOQ}\APACrefatitle {Explainable Artificial Intelligence: Objectives,
  Stakeholders, and Future Research Opportunities} {Explainable artificial
  intelligence: Objectives, stakeholders, and future research
  opportunities}.{\BBCQ}
\newblock
\APACjournalVolNumPages{Information Systems Management}{39}{1}{53--63}.
\newblock
\begin{APACrefDOI} \doi{10.1080/10580530.2020.1849465} \end{APACrefDOI}
\PrintBackRefs{\CurrentBib}

\bibitem [\protect \citeauthoryear {%
D.~Miller%
\ \protect \BOthers {.}}{%
D.~Miller%
\ \protect \BOthers {.}}{%
{\protect \APACyear {2016}}%
}]{%
Miller2016}
\APACinsertmetastar {%
Miller2016}%
\begin{APACrefauthors}%
Miller, D.%
, Johns, M.%
, Mok, B.%
, Gowda, N.%
, Sirkin, D.%
, Lee, K.%
\BCBL {}\ \BBA {} Ju, W.%
\end{APACrefauthors}%
\unskip\
\newblock
\APACrefYearMonthDay{2016}{sep}{}.
\newblock
{\BBOQ}\APACrefatitle {Behavioral Measurement of Trust in Automation}
  {Behavioral measurement of trust in automation}.{\BBCQ}
\newblock
\APACjournalVolNumPages{Proceedings of the Human Factors and Ergonomics Society
  Annual Meeting}{60}{1}{1849--1853}.
\newblock
\begin{APACrefDOI} \doi{10.1177/1541931213601422} \end{APACrefDOI}
\PrintBackRefs{\CurrentBib}

\bibitem [\protect \citeauthoryear {%
T.~Miller%
}{%
T.~Miller%
}{%
{\protect \APACyear {2022}}%
}]{%
Miller2022}
\APACinsertmetastar {%
Miller2022}%
\begin{APACrefauthors}%
Miller, T.%
\end{APACrefauthors}%
\unskip\
\newblock
\APACrefYearMonthDay{2022}{}{}.
\newblock
{\BBOQ}\APACrefatitle {Are we measuring trust correctly in explainability,
  interpretability, and transparency research?} {Are we measuring trust
  correctly in explainability, interpretability, and transparency
  research?}{\BBCQ}
\newblock
\APACjournalVolNumPages{ArXiv}{abs/2209.00651}{}{}.
\PrintBackRefs{\CurrentBib}

\bibitem [\protect \citeauthoryear {%
Ming~Yin%
}{%
Ming~Yin%
}{%
{\protect \APACyear {2019}}%
}]{%
MingYin2019}
\APACinsertmetastar {%
MingYin2019}%
\begin{APACrefauthors}%
Ming~Yin, H\BPBI W., Jennifer Wortman~Vaughan.%
\end{APACrefauthors}%
\unskip\
\newblock
\APACrefYearMonthDay{2019}{}{}.
\newblock
{\BBOQ}\APACrefatitle {Understanding the Effect of Accuracy on Trust in Machine
  Learning Models} {Understanding the effect of accuracy on trust in machine
  learning models}.{\BBCQ}.
\newblock
\begin{APACrefDOI} \doi{.org/10.1145/3290605.3300509} \end{APACrefDOI}
\PrintBackRefs{\CurrentBib}

\bibitem [\protect \citeauthoryear {%
Mohseni%
, Zarei%
\BCBL {}\ \BBA {} Ragan%
}{%
Mohseni%
\ \protect \BOthers {.}}{%
{\protect \APACyear {2021}}%
}]{%
Mohseni2021}
\APACinsertmetastar {%
Mohseni2021}%
\begin{APACrefauthors}%
Mohseni, S.%
, Zarei, N.%
\BCBL {}\ \BBA {} Ragan, E\BPBI D.%
\end{APACrefauthors}%
\unskip\
\newblock
\APACrefYearMonthDay{2021}{sep}{}.
\newblock
{\BBOQ}\APACrefatitle {A Multidisciplinary Survey and Framework for Design and
  Evaluation of Explainable AI Systems} {A multidisciplinary survey and
  framework for design and evaluation of explainable ai systems}.{\BBCQ}
\newblock
\APACjournalVolNumPages{ACM Trans. Interact. Intell. Syst.}{11}{3–4}{}.
\newblock
\begin{APACrefDOI} \doi{10.1145/3387166} \end{APACrefDOI}
\PrintBackRefs{\CurrentBib}

\bibitem [\protect \citeauthoryear {%
Mueller%
\ \protect \BOthers {.}}{%
Mueller%
\ \protect \BOthers {.}}{%
{\protect \APACyear {2021}}%
}]{%
Mueller2021}
\APACinsertmetastar {%
Mueller2021}%
\begin{APACrefauthors}%
Mueller, S\BPBI T.%
, Veinott, E\BPBI S.%
, Hoffman, R\BPBI R.%
, Klein, G.%
, Alam, L.%
, Mamun, T.%
\BCBL {}\ \BBA {} Clancey, W\BPBI J.%
\end{APACrefauthors}%
\unskip\
\newblock
\APACrefYearMonthDay{2021}{}{}.
\newblock
{\BBOQ}\APACrefatitle {Principles of Explanation in Human-AI Systems}
  {Principles of explanation in human-ai systems}.{\BBCQ}
\newblock
\APACjournalVolNumPages{CoRR}{abs/2102.04972}{}{}.
\newblock
\begin{APACrefURL} \url{https://arxiv.org/abs/2102.04972} \end{APACrefURL}
\PrintBackRefs{\CurrentBib}

\bibitem [\protect \citeauthoryear {%
Muir%
\ \BBA {} Moray%
}{%
Muir%
\ \BBA {} Moray%
}{%
{\protect \APACyear {1996}}%
}]{%
Muir1996}
\APACinsertmetastar {%
Muir1996}%
\begin{APACrefauthors}%
Muir, B\BPBI M.%
\BCBT {}\ \BBA {} Moray, N.%
\end{APACrefauthors}%
\unskip\
\newblock
\APACrefYearMonthDay{1996}{}{}.
\newblock
{\BBOQ}\APACrefatitle {Trust in automation. Part II. Experimental studies of
  trust and human intervention in a process control simulation} {Trust in
  automation. part ii. experimental studies of trust and human intervention in
  a process control simulation}.{\BBCQ}
\newblock
\APACjournalVolNumPages{Ergonomics}{39}{3}{429--460}.
\PrintBackRefs{\CurrentBib}

\bibitem [\protect \citeauthoryear {%
Nauta%
\ \protect \BOthers {.}}{%
Nauta%
\ \protect \BOthers {.}}{%
{\protect \APACyear {2022}}%
}]{%
Nauta2022}
\APACinsertmetastar {%
Nauta2022}%
\begin{APACrefauthors}%
Nauta, M.%
, Trienes, J.%
, Nguyen, E.%
, Peters, M.%
, Schmitt, Y.%
, Schlötterer, J.%
\BCBL {}\ \BBA {} Seifert, C.%
\end{APACrefauthors}%
\unskip\
\newblock
\APACrefYearMonthDay{2022}{1}{20}.
\newblock
{\BBOQ}\APACrefatitle {From Anecdotal Evidence to Quantitative Evaluation
  Methods: A Systematic Review on Evaluating Explainable AI} {From anecdotal
  evidence to quantitative evaluation methods: A systematic review on
  evaluating explainable ai}.{\BBCQ}
\newblock

\PrintBackRefs{\CurrentBib}

\bibitem [\protect \citeauthoryear {%
Nourani%
, Kabir%
, Mohseni%
\BCBL {}\ \BBA {} Ragan%
}{%
Nourani%
\ \protect \BOthers {.}}{%
{\protect \APACyear {2019}}%
}]{%
nourani2019effects}
\APACinsertmetastar {%
nourani2019effects}%
\begin{APACrefauthors}%
Nourani, M.%
, Kabir, S.%
, Mohseni, S.%
\BCBL {}\ \BBA {} Ragan, E\BPBI D.%
\end{APACrefauthors}%
\unskip\
\newblock
\APACrefYearMonthDay{2019}{}{}.
\newblock
{\BBOQ}\APACrefatitle {The Effects of Meaningful and Meaningless Explanations
  on Trust and Perceived System Accuracy in Intelligent Systems} {The effects
  of meaningful and meaningless explanations on trust and perceived system
  accuracy in intelligent systems}.{\BBCQ}
\newblock
\BIn{} \APACrefbtitle {Proceedings of the AAAI Conference on Human Computation
  and Crowdsourcing} {Proceedings of the aaai conference on human computation
  and crowdsourcing}\ (\BVOL~7, \BPGS\ 97--105).
\newblock
\begin{APACrefURL} \url{https://ojs.aaai.org/index.php/HCOMP/article/view/5284}
  \end{APACrefURL}
\PrintBackRefs{\CurrentBib}

\bibitem [\protect \citeauthoryear {%
Omeiza%
, Kollnig%
, Web%
, Jirotka%
\BCBL {}\ \BBA {} Kunze%
}{%
Omeiza%
\ \protect \BOthers {.}}{%
{\protect \APACyear {2021}}%
}]{%
Omeiza2021}
\APACinsertmetastar {%
Omeiza2021}%
\begin{APACrefauthors}%
Omeiza, D.%
, Kollnig, K.%
, Web, H.%
, Jirotka, M.%
\BCBL {}\ \BBA {} Kunze, L.%
\end{APACrefauthors}%
\unskip\
\newblock
\APACrefYearMonthDay{2021}{jul}{}.
\newblock
{\BBOQ}\APACrefatitle {Why Not Explain? Effects of Explanations on Human
  Perceptions of Autonomous Driving} {Why not explain? effects of explanations
  on human perceptions of autonomous driving}.{\BBCQ}
\newblock
\BIn{} \APACrefbtitle {2021 {IEEE} International Conference on Advanced
  Robotics and Its Social Impacts ({ARSO}).} {2021 {IEEE} international
  conference on advanced robotics and its social impacts ({ARSO}).}
\newblock
\APACaddressPublisher{}{{IEEE}}.
\newblock
\begin{APACrefDOI} \doi{10.1109/arso51874.2021.9542835} \end{APACrefDOI}
\PrintBackRefs{\CurrentBib}

\bibitem [\protect \citeauthoryear {%
Ooge%
\ \BBA {} Verbert%
}{%
Ooge%
\ \BBA {} Verbert%
}{%
{\protect \APACyear {2021}}%
}]{%
Ooge2021}
\APACinsertmetastar {%
Ooge2021}%
\begin{APACrefauthors}%
Ooge, J.%
\BCBT {}\ \BBA {} Verbert, K.%
\end{APACrefauthors}%
\unskip\
\newblock
\APACrefYearMonthDay{2021}{oct}{}.
\newblock
{\BBOQ}\APACrefatitle {Trust in Prediction Models: a Mixed-Methods Pilot Study
  on the Impact of Domain Expertise} {Trust in prediction models: a
  mixed-methods pilot study on the impact of domain expertise}.{\BBCQ}
\newblock
\BIn{} \APACrefbtitle {2021 {IEEE} Workshop on {TRust} and {EXpertise} in
  Visual Analytics ({TREX}).} {2021 {IEEE} workshop on {TRust} and {EXpertise}
  in visual analytics ({TREX}).}
\newblock
\APACaddressPublisher{}{{IEEE}}.
\newblock
\begin{APACrefDOI} \doi{10.1109/trex53765.2021.00007} \end{APACrefDOI}
\PrintBackRefs{\CurrentBib}

\bibitem [\protect \citeauthoryear {%
Ou%
\ \BBA {} Sia%
}{%
Ou%
\ \BBA {} Sia%
}{%
{\protect \APACyear {2010}}%
}]{%
Ou.2010}
\APACinsertmetastar {%
Ou.2010}%
\begin{APACrefauthors}%
Ou, C\BPBI X.%
\BCBT {}\ \BBA {} Sia, C\BPBI L.%
\end{APACrefauthors}%
\unskip\
\newblock
\APACrefYearMonthDay{2010}{}{}.
\newblock
{\BBOQ}\APACrefatitle {Consumer trust and distrust: An issue of website design}
  {Consumer trust and distrust: An issue of website design}.{\BBCQ}
\newblock
\APACjournalVolNumPages{International Journal of Human-Computer
  Studies}{68}{12}{913--934}.
\newblock
\begin{APACrefDOI} \doi{10.1016/j. ijhcs.2010.08.003} \end{APACrefDOI}
\PrintBackRefs{\CurrentBib}

\bibitem [\protect \citeauthoryear {%
Papenmeier%
, Englebienne%
\BCBL {}\ \BBA {} Seifert%
}{%
Papenmeier%
\ \protect \BOthers {.}}{%
{\protect \APACyear {2021}}%
}]{%
Papenmeier2021}
\APACinsertmetastar {%
Papenmeier2021}%
\begin{APACrefauthors}%
Papenmeier, A.%
, Englebienne, G.%
\BCBL {}\ \BBA {} Seifert, C.%
\end{APACrefauthors}%
\unskip\
\newblock
\APACrefYearMonthDay{2021}{}{}.
\newblock
{\BBOQ}\APACrefatitle {How model accuracy and explanation fidelity influence
  user trust in AI} {How model accuracy and explanation fidelity influence user
  trust in ai}.{\BBCQ}.
\PrintBackRefs{\CurrentBib}

\bibitem [\protect \citeauthoryear {%
Papenmeier%
, Kern%
, Englebienne%
\BCBL {}\ \BBA {} Seifert%
}{%
Papenmeier%
\ \protect \BOthers {.}}{%
{\protect \APACyear {2022}}%
}]{%
Papenmeier2022}
\APACinsertmetastar {%
Papenmeier2022}%
\begin{APACrefauthors}%
Papenmeier, A.%
, Kern, D.%
, Englebienne, G.%
\BCBL {}\ \BBA {} Seifert, C.%
\end{APACrefauthors}%
\unskip\
\newblock
\APACrefYearMonthDay{2022}{aug}{}.
\newblock
{\BBOQ}\APACrefatitle {It's Complicated: The Relationship between User Trust,
  Model Accuracy and Explanations in {AI}} {It's complicated: The relationship
  between user trust, model accuracy and explanations in {AI}}.{\BBCQ}
\newblock
\APACjournalVolNumPages{{ACM} Transactions on Computer-Human
  Interaction}{29}{4}{1--33}.
\newblock
\begin{APACrefDOI} \doi{10.1145/3495013} \end{APACrefDOI}
\PrintBackRefs{\CurrentBib}

\bibitem [\protect \citeauthoryear {%
Parasuraman%
\ \BBA {} Riley%
}{%
Parasuraman%
\ \BBA {} Riley%
}{%
{\protect \APACyear {1997}}%
}]{%
Parasuraman1997}
\APACinsertmetastar {%
Parasuraman1997}%
\begin{APACrefauthors}%
Parasuraman, R.%
\BCBT {}\ \BBA {} Riley, V.%
\end{APACrefauthors}%
\unskip\
\newblock
\APACrefYearMonthDay{1997}{}{}.
\newblock
{\BBOQ}\APACrefatitle {Humans and automation: Use, misuse, disuse, abuse}
  {Humans and automation: Use, misuse, disuse, abuse}.{\BBCQ}
\newblock
\APACjournalVolNumPages{Human factors}{39}{2}{230--253}.
\newblock
\begin{APACrefDOI} \doi{10.1518/001872097778543886} \end{APACrefDOI}
\PrintBackRefs{\CurrentBib}

\bibitem [\protect \citeauthoryear {%
Peters%
\ \BBA {} Visser%
}{%
Peters%
\ \BBA {} Visser%
}{%
{\protect \APACyear {2023}}%
}]{%
peters2023importance}
\APACinsertmetastar {%
peters2023importance}%
\begin{APACrefauthors}%
Peters, T\BPBI M.%
\BCBT {}\ \BBA {} Visser, R\BPBI W.%
\end{APACrefauthors}%
\unskip\
\newblock
\APACrefYearMonthDay{2023}{}{}.
\newblock
\APACrefbtitle {The Importance of Distrust in AI.} {The importance of distrust
  in ai.}
\PrintBackRefs{\CurrentBib}

\bibitem [\protect \citeauthoryear {%
Poortinga%
\ \BBA {} Pidgeon%
}{%
Poortinga%
\ \BBA {} Pidgeon%
}{%
{\protect \APACyear {2003}}%
}]{%
Poortinga.2003}
\APACinsertmetastar {%
Poortinga.2003}%
\begin{APACrefauthors}%
Poortinga, W.%
\BCBT {}\ \BBA {} Pidgeon, N\BPBI F.%
\end{APACrefauthors}%
\unskip\
\newblock
\APACrefYearMonthDay{2003}{}{}.
\newblock
{\BBOQ}\APACrefatitle {Exploring the dimensionality of trust in risk
  regulation} {Exploring the dimensionality of trust in risk
  regulation}.{\BBCQ}
\newblock
\APACjournalVolNumPages{Risk analysis : an official publication of the Society
  for Risk Analysis}{23}{5}{961--972}.
\newblock
\begin{APACrefDOI} \doi{10.1111/1539-6924.00373} \end{APACrefDOI}
\PrintBackRefs{\CurrentBib}

\bibitem [\protect \citeauthoryear {%
Posten%
\ \BBA {} Gino%
}{%
Posten%
\ \BBA {} Gino%
}{%
{\protect \APACyear {2021}}%
}]{%
Posten.2021}
\APACinsertmetastar {%
Posten.2021}%
\begin{APACrefauthors}%
Posten, A\BHBI C.%
\BCBT {}\ \BBA {} Gino, F.%
\end{APACrefauthors}%
\unskip\
\newblock
\APACrefYearMonthDay{2021}{}{}.
\newblock
{\BBOQ}\APACrefatitle {How trust and distrust shape perception and memory} {How
  trust and distrust shape perception and memory}.{\BBCQ}
\newblock
\APACjournalVolNumPages{Journal of Personality and Social
  Psychology}{121}{1}{43--58}.
\newblock
\begin{APACrefDOI} \doi{10.1037/pspa0000269} \end{APACrefDOI}
\PrintBackRefs{\CurrentBib}

\bibitem [\protect \citeauthoryear {%
Pynadath%
, Barnes%
, Wang%
\BCBL {}\ \BBA {} Chen%
}{%
Pynadath%
\ \protect \BOthers {.}}{%
{\protect \APACyear {2018}}%
}]{%
Pynadath2018}
\APACinsertmetastar {%
Pynadath2018}%
\begin{APACrefauthors}%
Pynadath, D\BPBI V.%
, Barnes, M\BPBI J.%
, Wang, N.%
\BCBL {}\ \BBA {} Chen, J\BPBI Y.%
\end{APACrefauthors}%
\unskip\
\newblock
\APACrefYearMonthDay{2018}{}{}.
\newblock
{\BBOQ}\APACrefatitle {Transparency communication for machine learning in
  human-automation interaction} {Transparency communication for machine
  learning in human-automation interaction}.{\BBCQ}
\newblock
\APACjournalVolNumPages{Human and Machine Learning: Visible, Explainable,
  Trustworthy and Transparent}{}{}{75--90}.
\newblock
\begin{APACrefURL}
  \url{https://link.springer.com/chapter/10.1007/978-3-319-90403-0\_5}
  \end{APACrefURL}
\PrintBackRefs{\CurrentBib}

\bibitem [\protect \citeauthoryear {%
Rechkemmer%
\ \BBA {} Yin%
}{%
Rechkemmer%
\ \BBA {} Yin%
}{%
{\protect \APACyear {2022}}%
}]{%
Rechkemmer2022}
\APACinsertmetastar {%
Rechkemmer2022}%
\begin{APACrefauthors}%
Rechkemmer, A.%
\BCBT {}\ \BBA {} Yin, M.%
\end{APACrefauthors}%
\unskip\
\newblock
\APACrefYearMonthDay{2022}{}{}.
\newblock
{\BBOQ}\APACrefatitle {When Confidence Meets Accuracy: Exploring the Effects of
  Multiple Performance Indicators on Trust in Machine Learning Models} {When
  confidence meets accuracy: Exploring the effects of multiple performance
  indicators on trust in machine learning models}.{\BBCQ}
\newblock
\BIn{} \APACrefbtitle {Proceedings of the 2022 CHI Conference on Human Factors
  in Computing Systems.} {Proceedings of the 2022 chi conference on human
  factors in computing systems.}
\newblock
\APACaddressPublisher{New York, NY, USA}{Association for Computing Machinery}.
\newblock
\begin{APACrefURL} \url{https://doi.org/10.1145/3491102.3501967}
  \end{APACrefURL}
\newblock
\begin{APACrefDOI} \doi{10.1145/3491102.3501967} \end{APACrefDOI}
\PrintBackRefs{\CurrentBib}

\bibitem [\protect \citeauthoryear {%
Ribeiro%
, Singh%
\BCBL {}\ \BBA {} Guestrin%
}{%
Ribeiro%
\ \protect \BOthers {.}}{%
{\protect \APACyear {2016}}%
}]{%
Ribeiro2016}
\APACinsertmetastar {%
Ribeiro2016}%
\begin{APACrefauthors}%
Ribeiro, M\BPBI T.%
, Singh, S.%
\BCBL {}\ \BBA {} Guestrin, C.%
\end{APACrefauthors}%
\unskip\
\newblock
\APACrefYearMonthDay{2016}{}{}.
\newblock
{\BBOQ}\APACrefatitle {"Why should i trust you?" Explaining the predictions of
  any classifier} {"why should i trust you?" explaining the predictions of any
  classifier}.{\BBCQ}
\newblock
\BIn{} \APACrefbtitle {Proceedings of the 22nd ACM SIGKDD international
  conference on knowledge discovery and data mining} {Proceedings of the 22nd
  acm sigkdd international conference on knowledge discovery and data mining}\
  (\BPGS\ 1135--1144).
\PrintBackRefs{\CurrentBib}

\bibitem [\protect \citeauthoryear {%
Riegelsberger%
, Sasse%
\BCBL {}\ \BBA {} McCarthy%
}{%
Riegelsberger%
\ \protect \BOthers {.}}{%
{\protect \APACyear {2005}}%
}]{%
Riegelsberger2005}
\APACinsertmetastar {%
Riegelsberger2005}%
\begin{APACrefauthors}%
Riegelsberger, J.%
, Sasse, M\BPBI A.%
\BCBL {}\ \BBA {} McCarthy, J\BPBI D.%
\end{APACrefauthors}%
\unskip\
\newblock
\APACrefYearMonthDay{2005}{mar}{}.
\newblock
{\BBOQ}\APACrefatitle {The mechanics of trust: A framework for research and
  design} {The mechanics of trust: A framework for research and design}.{\BBCQ}
\newblock
\APACjournalVolNumPages{International Journal of Human-Computer
  Studies}{62}{3}{381--422}.
\newblock
\begin{APACrefDOI} \doi{10.1016/j. ijhcs.2005.01.001} \end{APACrefDOI}
\PrintBackRefs{\CurrentBib}

\bibitem [\protect \citeauthoryear {%
Rohlfing%
\ \protect \BOthers {.}}{%
Rohlfing%
\ \protect \BOthers {.}}{%
{\protect \APACyear {2021}}%
}]{%
Rohlfing.2021}
\APACinsertmetastar {%
Rohlfing.2021}%
\begin{APACrefauthors}%
Rohlfing, K\BPBI J.%
, Cimiano, P.%
, Scharlau, I.%
, Matzner, T.%
, Buhl, H\BPBI M.%
, Buschmeier, H.%
\BDBL {}Wrede, B.%
\end{APACrefauthors}%
\unskip\
\newblock
\APACrefYearMonthDay{2021}{}{}.
\newblock
{\BBOQ}\APACrefatitle {Explanation as a Social Practice: Toward a Conceptual
  Framework for the Social Design of AI Systems} {Explanation as a social
  practice: Toward a conceptual framework for the social design of ai
  systems}.{\BBCQ}
\newblock
\APACjournalVolNumPages{IEEE Transactions on Cognitive and Developmental
  Systems}{13}{3}{717--728}.
\newblock
\begin{APACrefDOI} \doi{10.1109/tcds.2020.3044366} \end{APACrefDOI}
\PrintBackRefs{\CurrentBib}

\bibitem [\protect \citeauthoryear {%
Rudin%
}{%
Rudin%
}{%
{\protect \APACyear {2019}}%
}]{%
rudin2019stop}
\APACinsertmetastar {%
rudin2019stop}%
\begin{APACrefauthors}%
Rudin, C.%
\end{APACrefauthors}%
\unskip\
\newblock
\APACrefYearMonthDay{2019}{}{}.
\newblock
{\BBOQ}\APACrefatitle {Stop explaining black box machine learning models for
  high stakes decisions and use interpretable models instead} {Stop explaining
  black box machine learning models for high stakes decisions and use
  interpretable models instead}.{\BBCQ}
\newblock
\APACjournalVolNumPages{Nature Machine Intelligence}{1}{5}{206--215}.
\PrintBackRefs{\CurrentBib}

\bibitem [\protect \citeauthoryear {%
Rusk%
}{%
Rusk%
}{%
{\protect \APACyear {2018}}%
}]{%
Rusk.2018}
\APACinsertmetastar {%
Rusk.2018}%
\begin{APACrefauthors}%
Rusk, J\BPBI D.%
\end{APACrefauthors}%
\unskip\
\newblock
\APACrefYearMonthDay{2018}{}{}.
\newblock
{\BBOQ}\APACrefatitle {Trust and distrust scale development: Operationalization
  and instrument validation} {Trust and distrust scale development:
  Operationalization and instrument validation}.{\BBCQ}
\newblock

\PrintBackRefs{\CurrentBib}

\bibitem [\protect \citeauthoryear {%
Samek%
, Montavon%
, Lapuschkin%
, Anders%
\BCBL {}\ \BBA {} Müller%
}{%
Samek%
\ \protect \BOthers {.}}{%
{\protect \APACyear {2021}}%
}]{%
Samek2021}
\APACinsertmetastar {%
Samek2021}%
\begin{APACrefauthors}%
Samek, W.%
, Montavon, G.%
, Lapuschkin, S.%
, Anders, C\BPBI J.%
\BCBL {}\ \BBA {} Müller, K\BHBI R.%
\end{APACrefauthors}%
\unskip\
\newblock
\APACrefYearMonthDay{2021}{}{}.
\newblock
{\BBOQ}\APACrefatitle {Explaining Deep Neural Networks and Beyond: A Review of
  Methods and Applications} {Explaining deep neural networks and beyond: A
  review of methods and applications}.{\BBCQ}
\newblock
\APACjournalVolNumPages{Proceedings of the IEEE}{109}{3}{247--278}.
\newblock
\begin{APACrefDOI} \doi{10.1109/JPROC.2021.3060483} \end{APACrefDOI}
\PrintBackRefs{\CurrentBib}

\bibitem [\protect \citeauthoryear {%
Sarah~Bayer%
}{%
Sarah~Bayer%
}{%
{\protect \APACyear {2021}}%
}]{%
Bayer2021}
\APACinsertmetastar {%
Bayer2021}%
\begin{APACrefauthors}%
Sarah~Bayer, M\BPBI M., Henner~Gimpel.%
\end{APACrefauthors}%
\unskip\
\newblock
\APACrefYearMonthDay{2021}{}{}.
\newblock
{\BBOQ}\APACrefatitle {The role of domain expertise in trusting and following
  explainable AI decision support systems} {The role of domain expertise in
  trusting and following explainable ai decision support systems}.{\BBCQ}.
\newblock
\begin{APACrefDOI} \doi{.org/10.1080/12460125.2021.1958505} \end{APACrefDOI}
\PrintBackRefs{\CurrentBib}

\bibitem [\protect \citeauthoryear {%
Schaefer%
, Chen%
, Szalma%
\BCBL {}\ \BBA {} Hancock%
}{%
Schaefer%
\ \protect \BOthers {.}}{%
{\protect \APACyear {2016}}%
}]{%
Schaefer2016}
\APACinsertmetastar {%
Schaefer2016}%
\begin{APACrefauthors}%
Schaefer, K\BPBI E.%
, Chen, J\BPBI Y\BPBI C.%
, Szalma, J\BPBI L.%
\BCBL {}\ \BBA {} Hancock, P\BPBI A.%
\end{APACrefauthors}%
\unskip\
\newblock
\APACrefYearMonthDay{2016}{mar}{}.
\newblock
{\BBOQ}\APACrefatitle {A Meta-Analysis of Factors Influencing the Development
  of Trust in Automation} {A meta-analysis of factors influencing the
  development of trust in automation}.{\BBCQ}
\newblock
\APACjournalVolNumPages{Human Factors: The Journal of the Human Factors and
  Ergonomics Society}{58}{3}{377--400}.
\newblock
\begin{APACrefDOI} \doi{10.1177/0018720816634228} \end{APACrefDOI}
\PrintBackRefs{\CurrentBib}

\bibitem [\protect \citeauthoryear {%
Scharowski%
\ \BBA {} Perrig%
}{%
Scharowski%
\ \BBA {} Perrig%
}{%
{\protect \APACyear {2023}}%
}]{%
scharowski2023distrust}
\APACinsertmetastar {%
scharowski2023distrust}%
\begin{APACrefauthors}%
Scharowski, N.%
\BCBT {}\ \BBA {} Perrig, S\BPBI A.%
\end{APACrefauthors}%
\unskip\
\newblock
\APACrefYearMonthDay{2023}{}{}.
\newblock
{\BBOQ}\APACrefatitle {Distrust in {(X)AI} -- Measurement Artifact or Distinct
  Construct?} {Distrust in {(X)AI} -- measurement artifact or distinct
  construct?}{\BBCQ}
\newblock
\APACjournalVolNumPages{arXiv preprint arXiv:2303.16495}{}{}{}.
\PrintBackRefs{\CurrentBib}

\bibitem [\protect \citeauthoryear {%
Schlicker%
\ \BBA {} Langer%
}{%
Schlicker%
\ \BBA {} Langer%
}{%
{\protect \APACyear {2021}}%
}]{%
Schlicker2021}
\APACinsertmetastar {%
Schlicker2021}%
\begin{APACrefauthors}%
Schlicker, N.%
\BCBT {}\ \BBA {} Langer, M.%
\end{APACrefauthors}%
\unskip\
\newblock
\APACrefYearMonthDay{2021}{}{}.
\newblock
{\BBOQ}\APACrefatitle {Towards warranted trust: A model on the relation between
  actual and perceived system trustworthiness} {Towards warranted trust: A
  model on the relation between actual and perceived system
  trustworthiness}.{\BBCQ}
\newblock
\BIn{} \APACrefbtitle {Proceedings of Mensch und Computer 2021} {Proceedings of
  mensch und computer 2021}\ (\BPGS\ 325--329).
\PrintBackRefs{\CurrentBib}

\bibitem [\protect \citeauthoryear {%
Schmidt%
\ \BBA {} Biessmann%
}{%
Schmidt%
\ \BBA {} Biessmann%
}{%
{\protect \APACyear {2019}}%
}]{%
Schmidt2019}
\APACinsertmetastar {%
Schmidt2019}%
\begin{APACrefauthors}%
Schmidt, P.%
\BCBT {}\ \BBA {} Biessmann, F.%
\end{APACrefauthors}%
\unskip\
\newblock
\APACrefYearMonthDay{2019}{}{}.
\newblock
{\BBOQ}\APACrefatitle {Quantifying interpretability and trust in machine
  learning systems} {Quantifying interpretability and trust in machine learning
  systems}.{\BBCQ}
\newblock
\APACjournalVolNumPages{arXiv preprint arXiv:1901.08558}{}{}{}.
\PrintBackRefs{\CurrentBib}

\bibitem [\protect \citeauthoryear {%
Schmidt%
, Biessmann%
\BCBL {}\ \BBA {} Teubner%
}{%
Schmidt%
\ \protect \BOthers {.}}{%
{\protect \APACyear {2020}}%
}]{%
Schmidt2020}
\APACinsertmetastar {%
Schmidt2020}%
\begin{APACrefauthors}%
Schmidt, P.%
, Biessmann, F.%
\BCBL {}\ \BBA {} Teubner, T.%
\end{APACrefauthors}%
\unskip\
\newblock
\APACrefYearMonthDay{2020}{}{}.
\newblock
{\BBOQ}\APACrefatitle {Transparency and trust in artificial intelligence
  systems} {Transparency and trust in artificial intelligence systems}.{\BBCQ}
\newblock
\APACjournalVolNumPages{Journal of Decision Systems}{29}{4}{260--278}.
\newblock
\begin{APACrefURL} \url{https://doi.org/10.1080/12460125.2020.1819094}
  \end{APACrefURL}
\newblock
\begin{APACrefDOI} \doi{10.1080/12460125.2020.1819094} \end{APACrefDOI}
\PrintBackRefs{\CurrentBib}

\bibitem [\protect \citeauthoryear {%
Schoorman%
, Mayer%
\BCBL {}\ \BBA {} Davis%
}{%
Schoorman%
\ \protect \BOthers {.}}{%
{\protect \APACyear {2007}}%
}]{%
Schoorman.2007}
\APACinsertmetastar {%
Schoorman.2007}%
\begin{APACrefauthors}%
Schoorman, F\BPBI D.%
, Mayer, R\BPBI C.%
\BCBL {}\ \BBA {} Davis, J\BPBI H.%
\end{APACrefauthors}%
\unskip\
\newblock
\APACrefYearMonthDay{2007}{}{}.
\newblock
{\BBOQ}\APACrefatitle {An Integrative Model of Organizational Trust: Past,
  Present, and Future} {An integrative model of organizational trust: Past,
  present, and future}.{\BBCQ}
\newblock
\APACjournalVolNumPages{Academy of Management Review}{32}{2}{344--354}.
\newblock
\begin{APACrefDOI} \doi{10.5465/amr.2007.24348410} \end{APACrefDOI}
\PrintBackRefs{\CurrentBib}

\bibitem [\protect \citeauthoryear {%
Schweer%
, Vaske%
\BCBL {}\ \BBA {} Vaske%
}{%
Schweer%
\ \protect \BOthers {.}}{%
{\protect \APACyear {2009}}%
}]{%
Schweer.2009}
\APACinsertmetastar {%
Schweer.2009}%
\begin{APACrefauthors}%
Schweer, M.%
, Vaske, C.%
\BCBL {}\ \BBA {} Vaske, A\BHBI K.%
\end{APACrefauthors}%
\unskip\
\newblock
\APACrefYear{2009}.
\newblock
\APACrefbtitle {Zur Funktionalit{\"a}t und Dysfunktionalit{\"a}t von Misstrauen
  in virtuellen Organisationen} {Zur funktionalit{\"a}t und
  dysfunktionalit{\"a}t von misstrauen in virtuellen organisationen}.
\newblock
\begin{APACrefURL} \url{https://dl.gi.de/handle/20.500.12116/35191}
  \end{APACrefURL}
\PrintBackRefs{\CurrentBib}

\bibitem [\protect \citeauthoryear {%
Seckler%
, Heinz%
, Forde%
, Tuch%
\BCBL {}\ \BBA {} Opwis%
}{%
Seckler%
\ \protect \BOthers {.}}{%
{\protect \APACyear {2015}}%
}]{%
Seckler.2015}
\APACinsertmetastar {%
Seckler.2015}%
\begin{APACrefauthors}%
Seckler, M.%
, Heinz, S.%
, Forde, S.%
, Tuch, A\BPBI N.%
\BCBL {}\ \BBA {} Opwis, K.%
\end{APACrefauthors}%
\unskip\
\newblock
\APACrefYearMonthDay{2015}{}{}.
\newblock
{\BBOQ}\APACrefatitle {Trust and distrust on the web: User experiences and
  website characteristics} {Trust and distrust on the web: User experiences and
  website characteristics}.{\BBCQ}
\newblock
\APACjournalVolNumPages{Computers in Human Behavior}{45}{}{39--50}.
\newblock
\begin{APACrefDOI} \doi{10.1016/j. chb.2014.11.064} \end{APACrefDOI}
\PrintBackRefs{\CurrentBib}

\bibitem [\protect \citeauthoryear {%
Sedlmeier%
}{%
Sedlmeier%
}{%
{\protect \APACyear {2013}}%
}]{%
Sedlmeier.2013}
\APACinsertmetastar {%
Sedlmeier.2013}%
\begin{APACrefauthors}%
Sedlmeier, P.%
\end{APACrefauthors}%
\unskip\
\newblock
\APACrefYear{2013}.
\newblock
\APACrefbtitle {{Forschungsmethoden und Statistik f{\"u}r Psychologen und
  Sozialwissenschaftler}} {{Forschungsmethoden und Statistik f{\"u}r
  Psychologen und Sozialwissenschaftler}}.
\newblock
\APACaddressPublisher{}{Pearson Deutschland GmbH}.
\PrintBackRefs{\CurrentBib}

\bibitem [\protect \citeauthoryear {%
Shin%
}{%
Shin%
}{%
{\protect \APACyear {2021}}%
}]{%
Shin2021}
\APACinsertmetastar {%
Shin2021}%
\begin{APACrefauthors}%
Shin, D.%
\end{APACrefauthors}%
\unskip\
\newblock
\APACrefYearMonthDay{2021}{}{}.
\newblock
{\BBOQ}\APACrefatitle {The effects of explainability and causability on
  perception, trust, and acceptance: Implications for explainable AI} {The
  effects of explainability and causability on perception, trust, and
  acceptance: Implications for explainable ai}.{\BBCQ}
\newblock
\APACjournalVolNumPages{International Journal of Human-Computer
  Studies}{146}{}{102551}.
\newblock
\begin{APACrefURL}
  \url{https://www.sciencedirect.com/science/article/pii/S1071581920301531}
  \end{APACrefURL}
\newblock
\begin{APACrefDOI} \doi{https://doi.org/10.1016/j.ijhcs.2020.102551}
  \end{APACrefDOI}
\PrintBackRefs{\CurrentBib}

\bibitem [\protect \citeauthoryear {%
Siau%
\ \BBA {} Wang%
}{%
Siau%
\ \BBA {} Wang%
}{%
{\protect \APACyear {2018}}%
}]{%
Siau2018}
\APACinsertmetastar {%
Siau2018}%
\begin{APACrefauthors}%
Siau, K.%
\BCBT {}\ \BBA {} Wang, W.%
\end{APACrefauthors}%
\unskip\
\newblock
\APACrefYearMonthDay{2018}{}{}.
\newblock
{\BBOQ}\APACrefatitle {Building Trust in Artificial Intelligence,Machine
  Learning, and Robotics} {Building trust in artificial intelligence,machine
  learning, and robotics}.{\BBCQ}.
\PrintBackRefs{\CurrentBib}

\bibitem [\protect \citeauthoryear {%
Spain%
, Bustamante%
\BCBL {}\ \BBA {} Bliss%
}{%
Spain%
\ \protect \BOthers {.}}{%
{\protect \APACyear {2008}}%
}]{%
Spain.2008}
\APACinsertmetastar {%
Spain.2008}%
\begin{APACrefauthors}%
Spain, R\BPBI D.%
, Bustamante, E\BPBI A.%
\BCBL {}\ \BBA {} Bliss, J\BPBI P.%
\end{APACrefauthors}%
\unskip\
\newblock
\APACrefYearMonthDay{2008}{}{}.
\newblock
{\BBOQ}\APACrefatitle {Towards an Empirically Developed Scale for System Trust:
  Take Two} {Towards an empirically developed scale for system trust: Take
  two}.{\BBCQ}
\newblock
\APACjournalVolNumPages{Proceedings of the Human Factors and Ergonomics Society
  Annual Meeting}{52}{19}{1335--1339}.
\newblock
\begin{APACrefDOI} \doi{10.1177/154193120805201907} \end{APACrefDOI}
\PrintBackRefs{\CurrentBib}

\bibitem [\protect \citeauthoryear {%
Stanton%
\ \BBA {} Jensen%
}{%
Stanton%
\ \BBA {} Jensen%
}{%
{\protect \APACyear {2021}}%
}]{%
Stanton.2021}
\APACinsertmetastar {%
Stanton.2021}%
\begin{APACrefauthors}%
Stanton, B.%
\BCBT {}\ \BBA {} Jensen, T.%
\end{APACrefauthors}%
\unskip\
\newblock
\APACrefYearMonthDay{2021}{}{}.
\newblock
{\BBOQ}\APACrefatitle {Trust and Artificial Intelligence} {Trust and artificial
  intelligence}.{\BBCQ}
\newblock

\newblock
\begin{APACrefDOI} \doi{10.6028/nist.ir.8332-draft} \end{APACrefDOI}
\PrintBackRefs{\CurrentBib}

\bibitem [\protect \citeauthoryear {%
Suresh%
, Lao%
\BCBL {}\ \BBA {} Liccardi%
}{%
Suresh%
\ \protect \BOthers {.}}{%
{\protect \APACyear {2020}}%
}]{%
Suresh2020}
\APACinsertmetastar {%
Suresh2020}%
\begin{APACrefauthors}%
Suresh, H.%
, Lao, N.%
\BCBL {}\ \BBA {} Liccardi, I.%
\end{APACrefauthors}%
\unskip\
\newblock
\APACrefYearMonthDay{2020}{}{}.
\newblock
{\BBOQ}\APACrefatitle {Misplaced Trust: Measuring the Interference of Machine
  Learning in Human Decision-Making} {Misplaced trust: Measuring the
  interference of machine learning in human decision-making}.{\BBCQ}
\newblock
\BIn{} \APACrefbtitle {Proceedings of the 12th ACM Conference on Web Science}
  {Proceedings of the 12th acm conference on web science}\ (\BPG~315–324).
\newblock
\APACaddressPublisher{New York, NY, USA}{Association for Computing Machinery}.
\newblock
\begin{APACrefURL} \url{https://doi.org/10.1145/3394231.3397922}
  \end{APACrefURL}
\newblock
\begin{APACrefDOI} \doi{10.1145/3394231.3397922} \end{APACrefDOI}
\PrintBackRefs{\CurrentBib}

\bibitem [\protect \citeauthoryear {%
Thaler%
\ \BBA {} Schmid%
}{%
Thaler%
\ \BBA {} Schmid%
}{%
{\protect \APACyear {2021}}%
}]{%
Thaler2021}
\APACinsertmetastar {%
Thaler2021}%
\begin{APACrefauthors}%
Thaler, A\BPBI M.%
\BCBT {}\ \BBA {} Schmid, U.%
\end{APACrefauthors}%
\unskip\
\newblock
\APACrefYearMonthDay{2021}{}{}.
\newblock
{\BBOQ}\APACrefatitle {Explaining machine learned relational concepts in visual
  domains-effects of perceived accuracy on joint performance and trust}
  {Explaining machine learned relational concepts in visual domains-effects of
  perceived accuracy on joint performance and trust}.{\BBCQ}
\newblock
\BIn{} \APACrefbtitle {Proceedings of the Annual Meeting of the Cognitive
  Science Society} {Proceedings of the annual meeting of the cognitive science
  society}\ (\BVOL~43).
\PrintBackRefs{\CurrentBib}

\bibitem [\protect \citeauthoryear {%
Thiebes%
, Lins%
\BCBL {}\ \BBA {} Sunyaev%
}{%
Thiebes%
\ \protect \BOthers {.}}{%
{\protect \APACyear {2021}}%
}]{%
Thiebes.2021}
\APACinsertmetastar {%
Thiebes.2021}%
\begin{APACrefauthors}%
Thiebes, S.%
, Lins, S.%
\BCBL {}\ \BBA {} Sunyaev, A.%
\end{APACrefauthors}%
\unskip\
\newblock
\APACrefYearMonthDay{2021}{}{}.
\newblock
{\BBOQ}\APACrefatitle {Trustworthy artificial intelligence} {Trustworthy
  artificial intelligence}.{\BBCQ}
\newblock
\APACjournalVolNumPages{Electronic Markets}{31}{2}{447--464}.
\newblock
\begin{APACrefDOI} \doi{10.1007/s12525-020-00441-4} \end{APACrefDOI}
\PrintBackRefs{\CurrentBib}

\bibitem [\protect \citeauthoryear {%
Thielsch%
, Mee{\ss}en%
\BCBL {}\ \BBA {} Hertel%
}{%
Thielsch%
\ \protect \BOthers {.}}{%
{\protect \APACyear {2018}}%
}]{%
Thielsch.2018}
\APACinsertmetastar {%
Thielsch.2018}%
\begin{APACrefauthors}%
Thielsch, M\BPBI T.%
, Mee{\ss}en, S\BPBI M.%
\BCBL {}\ \BBA {} Hertel, G.%
\end{APACrefauthors}%
\unskip\
\newblock
\APACrefYearMonthDay{2018}{}{}.
\newblock
{\BBOQ}\APACrefatitle {Trust and distrust in information systems at the
  workplace} {Trust and distrust in information systems at the
  workplace}.{\BBCQ}
\newblock
\APACjournalVolNumPages{PeerJ}{6}{}{e5483}.
\newblock
\begin{APACrefDOI} \doi{10.7717/peerj.5483} \end{APACrefDOI}
\PrintBackRefs{\CurrentBib}

\bibitem [\protect \citeauthoryear {%
Tintarev%
\ \BBA {} Masthoff%
}{%
Tintarev%
\ \BBA {} Masthoff%
}{%
{\protect \APACyear {2012}}%
}]{%
Tintarev2012}
\APACinsertmetastar {%
Tintarev2012}%
\begin{APACrefauthors}%
Tintarev, N.%
\BCBT {}\ \BBA {} Masthoff, J.%
\end{APACrefauthors}%
\unskip\
\newblock
\APACrefYearMonthDay{2012}{Oct}{01}.
\newblock
{\BBOQ}\APACrefatitle {Evaluating the effectiveness of explanations for
  recommender systems} {Evaluating the effectiveness of explanations for
  recommender systems}.{\BBCQ}
\newblock
\APACjournalVolNumPages{User Modeling and User-Adapted
  Interaction}{22}{4}{399--439}.
\newblock
\begin{APACrefURL} \url{https://doi.org/10.1007/s11257-011-9117-5}
  \end{APACrefURL}
\newblock
\begin{APACrefDOI} \doi{10.1007/s11257-011-9117-5} \end{APACrefDOI}
\PrintBackRefs{\CurrentBib}

\bibitem [\protect \citeauthoryear {%
Toreini%
\ \protect \BOthers {.}}{%
Toreini%
\ \protect \BOthers {.}}{%
{\protect \APACyear {2020}}%
}]{%
Toreini2020}
\APACinsertmetastar {%
Toreini2020}%
\begin{APACrefauthors}%
Toreini, E.%
, Aitken, M.%
, Coopamootoo, K.%
, Elliott, K.%
, Zelaya, C\BPBI G.%
\BCBL {}\ \BBA {} Van~Moorsel, A.%
\end{APACrefauthors}%
\unskip\
\newblock
\APACrefYearMonthDay{2020}{}{}.
\newblock
{\BBOQ}\APACrefatitle {The relationship between trust in AI and trustworthy
  machine learning technologies} {The relationship between trust in ai and
  trustworthy machine learning technologies}.{\BBCQ}
\newblock
\BIn{} \APACrefbtitle {Proceedings of the 2020 conference on fairness,
  accountability, and transparency} {Proceedings of the 2020 conference on
  fairness, accountability, and transparency}\ (\BPGS\ 272--283).
\PrintBackRefs{\CurrentBib}

\bibitem [\protect \citeauthoryear {%
{van der Waa}%
, Nieuwburg%
, Cremers%
\BCBL {}\ \BBA {} Neerincx%
}{%
{van der Waa}%
\ \protect \BOthers {.}}{%
{\protect \APACyear {2021}}%
}]{%
vanderWaa2021}
\APACinsertmetastar {%
vanderWaa2021}%
\begin{APACrefauthors}%
{van der Waa}, J.%
, Nieuwburg, E.%
, Cremers, A.%
\BCBL {}\ \BBA {} Neerincx, M.%
\end{APACrefauthors}%
\unskip\
\newblock
\APACrefYearMonthDay{2021}{}{}.
\newblock
{\BBOQ}\APACrefatitle {Evaluating {XAI}: A comparison of rule-based and
  example-based explanations} {Evaluating {XAI}: A comparison of rule-based and
  example-based explanations}.{\BBCQ}
\newblock
\APACjournalVolNumPages{Artificial Intelligence}{291}{}{103404}.
\newblock
\begin{APACrefURL}
  \url{https://www.sciencedirect.com/science/article/pii/S0004370220301533}
  \end{APACrefURL}
\newblock
\begin{APACrefDOI} \doi{https://doi.org/10.1016/j.artint.2020.103404}
  \end{APACrefDOI}
\PrintBackRefs{\CurrentBib}

\bibitem [\protect \citeauthoryear {%
Vaske%
}{%
Vaske%
}{%
{\protect \APACyear {2016}}%
}]{%
Vaske.2016}
\APACinsertmetastar {%
Vaske.2016}%
\begin{APACrefauthors}%
Vaske, C.%
\end{APACrefauthors}%
\unskip\
\newblock
\APACrefYear{2016}.
\newblock
\APACrefbtitle {Misstrauen und Vertrauen} {Misstrauen und vertrauen}.
\newblock
\APACaddressPublisher{}{Universit{\"a}t Vechta}.
\PrintBackRefs{\CurrentBib}

\bibitem [\protect \citeauthoryear {%
Vilone%
\ \BBA {} Longo%
}{%
Vilone%
\ \BBA {} Longo%
}{%
{\protect \APACyear {2020}}%
}]{%
Vilone2020}
\APACinsertmetastar {%
Vilone2020}%
\begin{APACrefauthors}%
Vilone, G.%
\BCBT {}\ \BBA {} Longo, L.%
\end{APACrefauthors}%
\unskip\
\newblock
\APACrefYearMonthDay{2020}{10}{13}.
\newblock
{\BBOQ}\APACrefatitle {Explainable Artificial Intelligence: a Systematic
  Review} {Explainable artificial intelligence: a systematic review}.{\BBCQ}
\newblock

\PrintBackRefs{\CurrentBib}

\bibitem [\protect \citeauthoryear {%
Vilone%
\ \BBA {} Longo%
}{%
Vilone%
\ \BBA {} Longo%
}{%
{\protect \APACyear {2021}}%
}]{%
Vilone2021}
\APACinsertmetastar {%
Vilone2021}%
\begin{APACrefauthors}%
Vilone, G.%
\BCBT {}\ \BBA {} Longo, L.%
\end{APACrefauthors}%
\unskip\
\newblock
\APACrefYearMonthDay{2021}{dec}{25}.
\newblock
{\BBOQ}\APACrefatitle {Notions of explainability and evaluation approaches for
  explainable artificial intelligence} {Notions of explainability and
  evaluation approaches for explainable artificial intelligence}.{\BBCQ}
\newblock
\APACjournalVolNumPages{Information Fusion}{76}{}{89--106}.
\newblock
\begin{APACrefDOI} \doi{10.1016/j.inffus.2021.05.009} \end{APACrefDOI}
\PrintBackRefs{\CurrentBib}

\bibitem [\protect \citeauthoryear {%
Wang%
\ \BBA {} Yin%
}{%
Wang%
\ \BBA {} Yin%
}{%
{\protect \APACyear {2021}}%
}]{%
Wang2021}
\APACinsertmetastar {%
Wang2021}%
\begin{APACrefauthors}%
Wang, X.%
\BCBT {}\ \BBA {} Yin, M.%
\end{APACrefauthors}%
\unskip\
\newblock
\APACrefYearMonthDay{2021}{apr}{16}.
\newblock
{\BBOQ}\APACrefatitle {Are Explanations Helpful? A Comparative Study of the
  Effects of Explanations in {AI}-Assisted Decision-Making} {Are explanations
  helpful? a comparative study of the effects of explanations in {AI}-assisted
  decision-making}.{\BBCQ}
\newblock
\BIn{} \APACrefbtitle {26th International Conference on Intelligent User
  Interfaces.} {26th international conference on intelligent user interfaces.}
\newblock
\APACaddressPublisher{}{{ACM}}.
\newblock
\begin{APACrefDOI} \doi{10.1145/3397481.3450650} \end{APACrefDOI}
\PrintBackRefs{\CurrentBib}

\bibitem [\protect \citeauthoryear {%
Wang%
\ \BBA {} Yin%
}{%
Wang%
\ \BBA {} Yin%
}{%
{\protect \APACyear {2022}}%
}]{%
Wang2022}
\APACinsertmetastar {%
Wang2022}%
\begin{APACrefauthors}%
Wang, X.%
\BCBT {}\ \BBA {} Yin, M.%
\end{APACrefauthors}%
\unskip\
\newblock
\APACrefYearMonthDay{2022}{apr}{}.
\newblock
{\BBOQ}\APACrefatitle {Effects of Explanations in {AI}-Assisted Decision
  Making: Principles and Comparisons} {Effects of explanations in {AI}-assisted
  decision making: Principles and comparisons}.{\BBCQ}
\newblock
\APACjournalVolNumPages{{ACM} Transactions on Interactive Intelligent
  Systems}{}{}{}.
\newblock
\begin{APACrefDOI} \doi{10.1145/3519266} \end{APACrefDOI}
\PrintBackRefs{\CurrentBib}

\bibitem [\protect \citeauthoryear {%
Wang%
\ \BBA {} Yin%
}{%
Wang%
\ \BBA {} Yin%
}{%
{\protect \APACyear {2023}}%
}]{%
Wang2023}
\APACinsertmetastar {%
Wang2023}%
\begin{APACrefauthors}%
Wang, X.%
\BCBT {}\ \BBA {} Yin, M.%
\end{APACrefauthors}%
\unskip\
\newblock
\APACrefYearMonthDay{2023}{}{}.
\newblock
{\BBOQ}\APACrefatitle {Watch Out for Updates: Understanding the Effects of
  Model Explanation Updates in AI-Assisted Decision Making} {Watch out for
  updates: Understanding the effects of model explanation updates in
  ai-assisted decision making}.{\BBCQ}
\newblock
\BIn{} \APACrefbtitle {Proceedings of the 2023 CHI Conference on Human Factors
  in Computing Systems.} {Proceedings of the 2023 chi conference on human
  factors in computing systems.}
\newblock
\APACaddressPublisher{New York, NY, USA}{Association for Computing Machinery}.
\newblock
\begin{APACrefURL} \url{https://doi.org/10.1145/3544548.3581366}
  \end{APACrefURL}
\newblock
\begin{APACrefDOI} \doi{10.1145/3544548.3581366} \end{APACrefDOI}
\PrintBackRefs{\CurrentBib}

\bibitem [\protect \citeauthoryear {%
Wanner%
, Herm%
, Heinrich%
, Janiesch%
\BCBL {}\ \BBA {} Zschech%
}{%
Wanner%
\ \protect \BOthers {.}}{%
{\protect \APACyear {2020}}%
}]{%
Wanner2020}
\APACinsertmetastar {%
Wanner2020}%
\begin{APACrefauthors}%
Wanner, J.%
, Herm, L\BHBI V.%
, Heinrich, K.%
, Janiesch, C.%
\BCBL {}\ \BBA {} Zschech, P.%
\end{APACrefauthors}%
\unskip\
\newblock
\APACrefYearMonthDay{2020}{}{}.
\newblock
{\BBOQ}\APACrefatitle {White, Grey, Black: Effects of {XAI} Augmentation on the
  Confidence in AI-based Decision Support Systems Short Paper} {White, grey,
  black: Effects of {XAI} augmentation on the confidence in ai-based decision
  support systems short paper}.{\BBCQ}.
\PrintBackRefs{\CurrentBib}

\bibitem [\protect \citeauthoryear {%
F.~Yang%
, Huang%
, Scholtz%
\BCBL {}\ \BBA {} Arendt%
}{%
F.~Yang%
\ \protect \BOthers {.}}{%
{\protect \APACyear {2020}}%
}]{%
Yang2020}
\APACinsertmetastar {%
Yang2020}%
\begin{APACrefauthors}%
Yang, F.%
, Huang, Z.%
, Scholtz, J.%
\BCBL {}\ \BBA {} Arendt, D\BPBI L.%
\end{APACrefauthors}%
\unskip\
\newblock
\APACrefYearMonthDay{2020}{}{}.
\newblock
{\BBOQ}\APACrefatitle {How Do Visual Explanations Foster End Users' Appropriate
  Trust in Machine Learning?} {How do visual explanations foster end users'
  appropriate trust in machine learning?}{\BBCQ}
\newblock
\BIn{} \APACrefbtitle {Proceedings of the 25th International Conference on
  Intelligent User Interfaces} {Proceedings of the 25th international
  conference on intelligent user interfaces}\ (\BPG~189–201).
\newblock
\APACaddressPublisher{New York, NY, USA}{Association for Computing Machinery}.
\newblock
\begin{APACrefURL} \url{https://doi.org/10.1145/3377325.3377480}
  \end{APACrefURL}
\newblock
\begin{APACrefDOI} \doi{10.1145/3377325.3377480} \end{APACrefDOI}
\PrintBackRefs{\CurrentBib}

\bibitem [\protect \citeauthoryear {%
R.~Yang%
\ \BBA {} Wibowo%
}{%
R.~Yang%
\ \BBA {} Wibowo%
}{%
{\protect \APACyear {2022}}%
}]{%
Yang2022}
\APACinsertmetastar {%
Yang2022}%
\begin{APACrefauthors}%
Yang, R.%
\BCBT {}\ \BBA {} Wibowo, S.%
\end{APACrefauthors}%
\unskip\
\newblock
\APACrefYearMonthDay{2022}{Dec}{01}.
\newblock
{\BBOQ}\APACrefatitle {User trust in artificial intelligence: A comprehensive
  conceptual framework} {User trust in artificial intelligence: A comprehensive
  conceptual framework}.{\BBCQ}
\newblock
\APACjournalVolNumPages{Electronic Markets}{32}{4}{2053--2077}.
\newblock
\begin{APACrefURL} \url{https://doi.org/10.1007/s12525-022-00592-6}
  \end{APACrefURL}
\newblock
\begin{APACrefDOI} \doi{10.1007/s12525-022-00592-6} \end{APACrefDOI}
\PrintBackRefs{\CurrentBib}

\bibitem [\protect \citeauthoryear {%
Yu%
\ \protect \BOthers {.}}{%
Yu%
\ \protect \BOthers {.}}{%
{\protect \APACyear {2016}}%
}]{%
Yu2016}
\APACinsertmetastar {%
Yu2016}%
\begin{APACrefauthors}%
Yu, K.%
, Berkovsky, S.%
, Conway, D.%
, Taib, R.%
, Zhou, J.%
\BCBL {}\ \BBA {} Chen, F.%
\end{APACrefauthors}%
\unskip\
\newblock
\APACrefYearMonthDay{2016}{jul}{}.
\newblock
{\BBOQ}\APACrefatitle {Trust and Reliance Based on System Accuracy} {Trust and
  reliance based on system accuracy}.{\BBCQ}
\newblock
\BIn{} \APACrefbtitle {Proceedings of the 2016 Conference on User Modeling
  Adaptation and Personalization.} {Proceedings of the 2016 conference on user
  modeling adaptation and personalization.}
\newblock
\APACaddressPublisher{}{{ACM}}.
\newblock
\begin{APACrefDOI} \doi{10.1145/2930238.2930290} \end{APACrefDOI}
\PrintBackRefs{\CurrentBib}

\bibitem [\protect \citeauthoryear {%
Yu%
\ \protect \BOthers {.}}{%
Yu%
\ \protect \BOthers {.}}{%
{\protect \APACyear {2018}}%
}]{%
Yu2018}
\APACinsertmetastar {%
Yu2018}%
\begin{APACrefauthors}%
Yu, K.%
, Berkovsky, S.%
, Conway, D.%
, Taib, R.%
, Zhou, J.%
\BCBL {}\ \BBA {} Chen, F.%
\end{APACrefauthors}%
\unskip\
\newblock
\APACrefYearMonthDay{2018}{}{}.
\newblock
{\BBOQ}\APACrefatitle {Do I Trust a Machine? Differences in User Trust Based on
  System Performance} {Do i trust a machine? differences in user trust based on
  system performance}.{\BBCQ}
\newblock
\BIn{} \APACrefbtitle {Human and Machine Learning} {Human and machine
  learning}\ (\BPGS\ 245--264).
\newblock
\APACaddressPublisher{}{Springer International Publishing}.
\newblock
\begin{APACrefDOI} \doi{10.1007/978-3-319-90403-0\_12} \end{APACrefDOI}
\PrintBackRefs{\CurrentBib}

\bibitem [\protect \citeauthoryear {%
Yu%
\ \protect \BOthers {.}}{%
Yu%
\ \protect \BOthers {.}}{%
{\protect \APACyear {2017}}%
}]{%
Yu2017}
\APACinsertmetastar {%
Yu2017}%
\begin{APACrefauthors}%
Yu, K.%
, Berkovsky, S.%
, Taib, R.%
, Conway, D.%
, Zhou, J.%
\BCBL {}\ \BBA {} Chen, F.%
\end{APACrefauthors}%
\unskip\
\newblock
\APACrefYearMonthDay{2017}{mar}{}.
\newblock
{\BBOQ}\APACrefatitle {User Trust Dynamics: An Investigation Driven by
  Differences in System Performance} {User trust dynamics: An investigation
  driven by differences in system performance}.{\BBCQ}
\newblock
\BIn{} \APACrefbtitle {Proceedings of the 22nd International Conference on
  Intelligent User Interfaces.} {Proceedings of the 22nd international
  conference on intelligent user interfaces.}
\newblock
\APACaddressPublisher{}{{ACM}}.
\newblock
\begin{APACrefDOI} \doi{10.1145/3025171.3025219} \end{APACrefDOI}
\PrintBackRefs{\CurrentBib}

\bibitem [\protect \citeauthoryear {%
Yu%
, Berkovsky%
, Taib%
, Zhou%
\BCBL {}\ \BBA {} Chen%
}{%
Yu%
\ \protect \BOthers {.}}{%
{\protect \APACyear {2019}}%
}]{%
Yu2019}
\APACinsertmetastar {%
Yu2019}%
\begin{APACrefauthors}%
Yu, K.%
, Berkovsky, S.%
, Taib, R.%
, Zhou, J.%
\BCBL {}\ \BBA {} Chen, F.%
\end{APACrefauthors}%
\unskip\
\newblock
\APACrefYearMonthDay{2019}{}{}.
\newblock
{\BBOQ}\APACrefatitle {Do I Trust My Machine Teammate? An Investigation from
  Perception to Decision} {Do i trust my machine teammate? an investigation
  from perception to decision}.{\BBCQ}
\newblock
\BIn{} \APACrefbtitle {Proceedings of the 24th International Conference on
  Intelligent User Interfaces} {Proceedings of the 24th international
  conference on intelligent user interfaces}\ (\BPG~460–468).
\newblock
\APACaddressPublisher{New York, NY, USA}{Association for Computing Machinery}.
\newblock
\begin{APACrefURL} \url{https://doi.org/10.1145/3301275.3302277}
  \end{APACrefURL}
\newblock
\begin{APACrefDOI} \doi{10.1145/3301275.3302277} \end{APACrefDOI}
\PrintBackRefs{\CurrentBib}

\bibitem [\protect \citeauthoryear {%
Zhang%
, Liao%
\BCBL {}\ \BBA {} Bellamy%
}{%
Zhang%
\ \protect \BOthers {.}}{%
{\protect \APACyear {2020}}%
}]{%
Zhang2020}
\APACinsertmetastar {%
Zhang2020}%
\begin{APACrefauthors}%
Zhang, Y.%
, Liao, Q\BPBI V.%
\BCBL {}\ \BBA {} Bellamy, R\BPBI K\BPBI E.%
\end{APACrefauthors}%
\unskip\
\newblock
\APACrefYearMonthDay{2020}{jan}{}.
\newblock
{\BBOQ}\APACrefatitle {Effect of confidence and explanation on accuracy and
  trust calibration in {AI}-assisted decision making} {Effect of confidence and
  explanation on accuracy and trust calibration in {AI}-assisted decision
  making}.{\BBCQ}
\newblock
\BIn{} \APACrefbtitle {Proceedings of the 2020 Conference on Fairness,
  Accountability, and Transparency.} {Proceedings of the 2020 conference on
  fairness, accountability, and transparency.}
\newblock
\APACaddressPublisher{}{{ACM}}.
\newblock
\begin{APACrefDOI} \doi{10.1145/3351095.3372852} \end{APACrefDOI}
\PrintBackRefs{\CurrentBib}

\bibitem [\protect \citeauthoryear {%
Zhao%
, Benbasat%
\BCBL {}\ \BBA {} Cavusoglu%
}{%
Zhao%
\ \protect \BOthers {.}}{%
{\protect \APACyear {2019}}%
}]{%
Zhao2019}
\APACinsertmetastar {%
Zhao2019}%
\begin{APACrefauthors}%
Zhao, R.%
, Benbasat, I.%
\BCBL {}\ \BBA {} Cavusoglu, H.%
\end{APACrefauthors}%
\unskip\
\newblock
\APACrefYearMonthDay{2019}{}{}.
\newblock
{\BBOQ}\APACrefatitle {Do Users Always Want to Know More? Investigating the
  Relationship between System Transparency and Users' Trust in Advice-Giving
  Systems} {Do users always want to know more? investigating the relationship
  between system transparency and users' trust in advice-giving
  systems}.{\BBCQ}
\newblock
\BIn{} \APACrefbtitle {European Conference on Information Systems.} {European
  conference on information systems.}
\newblock
\begin{APACrefURL} \url{https://aisel.aisnet.org/ecis2019\_rip/42/}
  \end{APACrefURL}
\PrintBackRefs{\CurrentBib}

\bibitem [\protect \citeauthoryear {%
Zhou%
, Gandomi%
, Chen%
\BCBL {}\ \BBA {} Holzinger%
}{%
Zhou%
\ \protect \BOthers {.}}{%
{\protect \APACyear {2021}}%
}]{%
Zhou2021}
\APACinsertmetastar {%
Zhou2021}%
\begin{APACrefauthors}%
Zhou, J.%
, Gandomi, A\BPBI H.%
, Chen, F.%
\BCBL {}\ \BBA {} Holzinger, A.%
\end{APACrefauthors}%
\unskip\
\newblock
\APACrefYearMonthDay{2021}{}{}.
\newblock
{\BBOQ}\APACrefatitle {Evaluating the Quality of Machine Learning Explanations:
  A Survey on Methods and Metrics} {Evaluating the quality of machine learning
  explanations: A survey on methods and metrics}.{\BBCQ}.
\newblock
\begin{APACrefDOI} \doi{.org/10.3390/electronics10050593} \end{APACrefDOI}
\PrintBackRefs{\CurrentBib}

\bibitem [\protect \citeauthoryear {%
Zhou%
\ \protect \BOthers {.}}{%
Zhou%
\ \protect \BOthers {.}}{%
{\protect \APACyear {2019}}%
}]{%
Zhou2019}
\APACinsertmetastar {%
Zhou2019}%
\begin{APACrefauthors}%
Zhou, J.%
, Li, Z.%
, Hu, H.%
, Yu, K.%
, Chen, F.%
, Li, Z.%
\BCBL {}\ \BBA {} Wang, Y.%
\end{APACrefauthors}%
\unskip\
\newblock
\APACrefYearMonthDay{2019}{may}{}.
\newblock
{\BBOQ}\APACrefatitle {Effects of Influence on User Trust in Predictive
  Decision Making} {Effects of influence on user trust in predictive decision
  making}.{\BBCQ}
\newblock
\BIn{} \APACrefbtitle {Extended Abstracts of the 2019 {CHI} Conference on Human
  Factors in Computing Systems.} {Extended abstracts of the 2019 {CHI}
  conference on human factors in computing systems.}
\newblock
\APACaddressPublisher{}{{ACM}}.
\newblock
\begin{APACrefDOI} \doi{10.1145/3290607.3312962} \end{APACrefDOI}
\PrintBackRefs{\CurrentBib}

\bibitem [\protect \citeauthoryear {%
Zhou%
, Yu%
\BCBL {}\ \BBA {} Chen%
}{%
Zhou%
\ \protect \BOthers {.}}{%
{\protect \APACyear {2018}}%
}]{%
Zhou2018}
\APACinsertmetastar {%
Zhou2018}%
\begin{APACrefauthors}%
Zhou, J.%
, Yu, K.%
\BCBL {}\ \BBA {} Chen, F.%
\end{APACrefauthors}%
\unskip\
\newblock
\APACrefYearMonthDay{2018}{}{}.
\newblock
{\BBOQ}\APACrefatitle {Revealing User Confidence in Machine Learning-Based
  Decision Making} {Revealing user confidence in machine learning-based
  decision making}.{\BBCQ}
\newblock
\BIn{} J.~Zhou\ \BBA {} F.~Chen\ (\BEDS), \APACrefbtitle {Human and Machine
  Learning: Visible, Explainable, Trustworthy and Transparent} {Human and
  machine learning: Visible, explainable, trustworthy and transparent}\ (\BPGS\
  225--244).
\newblock
\APACaddressPublisher{Cham}{Springer International Publishing}.
\newblock
\begin{APACrefURL} \url{https://doi.org/10.1007/978-3-319-90403-0\_11}
  \end{APACrefURL}
\newblock
\begin{APACrefDOI} \doi{10.1007/978-3-319-90403-0\_11} \end{APACrefDOI}
\PrintBackRefs{\CurrentBib}

\end{thebibliography}

\end{document}